\documentclass[aps,prl,reprint,preprintnumbers,nofootinbib,superscriptaddress]{revtex4-2}
\input{header.tex}

\usepackage{fontawesome5}
\definecolor{color_git}{rgb}{0.098, 0.160, 0.345}
\newcommand{\gitlink}{\href{https://github.com/KensukeAkita/sterile-dm-lfa/tree/main}{\textsc{g}it\textsc{h}ub {\large\color{color_git}\faGithub}}}

\definecolor{forestgreen}{RGB}{34,139,34}

\begin{document}

\reportnum{-2}{CERN-TH-2025-149}

\title{Maximal parameter space of sterile neutrino dark matter with lepton asymmetries}

\author{Kensuke~Akita}
\email{kensuke@hep-th.phys.s.u-tokyo.ac.jp}
\affiliation{Department of Physics, The University of Tokyo, Bunkyo-ku, Tokyo 113-0033, Japan}
\author{Koichi Hamaguchi}
\email{hama@hep-th.phys.s.u-tokyo.ac.jp}
\affiliation{Department of Physics, The University of Tokyo, Bunkyo-ku, Tokyo 113-0033, Japan}
\author{Maksym~Ovchynnikov}
\email{maksym.ovchynnikov@cern.ch}
\affiliation{Theoretical Physics Department, CERN, 1211 Geneva 23, Switzerland}

\date{\today}

\begin{abstract}
We delineate the maximal parameter space of sterile neutrino dark matter in the presence of lepton flavor asymmetries. We focus on large flavor asymmetries with vanishing total lepton asymmetry, which are washed out by neutrino oscillations at MeV temperatures and hence are consistent with BBN and CMB constraints. We derive a semi-classical Boltzmann equation for sterile neutrinos applicable in this regime and validate it against quantum kinetic equations. For sterile neutrino masses up to 60 keV, the viable range of mixing angles extends by up to two orders of magnitude, with broad prospects for tests in forthcoming X-ray, CMB, and structure formation observations. We also release a public framework to compute the production of sterile neutrinos, and in particular their momentum distribution, enabling dedicated structure formation analyses.
\end{abstract}

\maketitle

{\bf\textit{Introduction.}}--- 
One of the outstanding issues in both particle physics and cosmology is the nature of dark matter (DM). Among many candidates for DM, the sterile neutrino, a putative massive fermion that is a singlet under the Standard Model (SM) gauge group, is an attractive candidate. 

Many mechanisms for producing the DM relic density of sterile neutrinos are testable by astrophysical observations. The simplest one, known as the Dodelson-Widrow (DW) mechanism~\cite{Dodelson:1993je}, produces sterile neutrinos through neutrino oscillations in the Early Universe, assuming the standard $\Lambda$CDM thermal history. Unfortunately, it is excluded by observations for X-rays \cite{Yuksel:2007xh,Boyarsky:2007ge,Horiuchi:2013noa,Perez:2016tcq,Ng:2019gch,Roach:2019ctw,Roach:2022lgo,Fischer:2022pse,Krivonos:2024yvm} and structure formation~\cite{Boyarsky:2009ix,Polisensky:2010rw,Baur:2017stq,Alvey:2020xsk,DES:2020fxi,Dekker:2021scf,Zelko:2022tgf}.
This promotes the search for alternative mechanisms such as resonant production in the presence of lepton asymmetry (Shi-Fuller mechanism)~\cite{Shi:1998km,Abazajian:2001nj,Kishimoto:2008ic,Laine:2008pg,Ghiglieri:2015jua,Venumadhav:2015pla,Kasai:2024diy,Shaposhnikov:2023hrx,Gorbunov:2025nqs,Vogel:2025aut}, production by the decays of scalars~\cite{Shaposhnikov:2006xi,Kusenko:2006rh,Petraki:2007gq,Khalil:2008kp,Merle:2013wta,Adulpravitchai:2014xna,Shuve:2014doa,Merle:2015oja,Roland:2014vba,Konig:2016dzg,Kusenko:2010ik}, thermal production with subsequent dilution~\cite{Bezrukov:2009th,Nemevsek:2012cd,Patwardhan:2015kga,Dror:2020jzy}, production in the presence of new active/sterile neutrino self-interactions~\cite{Hansen:2017rxr,Herms:2018ajr,DeGouvea:2019wpf,Kelly:2020pcy,Chichiri:2021wvw,Benso:2021hhh,Bringmann:2022aim,Fuyuto:2024oii,Benso:2024qrg,Fuller:2024noz,Dev:2025sah}.

The Shi-Fuller mechanism efficiently produces sterile neutrinos in the presence of primordial lepton flavor asymmetries $L_{\alpha}\equiv n_{L_\alpha}/s$, where $n_{L_\alpha}$ and $s$ are the net lepton number density and the entropy density, respectively, while $\alpha = e,\mu,\tau$ is the lepton flavor. The asymmetries induce a resonant enhancement of the mixing between sterile and active neutrinos. The mechanism is attractive because it does not modify the sterile neutrino interaction Lagrangian beyond the minimal model. However, to produce the sterile neutrino DM while evading all observational bounds, large asymmetries $|\sum_\alpha L_{\alpha}|\gtrsim 10^{-3}$ may be required. If surviving down to temperatures $T\lesssim 1\text{ MeV}$, such asymmetries would heavily modify Big Bang Nucleosynthesis (BBN) and Cosmic Microwave Background (CMB); hence, they are disfavored~\cite{Oldengott:2017tzj,Pitrou:2018cgg,Seto:2021tad,Matsumoto:2022tlr,Kumar:2022vee,Burns:2022hkq,Escudero:2022okz,Froustey:2024mgf,Domcke:2025lzg,Yanagisawa:2025mgx}.

The BBN/CMB constraints are, however, much weaker if the asymmetries were large at $T\gg 1 \text{ MeV}$, but later relaxed to zero. It is possible if the $L_{\alpha}$ pattern is such that the total asymmetry is tiny, $|\sum_{\alpha}L_{\alpha}|\lesssim 10^{-3}$. The relaxation may have happened because of active neutrino oscillations, which became effective at $T\simeq 15\text{ MeV}$ and mixed neutrinos of different flavors~\cite{Dolgov:2002ab,Pastor:2008ti,Mangano:2010ei,Castorina:2012md,Froustey:2021azz,Froustey:2024mgf,Domcke:2025lzg}. The recent study~\cite{Domcke:2025lzg} has demonstrated that in this scenario the allowed individual asymmetries may be as large as $|L_{\alpha}| \simeq 0.1$.

\begin{figure*}[t!]
\vskip-6pt
    \centering
    \includegraphics[width=1\columnwidth]{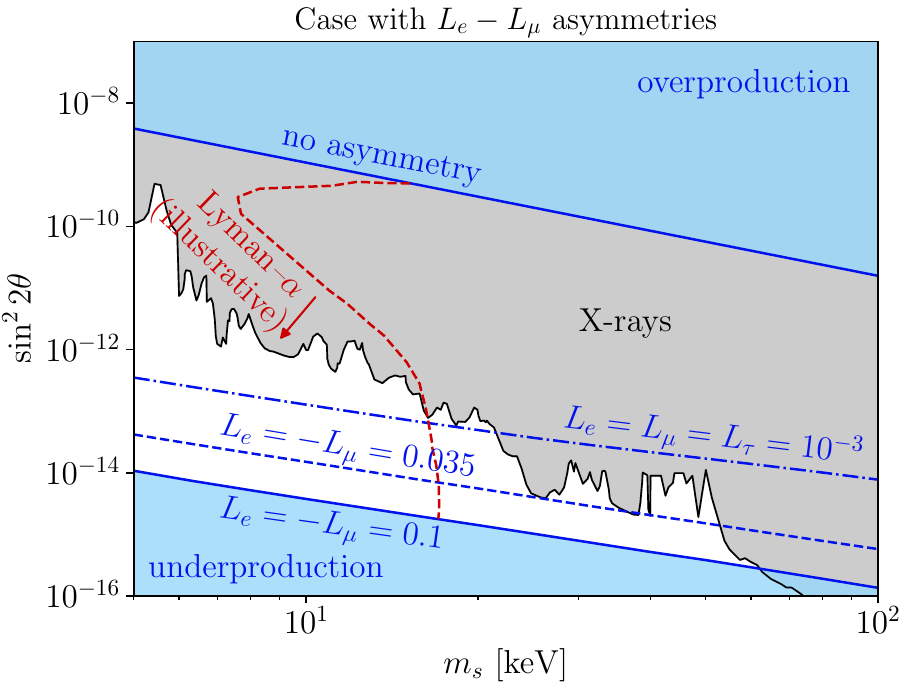}
    \includegraphics[width=1\columnwidth]{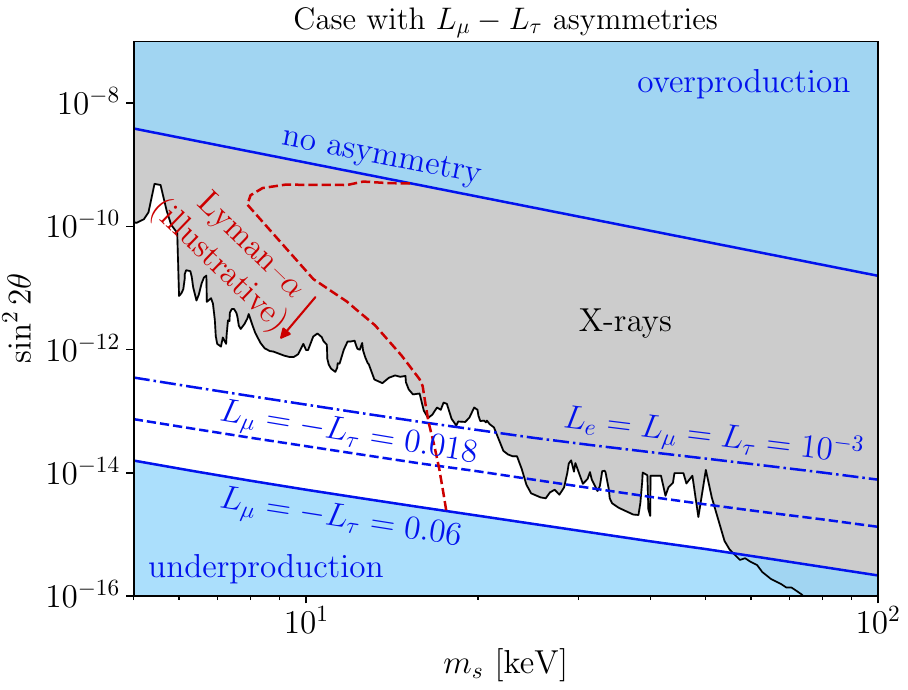}
    \vskip-6pt
    \caption{Parameter space of sterile neutrino mass $m_s$ and its mixing angle with active neutrinos $\sin^22\theta$. The white region is the domain where sterile neutrinos may be dark matter (DM) in the presence of lepton flavor asymmetry, as derived in this work. The lower boundary corresponds to the scenarios of net-zero lepton flavor asymmetry $\sum_{\alpha}L_{\alpha} = 0$. In the upper light blue-shaded region, sterile neutrinos are overproduced, whereas in the lower blue-shaded region, no reliable treatment of the impact of lepton flavor asymmetries on BBN and CMB exists (see text for details). {\it Left}: The case of $L_e=-L_\mu$ and $\nu_s$ mixing with $\nu_e$. {\it Right}: The case of $L_\mu=-L_\tau$ and $\nu_s$ mixing with $\nu_\mu$. The gray shaded region is excluded by X-ray observations~\cite{Yuksel:2007xh,Boyarsky:2007ge,Horiuchi:2013noa,Perez:2016tcq,Ng:2019gch,Roach:2019ctw,Roach:2022lgo, Fischer:2022pse,Krivonos:2024yvm}. The space below the contours for $L_e=-L_\mu= 0.035$ and $L_\mu=-L_\tau= 0.018$ (dashed lines) is the target sensitivity of the ongoing Simons Observatory~\cite{Domcke:2025lzg,SimonsObservatory:2018koc}, assuming normal neutrino mass ordering (see text for the details). The dot-dashed line explains all dark matter with $\nu_s$ mixing with $\nu_e$ and $L_e=L_\mu=L_\tau=10^{-3}$, which is the maximal magnitude for flavor-universal lepton asymmetry allowed by the BBN and CMB~\cite{Froustey:2024mgf}. Light sterile neutrinos may be in tension with structure formation observations. While we do not derive bounds from structure formation, we mark -- purely as an illustrative guide -- the region where it could become relevant, using the approximate one-parameter $m_{\text{WDM}}$ mapping of Lyman-$\alpha$ forest constraints from Refs.~\cite{Ballesteros:2020adh,Villasenor:2022aiy} (see text for details); it is shown to the left of the dashed red line.}
    \label{fig:parameterspace}
\end{figure*}

In this Letter, we show that large lepton flavor asymmetries with almost zero total lepton asymmetry open up a new parameter space for sterile neutrino dark matter, consistent with all current experimental bounds. To reveal this parameter space, we perform a precise calculation by solving the semi-classical kinetic equations for sterile neutrinos, improving and generalizing the approach developed by Ghiglieri and Laine~\cite{Ghiglieri:2015jua}, and Venumadhav et al.~\cite{Venumadhav:2015pla} to the large lepton asymmetries. Our results -- viable parameter space of sterile neutrino DM in terms of mass $m_{s}$ and coupling $\sin^{2}2\theta$, marginalized over lepton flavor asymmetries consistent with cosmological observations -- are summarized in Fig.~\ref{fig:parameterspace}. We find that the sterile neutrino DM and lepton asymmetries of interest may be comprehensively tested by future CMB, X-ray, and structure formation observations. 

While this work was in preparation, Ref.~\cite{Gorbunov:2025nqs} made a related qualitative remark for a single benchmark mass $m_s=7.1~\mathrm{keV}$. Here, we go substantially beyond by (i) delineating the maximally allowed region in the mass-coupling plane, (ii) marginalizing over cosmologically allowed lepton flavor asymmetries. In contrast to Ref.~\cite{Gorbunov:2025nqs}, we have found that increasing the lepton flavor asymmetry tends to produce a \emph{warmer} sterile neutrino spectrum, which is directly relevant for interpreting structure formation observations.

Large primordial lepton flavor asymmetries may be naturally generated in a class of new physics scenarios, in particular, by the Affleck-Dine (AD) mechanism~\cite{Affleck:1984fy,Dine:1995kz}. To motivate the scenario we consider, we propose the AD leptoflavorgenesis scenario, which can consistently generate large yet total-zero lepton flavor asymmetries. This is discussed in detail in our upcoming paper~\cite{Akita:2025zvq}.

Lastly, to reproduce our results and support further studies with any model featuring arbitrary lepton flavor asymmetry, we publicly release the framework \texttt{sterile-dm-lfa} on \gitlink~\cite{sterile-dm-lfa}: the \texttt{Python} code that traces the evolution of the Early Universe and production of sterile neutrinos using the unintegrated Boltzmann equations in full generality, and the \texttt{Mathematica} code that solves it quickly and accurately using the narrow width approximation in the case of a negligible back-reaction from sterile neutrinos on the lepton asymmetries.

Technical details and the extended discussion, such as the cross-checks and comparison with other studies (including Refs.~\cite{Gorbunov:2025nqs,Vogel:2025aut} that appeared while our study was in preparation), are provided in the Supplemental Material (SuM).


\bigskip

{\bf\textit{System of equations.}}---
First, we introduce the system of equations we will solve. 

The most reliable way to track the sterile neutrino production through the resonant oscillations would be solving the evolution equations for the density matrix of active and sterile neutrinos, called the quantum kinetic equations (QKEs) \cite{Harris:1980zi,Stodolsky:1986dx,Raffelt:1991ck,Raffelt:1992uj,Sigl:1993ctk}. However, fully solving QKEs is computationally expensive.

To save time, the semi-classical Boltzmann equation on the sterile neutrino distribution function $f_{\nu_{s}}(p,t)$ has been considered in the previous literature~\cite{Shi:1998km,Abazajian:2001nj,Kishimoto:2008ic,Laine:2008pg,Ghiglieri:2015jua,Venumadhav:2015pla,Kasai:2024diy,Shaposhnikov:2023hrx,Gorbunov:2025nqs}. Its central ingredient is how active-sterile oscillations are treated: they are averaged over the oscillation length. However, as discussed in Refs.~\cite{Shi:1998km,Abazajian:2001nj,Kishimoto:2008ic,Kasai:2024diy}, for very large lepton asymmetries, the resonance timescale becomes shorter than the oscillation timescale. As a result, sterile neutrinos might be produced through non-averaged oscillations. In that case, the semi-classical Boltzmann equation with averaged oscillations may no longer be valid.

To deal with this issue, we analytically generalize the Boltzmann equation to the case of non-averaged neutrino oscillations. The resulting equation is, in principle, applicable to arbitrary lepton asymmetries and in excellent agreement with the results of QKEs. Explicitly, for sterile neutrinos $\nu_s$ mixing with one flavor of active neutrinos $\nu_\alpha$ with the vacuum mixing angle $\theta$, it reads
\begin{align}
    &\left(\frac{\partial}{\partial t}-Hp\frac{\partial}{\partial p} \right)f_{\nu_s}(p,t) \nonumber \\
    &\ \ \ \ =\frac{\Gamma_{\alpha}(p,\mu)}{2}P_{\rm eff}(\nu_\alpha\rightarrow \nu_s)\left[f_{\nu_\alpha}(p,\mu)-f_{\nu_s}(p,t)\right]\,
    \label{eq:boltzmann-main}
\end{align}
Here, $H$ is the Hubble parameter, $p$ is momentum, $f_{\nu_\alpha}(p,\mu)$ is the Fermi-Dirac distribution for active neutrinos, with $\mu$ being chemical potential due to lepton asymmetries. Finally, $\Gamma_{\alpha}(p,\mu)$ is the interaction rate for active neutrinos, and $P_{\rm eff}(\nu_\alpha\rightarrow \nu_s)$ is the effective oscillation probability,
\begin{align}
    P_{\rm eff}(\nu_\alpha\rightarrow \nu_s)=\frac{1}{2}\frac{\Delta(p)^{2}\sin^22\theta}{\left[\Delta(p)\cos2\theta -V_\alpha(p,\mu)\right]^2+\left(\frac{\Gamma_{\alpha}}{2}\right)^2}\,,
    \label{eq:Peff}
\end{align}
with $\Delta(p)\equiv(m_s^2-m_{\nu_{\alpha}}^2)/(2p)\approx m_s^2/(2p)$ being the oscillation frequency in vacuum with sterile neutrino mass $m_s$, and $V_\alpha(p,\mu)$ the matter potential for active neutrinos. The evolution equation for sterile antineutrinos is the same as that for sterile neutrinos, with the replacement $\mu \rightarrow -\mu$.

$P_{\rm eff}(\nu_\alpha\!\to\!\nu_s)$ differs from the averaged-oscillation probability used in the literature (e.g.,~\cite{Abazajian:2001nj,Kishimoto:2008ic,Venumadhav:2015pla}) in that its denominator lacks the factor $\Delta^{2}\sin^{2}2\theta$. 
The reason of this absence is as follows: the oscillation probability is suppressed by the frequent collisions (quantum Zeno effects), $(\Gamma_\alpha/2)^{-1}$, which reset the active state; on the other hand, all neutrinos accumulated by free-streaming length, $(\Gamma_{\alpha}/2)^{-1}$, experience the resonance. Therefore, the resonance conversion is always computed over $|\Delta\cos\theta-V_\alpha|<\Gamma_{\alpha}/2$, and the averaged oscillation factor $\Delta^2\sin^2\theta$ does not appear (see the SuM for the details).

The condition $\Delta(p)\cos2\theta-V_\alpha(p,\mu)\approx0$ defines the domain of temperatures/momenta where sterile neutrinos may be resonantly produced. Using it, we may show that sterile neutrinos with mass $m_s\gtrsim 1\text{ keV}$ are produced before the flavor asymmetries were washed out by neutrino oscillations, which developed at temperatures $T<T_{\rm osc}\simeq 15\text{ MeV}$~\cite{Dolgov:2002ab,Pastor:2008ti,Mangano:2010ei,Castorina:2012md,Froustey:2021azz,Froustey:2024mgf,Domcke:2025lzg}. To this end, let us analytically estimate the resonance temperature $T_{\rm res}$. Neglecting less important $\mathcal{O}(G_{F}^{2})$ terms, the neutrino matter potential can be sketchy written as $V_{\alpha}\approx\sqrt{2}G_F L s$~\cite{Notzold:1987ik}, where $L$ is of the order of the maximal asymmetry in the system. Assuming $\theta \ll 1$, $T_{\rm res}$ is
\begin{align}
    &T_{\rm res}\sim 27\ {\rm MeV} \nonumber \\
    &\ \ \ \ \times \left(\frac{10.75}{g_{\ast}} \right)^{1/4}\left(\frac{3}{y} \right)^{1/4}\left(\frac{0.1}{L} \right)^{1/4}\left(\frac{m_s}{5\ {\rm keV}} \right)^{1/2},
    \label{eq:Tres}
\end{align}
where $g_\ast$ is the effective number of relativistic species, $p=yT$
and $y\simeq3$ corresponds to the average energy for neutrinos in thermal equilibrium. For $m_s\gtrsim 5\ {\rm keV}$ and $L\lesssim 0.1$, the resonance temperature is well above $T_{\rm osc}$.

We assume that all SM particles were in equilibrium at the epoch well before neutrino decoupling. Then, the remaining equations governing the evolution of the Universe are on the plasma temperature and particle-antiparticle asymmetries. The presence of lepton asymmetries as large as $L\simeq 0.1$ heavily changes both the neutrino production rate $\Gamma_{\alpha}$ and the thermodynamics of the Early Universe (i.e., the quantities $n, P,\rho,s$). In this work, we account for these effects for the first time in the context of sterile neutrino production.

The evolution of the plasma temperature is given by the continuity equation (the energy conservation law of the Universe)
\begin{align}
    \frac{d\rho}{dt}=-3H(\rho+P)\,,
\end{align}
with
\begin{align}
    \rho=\rho_{\rm SM}(T,\mu)+\rho_{\nu_s}(t)\,, \ \ P=P_{\rm SM}(T,\mu)+P_{\nu_s}(t)\,.
\end{align}
Here, $\rho_{\rm SM}$ and $P_{\rm SM}$ are the energy density and pressure in the SM, $\rho_{\nu_s}$ and $P_{\nu_s}$ are those for sterile neutrinos. The Hubble parameter is given by $H=\sqrt{8\pi\rho/(3m_P^{2})}$, with the Planck mass $m_P=1.22\times 10^{19}~{\rm GeV}$.

Let us now discuss particle-antiparticle asymmetries. We consider the scenario when, in the absence of sterile neutrinos, the lepton flavor asymmetries $L_{\alpha}$ are conserved at $T\gtrsim 15~{\rm MeV}$.\footnote{The scenario of varying $L_{\alpha}$ and its impact on the sterile neutrino abundance is discussed in the SuM.} Since neutrinos and charged leptons are in thermal and chemical equilibrium with the other species, $L_{\alpha}$s induce chemical potentials for neutrinos, $\mu_{\nu_\alpha}$, as well as baryon and electric charge chemical potentials, $\mu_B$ and $\mu_Q$. As the Universe cools, some of these particles become non-relativistic and annihilate into lighter species, thereby redistributing their asymmetry under the conserved asymmetries~\cite{Ghiglieri:2015jua,Venumadhav:2015pla,Wygas:2018otj,Middeldorf-Wygas:2020glx}.\footnote{Examples of the redistributing processes are $\nu_\alpha +l_{\beta}^-\leftrightarrow \nu_\beta +l_{\alpha}^-$, $\nu_\alpha + l_{\alpha}^+\leftrightarrow U + \bar{D}$ and $\nu_\alpha + \pi^- \leftrightarrow l_{\alpha}^- + \pi^0$, where $U$ and $D$ are quarks with electric charge of $+2/3$ and $-1/3$, while $\pi^-$ and $\pi^0$ are negatively charged and neutral pions.} This redistribution is characterized by five equations for the conservation of asymmetries,
\begin{align}
    \frac{\Delta n_{\nu_{\alpha}}+\Delta n_\alpha}{s}&=L_\alpha\ \ \ \ (\alpha=e,\ \mu,\ \tau)\,, \\
    \sum_i\frac{b_i\Delta n_{i}}{s}&=B,\, \\
    \sum_i\frac{q_i \Delta n_{i}}{s}&=0\,,
    \label{eq:as-red}
\end{align}
where $\Delta n_i=n_i-n_{\bar{i}}$ is the number density asymmetry, $s(T,\mu)$ is the total entropy density, $L_\alpha$ is the conserved lepton flavor asymmetries, and $B=8.75\times 10^{-11}$~\cite{Planck:2018vyg} is the observed baryon asymmetry. $b_i$ and $q_i$ are the baryon number and electric charge for species $i$. Solving these equations, we can trace the evolution of $\mu_{\nu_\alpha}$, $\mu_B$, and $\mu_Q$. In this part, we mainly follow Ref.~\cite{Middeldorf-Wygas:2020glx}. 
 
The presence of sterile neutrinos mixing with the lepton flavor $\alpha$ modifies the conservation law of the corresponding lepton asymmetry to $L_{\alpha}+L_{\nu_{s}} = \text{const}$, where $L_{\nu_{s}} \equiv (n_{\nu_{s}}-n_{\bar{\nu}_{s}})/s$. We have numerically incorporated this modification in the differential form: 
\begin{align}
    \frac{d}{dt}L_\alpha= -\frac{1}{s(T,\mu)}\int \frac{dp}{2\pi^{2}}\ p^2 \frac{d}{dt}\left[f_{\nu_s}(p,t)-f_{\bar{\nu}_s}(p,t)\right]\,.
    \label{eq:backreaction}
\end{align}
Assuming that sterile neutrinos populate all the dark matter, we may get the upper bound on the back-reaction: $|\Delta L_\alpha|\lesssim10^{-4}(5\text{ keV}/m_{s})$, which is negligible compared to the magnitude of the lepton asymmetries considered in our study. In addition, a nonzero total lepton asymmetry induced by $L_{\nu_{s}}$ is well below the upper bound on the flavor-universal asymmetry $L_{\alpha}\sim 10^{-3}$ imposed by BBN and CMB~\cite{Froustey:2024mgf}. As a result, we may safely neglect the impact of $\nu_{s}$-driven $L_{\alpha}$ non-conservation on the BBN and CMB, and use the results of~\cite{Domcke:2025lzg} for the evolution of the lepton asymmetries at $T<T_{\rm osc}$.

On the other hand, even tiny dynamical changes in $L_{\alpha}$ may influence the abundance of sterile neutrinos, because of the dependence of the resonance on the asymmetry. We have confirmed that the $L_\alpha$ evolution only changes the sterile neutrino abundance by $< 10\%$, though. 

\bigskip

{\bf\textit{Parameter space and limitations of our study.}}--- We begin by reviewing current observational bounds on primordial lepton–flavor asymmetries, which serve as input parameters in this study. BBN and CMB data constrain various linear combinations of $L_\alpha$, although certain directions remain weakly constrained~\cite{Froustey:2024mgf, Domcke:2025lzg, Domcke:2025jiy}.\footnote{Additional bounds may arise from overproduction of the baryon asymmetry via the chiral plasma instability~\cite{Domcke:2022uue}, but these can be evaded if the asymmetries are generated below $T\lesssim 10^6~\mathrm{GeV}$.}
In particular, Refs.~\cite{Domcke:2025lzg,Domcke:2025jiy} identify $L_e\simeq -L_\mu$ and $L_\mu\simeq -L_\tau$ as comparatively weakly constrained directions.
However, the numerical framework used in those analyses ceases to be reliable for $|L_\alpha|\gtrsim 0.1$.\footnote{Constraints are often quoted in terms of the degeneracy parameter $\xi_\alpha\equiv \mu_\alpha/T$. At leading order in $\mu_\alpha/T$, one has $\xi_\alpha=\frac{4\pi^2}{15}\,g_\ast L_\alpha \simeq 28.3\,L_\alpha$ for $g_\ast=10.75$ at $T\simeq 10~\mathrm{MeV}$.} For this reason, in Fig.~\ref{fig:parameterspace} we only marginalize over $|L_\alpha|\leq 0.1$, which we take as the domain within which current BBN and CMB constraints are robust.

Specifically, we consider two setups to explore the allowed sterile neutrino parameter space under the condition of zero total lepton asymmetry: (i) the least constrained direction $L_e= -L_\mu\leq 0.1$, with $\nu_{s}$ mixing with electron neutrinos, and (ii) $L_\mu= -L_\tau\leq 0.06$, with $\nu_{s}$ mixing with muon neutrinos.\footnote{Strictly speaking, a slightly misaligned direction of $L_e=-L_\mu$ is unconstrained by the observations~\cite{Domcke:2025lzg}. In practice, however, we find no changes in the parameter space between the exact and misaligned spaces.}

Now, let us summarize Fig.~\ref{fig:parameterspace}. The solid line defining the upper blue domain corresponds to the production of sterile neutrino DM in the case $L_{\alpha} = 0$. The parameter space below it, down to the lower blue region, is the domain of allowed masses and couplings in the presence of net-zero lepton flavor asymmetry, as derived in this work. It heavily extends over the parameter space for flavor-universal lepton asymmetry. To highlight this, we show the dot-dashed contour, denoting the lowest mixing angle explaining all dark matter with $\nu_s$ mixing with $\nu_e$ and  $L_e=L_\mu=L_\tau=10^{-3}$, which is the maximal magnitude for flavor-universal lepton asymmetry allowed by the BBN and CMB~\cite{Froustey:2024mgf}. 

From the figure, we see that the mixing pattern of $\nu_{s}$ affects the allowed parameter space only weakly. This is related to the fact that the lower boundary on the mixing angle of the sterile neutrino DM shows a monotonic dependence on both
the modulus of $L_{\alpha}$ and $m_{s}$. In particular, for the considered asymmetry patterns, the lower bound is found to scale as $\sin^{2}2\theta \propto |L_{\alpha}|^{-1.25}m_{s}^{-1.4}$.

Importantly, at the floor of allowed sterile neutrino couplings, $\nu_{s}$ particles are produced at temperatures well below the QCD transition, so the associated uncertainties in the dynamics of strongly interacting matter -- such as possible charged-pion condensation and the limited validity of the Taylor expansion of the QCD equation of state in $\mu_Q/T$~\cite{Middeldorf-Wygas:2020glx,Vovchenko:2020crk,Ferreira:2025zeu} -- do not affect our result. At higher couplings, production may happen during the transition, but the uncertainties are under control, within tens of percent influence on the abundances. See the detailed discussion, including the description of uncertainties, in the SuM.

Let us now summarize constraints. The gray region is excluded by X-ray observations~\cite{Yuksel:2007xh,Boyarsky:2007ge,Horiuchi:2013noa,Perez:2016tcq,Ng:2019gch,Roach:2019ctw,Roach:2022lgo,Fischer:2022pse,Krivonos:2024yvm}. Additionally, light sterile neutrinos, especially in the keV-mass range, may be constrained by small-scale structure probes; the reason is that such light sterile neutrinos carry non-negligible momenta, suppressing small-scale growth. Such probes include phase-space densities in DM subhalos~\cite{Alvey:2020xsk}, Lyman-$\alpha$ forest~\cite{Baur:2017stq}, Milky Way satellite counts~\cite{DES:2020fxi,Dekker:2021scf}, and strong lensing~\cite{Zelko:2022tgf} (see also~\cite{Gorbunov:2025nqs,Vogel:2025aut}).

To flag where structure formation observations may matter in the sterile neutrino parameter space, we use the approximate ``equivalent-$m_{\rm WDM}$'' mapping of Ref.~\cite{Ballesteros:2020adh} and mark its intersection with the 95\% C.L.\ thermal-relic limit $m_{\rm WDM}>3.1~\mathrm{keV}$ from Ref.~\cite{Villasenor:2022aiy} (see details in the SuM). The corresponding region is shown in Fig.~\ref{fig:parameterspace}. This construction is only meant as a rough guide: for non-thermal dark matter, Lyman-$\alpha$ constraints cannot be reduced to a single number, since they depend on the detailed shape of the small-scale cutoff and on astrophysical/systematic assumptions~\cite{Boyarsky:2008xj,Boyarsky:2008mt,Murgia:2018now,Baur:2017stq}. Consistently, several studies find that a $7.1~\mathrm{keV}$ resonantly produced sterile neutrino can remain compatible with Lyman-$\alpha$ and Milky Way satellite data, subject to modeling choices~\cite{Garzilli:2019qki,Enzi:2020ieg,Lovell:2023olv,Newton:2024jsy}.
Moreover, Ref.~\cite{Villasenor:2022aiy} reports large analysis-dependent uncertainties in the inferred thermal mass, underscoring that a dedicated study is required for robust bounds.
We therefore include the $m_{\rm WDM}$ curve only as a guide. Nevertheless, our framework provides sterile neutrino spectra enabling future dedicated structure formation analyses.

Combining these results, we find that sterile neutrinos with masses $m_s\lesssim 60\ {\rm keV}$ and large lepton flavor asymmetries may explain the observed dark matter abundance without conflicting with various cosmological and astrophysical constraints.

Extending the analyses of Refs.~\cite{Domcke:2025lzg,Domcke:2025jiy} to larger asymmetries, $|L_\alpha|>0.1$, could further lower the allowed floor on sterile-neutrino couplings. Given that the production of sterile neutrinos in these scenarios happens below the QCD transition, our method may be straightforwardly extended to such asymmetries; we leave this to future work.


\bigskip

{\bf\textit{Observational prospects.}}---
The future X-ray experiments eROSITA~\cite{Dekker:2021bos}, Athena~\cite{Ando:2021fhj}, and eXTP~\cite{Malyshev:2020hcc} will test the smaller mixing angle. In particular, eXTP may significantly improve the current X-ray constraints for $m_s\leq100~{\rm keV}$~\cite{Malyshev:2020hcc}. However, the systematic uncertainty of eXTP is not yet well known. Predictions for future observations of the structure formation are less clear, but these observations might test even heavier masses of sterile neutrino dark matter.

Lepton flavor asymmetries will be further tested by future CMB/BBN observations~\cite{Domcke:2025lzg}. For normal neutrino mass ordering, the Simons Observatory~\cite{SimonsObservatory:2018koc} can potentially test $L_e=-L_\mu\gtrsim 0.035$ and $L_\mu=-L_\tau\gtrsim 0.018$. For inverted neutrino mass ordering, it may not improve the current constraints. In the near future, the DESI and CMB observations would more precisely measure the sum of neutrino masses and thereby may explore neutrino mass ordering~\cite{DESI:2024mwx,Wang:2024hen,Naredo-Tuero:2024sgf,Jiang:2024viw,RoyChoudhury:2024wri,RoyChoudhury:2025dhe,Chebat:2025kes,Du:2025xes}.


\bigskip

{\bf\textit{Origin of lepton flavor asymmetries.}}---
In this scenario, large lepton flavor asymmetries with zero total lepton asymmetry must exist in the Early Universe prior to the sterile neutrino production.

There are potentially several mechanisms for generating lepton flavor asymmetries in the Early Universe~\cite{Kuzmin:1987wn,March-Russell:1999hpw,Chiba:2003vp,Takahashi:2003db,Shu:2006mm,Gu:2010dg,Mukaida:2021sgv,Akita:2025zvq}. In particular, the Affleck-Dine (AD) mechanism~\cite{Affleck:1984fy,Dine:1995kz} is one of the promising scenarios that naturally explains the origins of large lepton flavor asymmetries. In the supersymmetric theory, there are flat directions in the scalar potential that have no total lepton charge but lepton flavor charge (e.g., $Q\bar{u}L_\alpha\bar{e}_\beta$). 
Scalar fields can have large expectation values along the flat direction, generating large lepton flavor asymmetries.  

Large lepton flavor asymmetries can also offer a natural explanation of the small baryon asymmetry. The sphaleron process preserves the quantity $(B/3-L_\alpha)$ for each lepton flavor $\alpha$ but violates $B+L$, where $L=\sum_{\alpha}L_\alpha$.
If lepton flavor asymmetries with $B-L=0$ are generated before the sphaleron transition, the conversion from the flavor asymmetries to baryon asymmetry cancels out, but not completely~\cite{March-Russell:1999hpw,Khlebnikov:1988sr, Laine:1999wv}, suggesting that large lepton flavor asymmetries may underlie the observed small baryon asymmetry.
In addition, in the AD mechanism, these scalar fields can deform into non-topological solitons called Q-balls, where the $B+L$ charge is protected from the sphaleron processes, thus allowing even larger lepton asymmetries without overproducing the baryon asymmetry.

The AD mechanism with the $Q\bar{u}L_\alpha \bar{e}_\beta$ direction can successfully produce large yet total-zero lepton asymmetries at $T\gtrsim 1~{\rm GeV}$, which is much higher than the resonance temperature $T_{\rm res}$, Eq.~\eqref{eq:Tres}, where sterile neutrinos are resonantly produced. Detailed discussions are devoted to the upcoming paper~\cite{Akita:2025zvq}.


\bigskip

{\bf\textit{Conclusion.}}---
keV-mass sterile neutrinos were proposed as one of the excellent dark matter (DM) candidates. However, the minimal realizations of sterile neutrino DM are severely constrained by the observations of X-rays and structure formation.  

We have demonstrated that lepton flavor asymmetries with zero total lepton asymmetry, loosely constrained by the current BBN and CMB observations, open up a new parameter space for sterile neutrino DM. To this end, we have performed a precise calculation of the resonant production of sterile neutrinos, including, for the first time, the impact of the large lepton asymmetries on the neutrino interaction rates and thermodynamics of the Universe. The semi-classical Boltzmann equations with non-averaged neutrino oscillations we used are confirmed by quantum kinetic equations for various regimes where oscillations may or may not be averaged over the oscillation length. 

Widely marginalizing over the lepton flavor asymmetries, we have estimated the maximal parameter space to explain all DM in the mass range $m_{s}\lesssim 60\text{ keV}$, and found that the allowed sterile neutrino squared couplings may cover up to two orders of magnitude, depending on mass. The newly opened parameter space is highly testable by future X-ray, structure formation, and CMB searches.


\bigskip

{\bf\textit{Acknowledgements.}}---
The authors thank Miguel Escudero for contributing to the early stage of this project, carefully reading the manuscript, and for valuable discussions regarding lepton asymmetries, and Yotam Soreq for useful discussions. This work has received support from JSPS Grant-in-Aid for Scientific Research KAKENHI Grant No.~24KJ0060, No.~24H02244, and No.~24K07041, and from the European Union's Horizon Europe research and innovation programme under the Marie Sklodowska-Curie grant agreement No~101204216.

\bibliography{bib.bib}

\clearpage
\onecolumngrid
\twocolumngrid 
\onecolumngrid 

\appendix

\setcounter{equation}{0}
\setcounter{figure}{0}
\setcounter{table}{0}
\setcounter{page}{1}
\makeatletter
\renewcommand{\thefigure}{S\arabic{figure}}
\renewcommand{\thepage}{S\arabic{page}}
\setcounter{secnumdepth}{2}
\renewcommand{\thesection}{\Alph{section}}

\begin{center}
\textbf{\large Supplemental Material for\\[0.5ex]
Maximal parameter space of sterile neutrino dark matter with lepton asymmetries}
\end{center}

\begin{center}
\text{\large Kensuke Akita,\ \ Koichi Hamaguchi,\ \ Maksym Ovchynnikov}
\end{center}


We summarize some details about a precise calculation of the resonant production of sterile neutrino dark matter with large lepton flavor asymmetries.

The central part of the approach is constructing the effective Boltzmann equation for arbitrarily large lepton asymmetries. As discussed in Refs.~\cite{Shi:1998km,Abazajian:2001nj,Kishimoto:2008ic,Kasai:2024diy}, for large lepton asymmetries of $|L_\alpha|\gtrsim 5\times 10^{-3}$, the resonance time scale is shorter than the neutrino oscillation length. In such a case, the sterile neutrino production through neutrino oscillations may be significantly suppressed by short resonance times. However, simultaneously, we find an enhancement factor due to the fact that neutrinos produced cumulatively over the mean free path can experience the resonance. If a typical resonance scale is shorter than the neutrino mean free path, the resonance scale is effectively extended to the mean free path. Thus, even at very short resonance times, sterile neutrinos can still be efficiently produced through sizable active-sterile neutrino oscillations. Being embedded in the Boltzmann formalism, this description is confirmed by Quantum Kinetic Equations (QKEs).
We also fully include chemical potentials due to large lepton asymmetries in the Boltzmann system for sterile neutrino production for the first time.

The Supplemental Material is organized as follows. In Section~\ref{sec:precise_cal}, we outline the system of equations governing the evolution of the Universe with large lepton asymmetries and sterile neutrinos.
First, we show the evolution equations for the system of active and sterile neutrinos, and the electroweak plasma:
Subsection~\ref{sec:kinetic} for the full kinetic equations for sterile neutrinos, Subsection~\ref{sec:evolution_asy} for the equations for asymmetry that include effects of sterile neutrino production and the asymmetries redistribution, Subsection~\ref{sec:evolution_temperature} for the equation for the plasma temperature. Here, we include chemical potentials due to large lepton flavor asymmetries in the neutrino interaction rate and thermodynamic quantities to estimate the sterile neutrino production for the first time. 
In Subsection~\ref{app:QCD-transition}, we explain our treatment of the quark-hadron transition.
In Subsection~\ref{sec:neutrino-interaction-rate}, we discuss the neutrino interaction rate, including chemical potentials due to large lepton asymmetries.
We found effects of chemical potentials on the interaction rate are sizable, as shown in Figure~\ref{fig:Rate_01}. In subsection~\ref{app:QCD-uncertainties-impact}, we summarize the impact of uncertainties and approximations in governing the dynamics of strongly interacting particles on the production of sterile neutrinos.
In Subsection~\ref{sec:results_sterile_nu}, we present some detailed results for sterile neutrino momentum distributions and the evolution of lepton asymmetries. 
In Subsection~\ref{sec:numerical_calculation}, we present details of our numerical setup and discuss the numerical convergence in our results.

Section~\ref{app:analytic} is devoted to revisiting the analytic behavior of the resonant production of sterile neutrinos and constructing semi-classical kinetic equations with non-averaged oscillations for sterile neutrinos.

Section~\ref{app:backreaction} qualitatively discusses the impact of sterile neutrinos on the evolution of the asymmetry $L_{\alpha}$.

In Section~\ref{app:cross-checks}, we numerically test the results of the constructed effective kinetic equations, comparing them with those obtained using the QKEs, reproducing thermodynamic identities, and checking against a simplified approach to solve the sterile neutrino Boltzmann equation from Section~\ref{app:simplified-approach}. These results are in excellent agreement, as shown in Figures~\ref{fig:QKEs} and~\ref{fig:deviation}.

Section~\ref{app:simplified-approach} is devoted to solving the sterile neutrino Boltzmann equation under the assumptions of negligible back-reaction and narrow width approximation for the oscillation probability, which allows for quickly and accurately scanning the parameter space in the case of large lepton asymmetries. 

Section~\ref{app:sterile-spectrum} discusses warming of sterile neutrino population in the presence of lepton flavor asymmetries, and also summarizes our method to calculate the indicative Lyman-$\alpha$ domain we show in Fig.~\ref{fig:parameterspace}.

Finally, in Section~\ref{app:compare}, we compare our study with the relevant previous literature.

Together with the study, we provided two codes to study the production of sterile neutrinos. The first code utilizes the comoving momentum binning approach to solve the Boltzmann equation. The second code uses the narrow width approximation in the case of a negligible back-reaction on sterile neutrinos. The codes \texttt{sterile-dm-lfa} are available on \gitlink~\cite{sterile-dm-lfa}.

\clearpage

\section{A precise calculation of the sterile neutrino production with large lepton asymmetries}
\label{sec:precise_cal}


\subsection{Kinetic equation for sterile neutrinos}
\label{sec:kinetic}

The semi-classical kinetic equation with neutrino oscillations, called the semi-classical Boltzmann equation, for the sterile neutrino momentum distribution $f_{\nu_s}$ mixing with one flavor of active neutrinos $\nu_\alpha$ in the homogeneous and isotropic Universe is~\cite{Venumadhav:2015pla}
\begin{align}
    \left(\frac{\partial}{\partial t}-Hp\frac{\partial}{\partial p} \right)f_{\nu_s}(p,t) &= \frac{1}{2p}\sum_{\nu_s+a+\cdots\rightarrow i+\cdots} \int \frac{d^3p_a}{(2\pi)^32E_a}\cdots\frac{d^3p_i}{(2\pi)^32E_i}\cdots(2\pi)^4\delta^{4}(p+p_a+\cdots-p_i-\cdots)
    \nonumber \\
    &\times \frac{1}{2}\biggl[P_{\rm eff}(\nu_\alpha\rightarrow \nu_s; p,\mu)\,(1-f_{\nu_s})\sum|\mathcal{M}|^2_{i+\cdots\rightarrow\nu_\alpha+a+\cdots}f_i\cdots(1\mp f_a)(1-f_{\nu_\alpha})\cdots \nonumber \\
    &\hspace{2.9cm}-P_{\rm eff}(\nu_s\rightarrow \nu_\alpha; p,\mu)\,f_{\nu_s}(1-f_{\nu_\alpha})\sum |\mathcal{M}|^2_{\nu_\alpha+a+\cdots\rightarrow i+\cdots}f_a\cdots (1\mp f_i)\cdots\biggr].
\end{align}
Here, $f_{\nu_\alpha}$ is the distribution function for active neutrinos mixing with $\nu_s$. We assume that all the SM particles, including active neutrinos, are in thermal and chemical equilibrium; i.e., for fermionic/bosonic SM particles, the distribution follows the Fermi-Dirac/Bose-Einstein shape:
\begin{align}
    f_{i}(E,\mu_i)=\frac{1}{e^{(E-\mu_i)/T}\pm1}\,,
\end{align}
where $E$ is the energy and $\mu_i$ is the chemical potential of species $i$. The factors $(1- f)$ and $(1+ f)$ are the Pauli blocking and Bose enhancement factors, respectively. 
$\sum|\mathcal{M}|^2$ denotes the squared matrix elements of the process producing or depleting the active neutrino $\nu_{\alpha}$, summed over spins of all particles (see Section~\ref{sec:neutrino-interaction-rate} for the discussion). Finally, $P_{\rm eff}(\nu_\alpha\rightarrow\nu_s; p,\mu)$ is the effective oscillation probability~\eqref{eq:Peff} for $\nu_{\alpha}\leftrightarrow \nu_{s}$,
\begin{align}
    P_{\rm eff}(\nu_\alpha\rightarrow \nu_s; p,\mu)=\frac{1}{2}\,
    \frac{\Delta(p)^2\sin^22\theta}{ \left[\Delta(p)\cos2\theta -V_\alpha(p,\mu)\right]^2+\left(\frac{\Gamma_\alpha(p,\mu)}{2}\right)^2}\,.
    \label{eq:prob-eff}
\end{align}
In this expression, $\Delta(p) = \big(m_{s}^{2}-m_{
\nu_{\alpha}}^{2}\big)/(2p) \approx m_{s}^{2}/(2p)$ is the vacuum oscillation frequency for sterile neutrino mass $m_{s}$.

In the short resonance regime, frequent collisions reset the active state (quantum Zeno damping), yet neutrinos produced over a mean free path still traverse the resonance; accounting for this accumulation over the free-stream time yields the effective in-medium conversion probability above with the usual Lorentzian line shape, \emph{without} the extra $\Delta^2\sin^22\theta$ term in the denominator that appears under naive time averaging. We derive the effective oscillation probability~\eqref{eq:prob-eff} in Sec.~\ref{app:analytic} and validate the resulting Boltzmann equation against quantum kinetic equations (QKEs) in both averaged and non-averaged regimes, finding an excellent agreement (see Sec.~\ref{app:QKEs}).



Here, we have introduced the interaction rate for active neutrinos,
\begin{align}
    \Gamma_\alpha(p,\mu)&=\frac{1}{2p}\sum_{\nu_s+a+\cdots\rightarrow i+\cdots} \int \frac{d^3p_a}{(2\pi)^32E_a}\cdots\frac{d^3p_i}{(2\pi)^32E_i}\cdots(2\pi)^4\delta^{4}(p+p_a+\cdots-p_i-\cdots) \nonumber \\
    &\ \ \ \ \times \sum |\mathcal{M}|^2_{\nu_\alpha+a+\cdots\rightarrow i+\cdots}f_a\cdots (1\mp f_i)\cdots,
    \label{Nu_rate}
\end{align}
and $V_{\alpha}$ is the matter potential for active neutrinos induced by their forward scattering with thermal plasma background~\cite{Notzold:1987ik},
\begin{align}
    V_{\alpha}(p,\mu) &= \sqrt{2}G_F\left[\Delta n_{\nu_\alpha} + \Delta n_\alpha + \sum_{\beta=e,\mu,\tau}\left[\Delta n_{\nu_\beta} + \left(-\frac{1}{2} + 2\sin^2\theta_W \right)\Delta n_\beta \right] - \frac{1}{2}\Delta n_B + (1-2\sin^2\theta_W)\Delta n_Q \right] \nonumber \\
&\ \ \ \ -\frac{8\sqrt{2}G_Fp}{3}\left[\frac{\rho_{\nu_\alpha}}{m_Z^2} + \frac{\rho_\alpha}{m_W^2} \right],
\label{eq:Valpha}
\end{align}
where $\theta_W$ is the weak mixing angle, $m_{Z,W}$ is the mass of the weak gauge bosons. 

Let us discuss the structure of the potential in more detail. It contains two groups of summands: $\mathcal{O}(G_{F})$, coming from the particle-antiparticle asymmetries $\Delta n_{i} \equiv n_{i} - n_{\bar{i}}$, and $\mathcal{O}(G_{F}^{2})$, which as well exists in the system with zero asymmetries. 
$\Delta n_{\nu_\alpha}, \Delta n_{\alpha},\Delta n_B,\Delta n_Q$ are the asymmetries of neutrino and charged lepton of the flavor $\alpha$, baryon, and electric charge densities. $\rho_{\nu_\alpha}$ and $\rho_\alpha$ are the energy densities of the neutrinos and the charged lepton. The baryon number asymmetry is small compared to lepton asymmetries of interest~\cite{Planck:2018vyg}, and we neglect it. $\Delta n_{\nu_\alpha},\ \Delta n_\alpha$ and $\Delta n_Q$ are redistributed under the conserved baryon and lepton flavor asymmetries and the charge neutrality at $T\gtrsim 15~{\rm MeV}$ as the Universe cools, as discussed in Refs.~\cite{Ghiglieri:2015jua,Venumadhav:2015pla,Wygas:2018otj,Middeldorf-Wygas:2020glx} and in the next section.

We can simplify the kinetic equation using the detailed balance to equate the forward and backward reaction rates of active neutrinos. The resultant kinetic equation is, assuming $1-f_{\nu_s}\simeq 1$ and $f_{\nu_s}(1-f_{\nu_\alpha})\simeq f_{\nu_s}\ll f_{\nu_\alpha}$,
\begin{align}
    &\left(\frac{\partial}{\partial t}-Hp\frac{\partial}{\partial p} \right)f_{\nu_s}(p,t) = \frac{\Gamma_\alpha(p,\mu)}{2}P_{\rm eff}(\nu_\alpha\rightarrow \nu_s)\left[f_{\nu_\alpha}(p,\mu)-f_{\nu_s}(p,t) \right].
    \label{System_eq_1}
\end{align}

Similarly, the semi-classical Boltzmann equation for sterile antineutrinos is
\begin{align}
    &\left(\frac{\partial}{\partial t}-Hp\frac{\partial}{\partial p} \right)f_{\bar{\nu}_s}(p,t) = \frac{\bar{\Gamma}_{\alpha}(p,\mu)}{2}P_{\rm eff}(\bar{\nu}_\alpha\rightarrow\bar{ \nu}_s)\left[f_{\bar{\nu}_\alpha}(p,\mu)-f_{\bar{\nu}_s}(p,t) \right],
    \label{System_eq_1_2}
\end{align}
where $\bar{\Gamma}_{\alpha}(p,\mu)=\Gamma_{\alpha}(p,-\mu)$, $P_{\rm eff}(\bar{\nu}_\alpha\rightarrow\bar{\nu}_s;p,\mu)=P_{\rm eff}(\nu_\alpha\rightarrow\nu_s;p,-\mu)$ and $f_{\bar{\nu}_\alpha}(p,\mu)=f_{\nu_\alpha}(p,-\mu)$. In particular, $P_{\rm eff}(\bar{\nu}_\alpha\rightarrow\bar{\nu}_s)$ is explicitly given by
\begin{align}
    P_{\rm eff}(\bar{\nu}_\alpha\rightarrow\bar{\nu}_s)=\frac{1}{2}\frac{\Delta(p)^2\sin^22\theta}{ \left[\Delta(p)\cos2\theta -\bar{V}_\alpha\right]^2+\left(\frac{\bar{\Gamma}_{\alpha}}{2}\right)^2}
\end{align}
with
\begin{align}
    \bar{V}_{\alpha}(p,\mu) &= -\sqrt{2}G_F\left[\Delta n_{\nu_\alpha} + \Delta n_\alpha + \sum_{\beta=e,\mu,\tau}\left[\Delta n_{\nu_\beta} + \left(-\frac{1}{2} + 2\sin^2\theta_W \right)\Delta n_\beta \right] - \frac{1}{2}\Delta n_B + (1-2\sin^2\theta_W)\Delta n_Q \right] \nonumber \\
&\ \ \ \ -\frac{8\sqrt{2}G_Fp}{3}\left[\frac{\rho_{\nu_\alpha}}{m_Z^2} + \frac{\rho_\alpha}{m_W^2} \right].
\label{eq:Vantialpha}
\end{align}


\subsection{Time evolution of asymmetries and chemical potentials}
\label{sec:evolution_asy}

At $T\gtrsim 15\ {\rm MeV}$, neutrino oscillations are negligible. Then, lepton flavor asymmetries, baryon asymmetry, and electric charge are conserved.
However, the weak interaction processes couple neutrinos, charged leptons, and quarks/hadrons. 
As the Universe cools and associated particles become non-relativistic, each particle asymmetry is redistributed under the conserved asymmetries~\cite{Ghiglieri:2015jua,Venumadhav:2015pla,Wygas:2018otj,Middeldorf-Wygas:2020glx} through, e.g., $\nu_\alpha +\beta^-\leftrightarrow \nu_\beta +\alpha^-$, $\nu_\alpha + \alpha^+\leftrightarrow a + \bar{b}$ and $\nu_\alpha + \pi^- \leftrightarrow \alpha^- + \pi^0$, where $a$ and $\bar{b}$ are quarks with electric charge of $+2/3$ and $-1/3$.
For example, the ratio for electron neutrino asymmetry and electron asymmetry can be changed within the conserved electron flavor asymmetry as the Universe cools.

The equations for the conserved lepton flavor, baryon, and electric charge asymmetries are
\begin{align}
    \frac{\Delta n_{\nu_{\alpha}}+\Delta n_\alpha}{s}&=L_\alpha\ \ \ \ (\alpha=e,\ \mu,\ \tau), \label{eq_redist_1} \\
    \sum_i\frac{b_i\Delta n_{i}}{s}&=B \label{eq_redist_2} \\
    \sum_i\frac{q_i \Delta n_{i}}{s}&=0,
    \label{eq_redist_3}
\end{align}
where $s(T,\mu)$ is the entropy density, $L_\alpha$ is an input lepton flavor asymmetry and $B=8.75\times10^{-11}$~\cite{Planck:2018vyg} is the observed baryon asymmetry. $b_i$ and $q_i$ are the baryon number and electric charge for species $i$. 
We assume that all reactions in the SM are in thermal and chemical equilibrium.
For photons and gluons, their chemical potentials are zero, $\mu_\gamma=\mu_g=0$, through a process such as $\alpha^++\alpha^-\leftrightarrow \gamma$.
The chemical equilibrium for the processes such as $\alpha^++\alpha^-\leftrightarrow \gamma$ and $\nu_\alpha+\alpha^+\leftrightarrow a + \bar{b}$ also implies
\begin{align}
    &\mu_i = -\mu_{\bar{i}},\\
    &\mu_{\nu_\alpha}-\mu_{\alpha^-}-\mu_Q=0,
\end{align}
where $\mu_i$ and $\mu_{\bar{i}}$ are chemical potentials for species $i$ and their antiparticles.
When one solves five Eqs.~\eqref{eq_redist_1}--\eqref{eq_redist_3} at a fixed $L_\alpha$ with a fixed $T$, one can find all of the chemical potentials in the plasma, $\mu_{\nu_\alpha}$, $\mu_B$, and $\mu_Q$. Then, one can compute thermodynamic quantities in thermal and chemical equilibrium.

The evolution of lepton flavor asymmetry mixed with the sterile state is also related to the evolution of sterile neutrinos because the resonance induced by lepton asymmetries produces either only $\nu_{s}$ or $\bar{\nu}_{s}$. The evolution equation for the lepton flavor asymmetry is, using the modified conservation law of the lepton asymmetry of $L_\alpha+L_{\nu_s}={\rm const}$,
\begin{align}
    \frac{d}{dt}L_\alpha= -\frac{1}{s}\int \frac{dp}{2\pi^2}\ p^2 \frac{d}{dt}\left[f_{\nu_s}(p,t)-f_{\bar{\nu}_s}(p,t)\right].
    \label{System_eq_2}
\end{align}

We calculate the number and entropy density, including chemical potentials. We estimate the total entropy as
\begin{align}
    s(T,\mu)=s_0(T)+\delta s(T,\mu),
\end{align}
where
\begin{align}
    s_0(T)=\frac{2\pi^2}{45}g_{\ast,s}(T)T^3,\ \ \ \ \ \ \delta s(T,\mu)=s(T,\mu)-s(T,0).
    \label{entropy_density}
\end{align}
$g_{\ast,s}$ is the effective number of relativistic degrees of freedom for the entropy density (with no asymmetries); to describe its behavior with $T$, we use the fitting formula in Ref.~\cite{Saikawa:2018rcs}.
For leptons, we estimate their contributions to $\delta s(T,\mu)$ in the ideal gas limit. 
For the quark-hadron sector, we have to estimate them, accounting for the confinement of quarks into hadrons around $T_{\rm QCD}\sim 150\ {\rm MeV}$. Details of this treatment are discussed in Section~\ref{app:QCD-transition}.


\subsection{Time-temperature relation}
\label{sec:evolution_temperature}
The evolution of temperature is characterized by the continuity equation (the energy conservation law),
\begin{align}
    \frac{d\rho}{dt}=-3H(\rho+P),
\end{align}
where $\rho$ and $P$ are the total energy density and pressure, which are decomposed as
\begin{align}
    \rho=\rho_{\rm SM}+\rho_{\nu_s},\ \ \ \ P=P_{\rm SM}+P_{\nu_s}.
\end{align}
Here, $\rho_{\rm SM}(T,\mu)$ and $P_{\rm SM}(T,\mu)$ are the quantities for the SM and $\rho_{\nu_s}$ and $P_{\nu_S}$ are the quantities for sterile neutrinos. $H$ is the Hubble parameter, which is calculated as,
\begin{align}
    H=\sqrt{\frac{8\pi}{3m_{P}^2}\rho}\simeq\sqrt{\frac{8\pi}{3m_{ P}^2}\rho_{\rm SM}},
\end{align}
where $m_P=1.22\times10^{19}\ {\rm GeV}$ is the Planck mass.
The continuity equation is rewritten as
\begin{align}
    \frac{dT}{dt}=-\frac{3H(\rho_{\rm SM}+P_{\rm SM})+\delta\rho_{\nu_s}/\delta t}{d\rho_{\rm SM}/dT},
    \label{System_eq_3}
\end{align}
where $\delta\rho_{\nu_s}/\delta t$ is
\begin{align}
\frac{\delta \rho_{\nu_s}}{\delta t}\equiv \frac{1}{2\pi^2}\int dp\ p^2\sqrt{p^2+m_s^2}\frac{d}{dt}\left[f_{\nu_s}(p,t)+f_{\bar{\nu}_s}(p,t) \right].
\end{align}
In practice, at temperatures $T\gtrsim 15\text{ MeV}$, $\rho_{\nu_s}$ gives a negligible contribution to the energy density; we include it for completeness.

$\rho_{\rm SM}(T,\mu)$ and $d\rho_{\rm SM}/dT$ are calculated as
\begin{align}
    \rho_{\rm SM}(T,\mu)&=\rho_{\rm SM,0}(T)+\delta \rho_{\rm SM}(T,\mu), \\
    \frac{d\rho_{\rm SM}}{dT}&=\frac{d\rho_{\rm SM,0}}{dT}+\frac{d(\delta\rho_{\rm SM})}{dT}
\end{align}
where 
\begin{align}
    \rho_{\rm SM,0}(T)=\frac{\pi^2}{30}g_{\ast,\rho}(T)T^4,\ \ \ \ \ \ \delta\rho_{\rm SM}(T,\mu)=\rho_{\rm SM}(T,\mu) - \rho_{\rm SM}(T,0).
\end{align}
$g_{\ast,\rho}$ is the effective number of relativistic degrees of freedom for the energy density (with no asymmetries), obtained using the fitting formula in Ref.~\cite{Saikawa:2018rcs}. $\delta \rho_{\rm SM}$ is calculated in the same way as $\delta s(T,\mu)$ in Eq.~\eqref{entropy_density}.
We estimate $d(\delta \rho_{\rm SM})/dT$ numerically,
\begin{align}
    \frac{d(\delta \rho_{\rm SM})}{dT}=\frac{\delta \rho_{\rm SM}\left(T', \mu(T')\right)-\delta\rho_{\rm SM}\left(T,\mu(T)\right)}{T'-T},
\end{align}
where $T'=T+h$. We set $h=10^{-5}~{\rm MeV}$ and confirm that the results are numerically well converged. 

Finally, we calculate the pressure $P_{\rm SM}$ using the standard relation, $P=\rho -Ts+\sum_{i}\mu_{i} n_{i}$, where $s$ and $n$ are calculated as in Sections~\ref{sec:evolution_asy} and~\ref{app:QCD-transition}.


\subsection{Quark-hadron transition in thermodynamic quantities}
\label{app:QCD-transition}

The resonance production of sterile neutrinos may occur around the QCD transition, $T_{\rm QCD}\sim 150\ {\rm MeV}$, as shown in Eq.~\eqref{eq:Tres}.
We need to calculate the thermodynamic quantities in the quark-hadron sector, accounting for the confinement of quarks into hadrons, to estimate the production of sterile neutrinos. 
For this purpose, we divide the temperature range into three different regimes as follows, based on Refs.~\cite{Venumadhav:2015pla,Wygas:2018otj,Middeldorf-Wygas:2020glx}. 
At $T\gg T_{\rm QCD}$, the QCD thermodynamic quantities consist of quarks and gluons, which are computed using the standard perturbative approach. At $T\simeq T_{\rm QCD}$, we compute them with the help of the results of the lattice calculations. At $T\ll T_{\rm QCD}$, they consist of hadrons and we compute them using the hadron resonance gas (HRG) model~\cite{Hagedorn:1984hz,Megias:2012hk}, where all known hadrons are approximated as ideal gas particles. For the actual computation, we divide three temperature ranges: $T<120\ {\rm MeV}$, $120\ {\rm MeV}<T<280\ {\rm MeV}$, and $280\ {\rm MeV}<T$.
\\
\\
{\bf 1.~Quark-gluon plasma at $T\gg T_{\rm QCD}$}

We treat quarks and gluons as an ideal gas at leading order, including chemical potentials of quarks. Due to sizable strong gluonic interactions, we include finite temperature QCD corrections perturbatively, following Refs.~\cite{Kajantie:1997tt,Laine:2006cp,Saikawa:2018rcs}.
For the total entropy, energy density, and pressure in the SM, we use the fitting formula for the effective numbers of degrees of relativistic freedom in Ref.~\cite{Saikawa:2018rcs}. For number density asymmetries in Section~\ref{sec:evolution_asy}, we calculate the QCD corrections up to $\mathcal{O}(g_s^2)$, where $g_s$ is the strong gauge coupling constant, following Ref.~\cite{Kajantie:1997tt,Laine:2006cp}. We neglect chemical potentials in the QCD corrections because effects of chemical potentials on thermodynamic quantities may still be subdominant for the resonance temperature of $T_{\rm res}\gg T_{\rm QCD}$ (see Eq.~\eqref{eq:Tres}).
\\
\\
{\bf 2.~QCD phase at $T\simeq T_{\rm QCD}$}

Quarks start to confine into hadrons, and the perturbative QCD approach is no longer valid. Following Refs.~\cite{Venumadhav:2015pla,Wygas:2018otj,Middeldorf-Wygas:2020glx}, we perform a Taylor expansion of the QCD pressure with chemical potential and use the susceptibilities $\chi^{ab}$ at zero chemical potentials studied in the lattice QCD calculations~\cite{Borsanyi:2011sw,HotQCD:2012fhj} to obtain the value of the QCD pressure,
\begin{align}
    p^{\rm QCD}(T,\mu)=p^{\rm QCD}(T,0)+\frac{1}{2}\mu_a\chi^{ab}(T)\mu_b + \mathcal{O}(\mu^4),
\end{align}
where $a,b$ are implicitly summed over $(a,b=B,Q)$ and 
\begin{align}
    \chi^{ab}(T)=\frac{\partial^2 p^{\rm QCD}}{\partial\mu_a\partial\mu_b}\Biggl|_{\mu_a,\mu_b=0}.
\end{align}
Such an expansion is originally used to avoid the sign problem in lattice QCD calculations with non-zero chemical potentials for heavy ion collision experiments \cite{Stephanov:2004wx,Philipsen:2007rm,Schmidt:2006kfp}. We do believe that such an approximation of the QCD transition accurately captures the main aspects important for understanding the sterile neutrino production -- the dynamics of chemical potentials. Less trivial properties of the transition (see, e.g.,~\cite{Bonanno:2023thi}) are unimportant as they do not enter the neutrino potential.

The off-diagonal term characterizes the fluctuations of the conserved baryon number and electric charge.
The pressure and energy density for the QCD plasma are given by the QCD partition function $Z^{\rm QCD}$,
\begin{align}
    p^{\rm QCD}(T,\mu)&=\frac{T}{V}\ln Z^{\rm QCD}(V,T,\mu_B,\mu_Q), \\
    \rho^{\rm QCD}(T,\mu)&=\frac{T^2}{V}\frac{\partial \ln Z^{\rm QCD}}{\partial T}=-p^{\rm QCD}+T\frac{\partial p^{\rm QCD}}{\partial T},
\end{align}
where $V$ is the volume of the system.
The baryon and electric charge number densities in the QCD plasma are
\begin{align}
    n_a^{\rm QCD}(T,\mu)=\frac{\partial p^{\rm QCD}(T,\mu)}{\partial\mu_a}=\chi^{ab}\mu_b+\mathcal{O}(\mu^3).
\end{align}
The entropy density of the QCD plasma is, using the standard relation, $\rho^{\rm QCD}=Ts^{\rm QCD}-p^{\rm QCD}+\mu_an_a^{\rm QCD}$,
\begin{align}
    Ts^{\rm QCD}(T,\mu)=T\frac{\partial p^{\rm QCD}}{\partial T}-\mu_a\frac{\partial p^{\rm QCD}}{\partial \mu_a}.
\end{align}
The energy and entropy densities can be written as
\begin{align}
    \rho^{\rm QCD}(T,\mu)-\rho^{\rm QCD}(T,0)&=\frac{1}{2}\left(-\chi^{ab} + T\frac{d\chi^{ab}}{dT}\right)\mu_a\mu_b  \\
    s^{\rm QCD}(T,\mu)-s^{\rm QCD}(T,0)&=\left(\frac{1}{2}\frac{d\chi^{ab}}{dT}-\frac{1}{T}\chi^{ab}\right)\mu_a\mu_b.
\end{align}
Let us outline the limitations of this description~\cite{Middeldorf-Wygas:2020glx,Vovchenko:2020crk}:
\begin{itemize}
\item The approach is only valid for 
\begin{equation}
\frac{1}{2}\mu_a\chi^{ab}(T)\mu_b/P^{\rm QCD}(T,0) = \chi \ll 1,
\label{eq:EOS-perturbativity}
\end{equation}
where $P^{\rm QCD}(T,0)$ is the QCD pressure at zero asymmetry. For the flavor direction $L_\mu=-L_\tau,\ L_e=0$, conservatively requiring $\chi = 0.1$, this limits the asymmetries by $|L_\mu|\lesssim 0.04$ \cite{Middeldorf-Wygas:2020glx}. 
\item Additionally, the approach breaks down if a condensation of charged pions occurs. Ref.~\cite{Middeldorf-Wygas:2020glx} has shown that this happens in the scenarios $L_{\mu}= -L_{\tau}$ with $L_{\mu}\gtrsim 0.06$.
\item For the orthogonal direction $L_e=-L_\mu,\ L_\tau=0$, the pion condensation criterion is not triggered; moreover, $\mu_Q\simeq 0$ and $\mu_B\ll \mu_Q$ in the temperature range of interest \cite{Vovchenko:2020crk}. As a result, the susceptibility expansion remains controlled even for much larger $L_{\alpha}>0.1$.
\end{itemize}
For the susceptibilities from lattice QCD, we use the continuum-extrapolated (2+1)-flavor results of the Wuppertal-Budapest (WB) collaboration~\cite{Borsanyi:2011sw} and HotQCD~\cite{HotQCD:2012fhj}, following Ref.~\cite{Venumadhav:2015pla}. These results agree with the HRG description at temperatures $T\lesssim 150~\mathrm{MeV}$ and with perturbative QCD by $250~\mathrm{MeV}\lesssim T \lesssim 300~\mathrm{MeV}$. In particular, using (2+1+1)-flavor susceptibilities~\cite{Bazavov:2014yba,Mukherjee:2015mxc}, Ref.~\cite{Wygas:2018otj} found neutral-lepton and electric-charge chemical potentials essentially identical to the (2+1)-flavor case for $120~\mathrm{MeV}<T<280~\mathrm{MeV}$, which is the window we utilize for the lattice description. Moreover, $\mu_B<\mu_Q$ is numerically negligible for our purposes~\cite{Wygas:2018otj,Middeldorf-Wygas:2020glx}.

WB and HotQCD agree well despite different staggered actions. In the continuum limit, the susceptibilities $\chi^{BB}$ and $\chi^{QQ}$ are consistent at the few-tens-percent level, while $\chi^{BQ}$ is less precise but mutually consistent~\cite{Borsanyi:2011sw,HotQCD:2012fhj} (see also~\cite{Venumadhav:2015pla}). Given the typical hierarchy $\chi^{QQ}\gg\chi^{BB}>\chi^{BQ}$ and $\mu_Q>\mu_B$~\cite{Wygas:2018otj,Middeldorf-Wygas:2020glx}, uncertainties in $\chi^{BQ}$ and in $\mu_B$ have a negligible impact on sterile neutrino production.
\\
\\
{\bf 3.~Hadron resonance gas at $T\ll T_{\rm QCD}$}

We assume an ideal gas of hadron resonances. We take into account only pions, protons, and neutrons.


\subsection{Neutrino interaction rate}
\label{sec:neutrino-interaction-rate}

We calculate the weak interaction rate for active neutrinos $\Gamma_\alpha$ in Eq.~\eqref{Nu_rate} with the approximation of the four Fermi-interaction processes, integrating out the massive $Z^0$ and $W^\pm$ gauge bosons.
We consider neutrino interactions with leptons and quarks/hadrons, accounting for the confinement of quarks into hadrons.
Our calculation method follows Ref.~\cite{Venumadhav:2015pla}, but we include effects of chemical potentials due to large asymmetries, that is, effects of degenerate particles in $\Gamma_\alpha(p,\mu)$ for the first time. 

Neutrinos may interact with leptons and strongly interacting particles, such as quarks and their bound states, hadrons. We consider all flavors of neutrinos and charged leptons; the interactions may be easily obtained using the Lagrangian of weak interactions. The strongly interacting sector is non-trivial: at large temperatures $T\gg \Lambda_{\text{QCD}}$, it comprises quarks and gluons, whereas at lower temperatures, we deal with hadrons.

To handle this complexity, we first define the confinement domain by $150\ {\rm MeV}<T<250\ {\rm MeV}$. Above, we only consider quarks, while well below, at $T< 120 \text{ MeV}$, we formulate the interactions in terms of hadrons. However, even at temperatures below the confinement scale, quarks may still contribute to neutrino reactions for large momentum transfer $Q \gg \Lambda_{\text{QCD}}$. To account for this, we follow Ref.~\cite{Venumadhav:2015pla} and consider the contribution of free quarks instead of hadrons at the center of mass energy of $> 4\pi f_{\pi}\sim 1\ {\rm GeV}$ when $T<150\ {\rm MeV}$, for the processes that go via the s-channel. 

Unfortunately, during the confinement stage ($150\ {\rm MeV}<T<250\ {\rm MeV}$), there is no reliable way to calculate the neutrino interaction rate. As a workaround, we interpolate the rate in between with the help of the cubic spline method. We motivate this simple treatment by the fact that the rates do not have resonant enhancement (which cannot be captured by interpolation).

For neutrino-quark interactions, we incorporate $u,d,c, s$ quarks and neglect the heavier $b,t$.
For neutrino-hadron interactions, we incorporate the contributions of $\pi$, $K$, $\eta$, $\rho$, $\omega$-mesons and neglect other hadrons, as they mostly have $m/T \gg 1$ at $T<150\ {\rm MeV}$ and hence negligibly contribute to the production. 
We use three-flavor chiral perturbation theory ($\chi$PT) \cite{Srednicki_2007} to obtain the meson currents coupled to the $Z^0$ and $W^\pm$ bosons and their contributions to neutrino interaction rates.
The relevant processes and their squared matrix elements are reported in Refs.~\cite{Venumadhav:2015pla,Dolgov:1997mb}.

The most time-consuming part of the calculations of $\Gamma_\alpha$ is to perform the integration in Eq.~\eqref{Nu_rate}. To reduce the dimensionality of the integration, we use some analytic methods proposed in Refs.~\cite{Venumadhav:2015pla,Dolgov:1997mb,Hannestad:1995rs}. For the $2\leftrightarrow2$ and 3-body fusion processes involving leptons and quarks, such as $\nu_e^++e^+\rightarrow u+\bar{d}$, the 9 integrals can be analytically reduced to 2 integrals, following Ref.~\cite{Dolgov:1997mb}.
For the $2\leftrightarrow2$ and 3-body fusion processes involving mesons such as $\nu_e+\pi^0\rightarrow e^- + \pi^+$, the 9 integrals can be analytically reduced to 3 integrals, following Ref.~\cite{Hannestad:1995rs}.
For the 2-body fusion processes such as $\nu_\mu + \mu^+\rightarrow \pi^+$, the 6 integrals can be analytically performed, using the method discussed in Appendix~B.2 in Ref.~\cite{Venumadhav:2015pla}.

Figure~\ref{fig:Rates_Noasy} shows the neutrino interaction rate $\Gamma_\alpha$ in Eq.~\eqref{Nu_rate} for various temperatures and momenta with no lepton asymmetry. The results in figure~\ref{fig:Rates_Noasy} are in excellent agreement with figure~9 in Ref.~\cite{Venumadhav:2015pla} and very good agreement with Ref.~\cite{Asaka:2006nq}.
$\Gamma_\tau$ is much smaller than $\Gamma_{e,\mu}$ at $T\lesssim m_\tau \simeq 2\ {\rm GeV}$ because charged current processes involving taus are suppressed below this temperature. 
$\Gamma_e$ is slightly larger for high momentum and slightly smaller for low momentum than $\Gamma_\mu$. 
Charged current processes involving muons are suppressed at $T\lesssim m_\mu\simeq 100\ {\rm MeV}$. On the other hand, the process of $\nu_\mu + \mu^+\rightarrow \pi^+$ compensates for $\Gamma_\mu$ with low momentum while the helicity-suppressed process of $\nu_e + e^+\rightarrow \pi^+$ less compensates for $\Gamma_e$. Bumps in $\Gamma_{\mu}$ with $p/T=0.25$ and $p/T=1$ around $T\sim 10\text{--}100\ {\rm MeV}$ in figure~\ref{fig:Rates_Noasy} stems from $\nu_\mu + \mu^+\rightarrow \pi^+$. We also observe a small bump in $\Gamma_e$ with $p/T=0.25$ due to $\nu_e + e^+\rightarrow \pi^+$

Figure~\ref{fig:Rate_01} shows $\Gamma_\alpha$ for electron neutrinos with lepton flavor asymmetries of $L_e=-L_\mu=0.1,~L_\tau=0$.
Here we neglect a small reduction in lepton asymmetries due to the production of sterile neutrino DM (see Figure~\ref{fig:Lepton_evo} in the next subsection~\ref{sec:results_sterile_nu} and Section~\ref{app:backreaction}).
For small momentum, the interaction rates are considerably suppressed due to the Pauli blocking effects. On the other hand, for larger momentum, the interaction rates are significantly enhanced because particles with larger momentum are populated in the thermal plasma due to the Pauli exclusion principle.  
To precisely estimate the abundance of sterile neutrinos with very large lepton asymmetries, it is very important to include chemical potentials in the neutrino interaction rate.

\begin{figure*}
    \centering
    \includegraphics[width=0.32\textwidth]{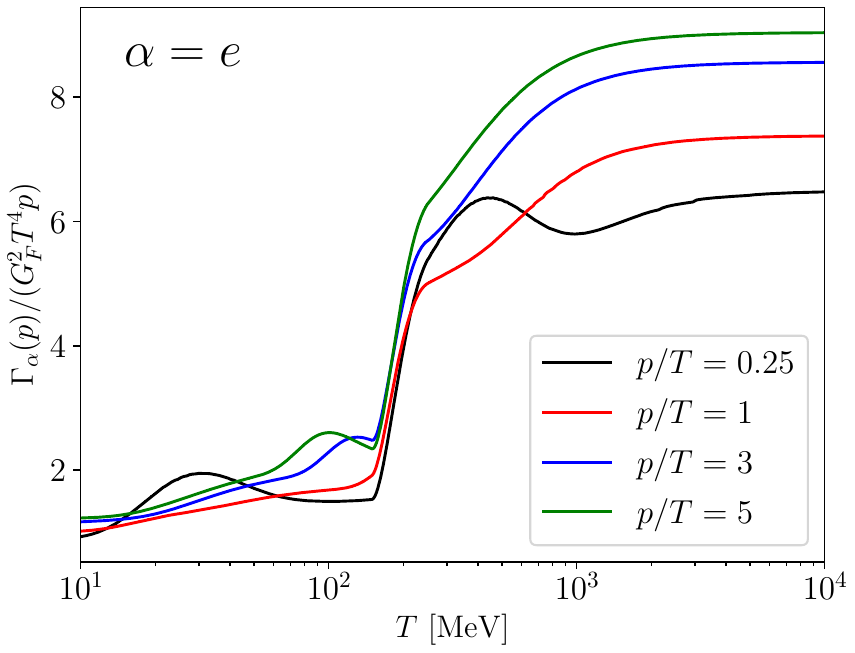}
    \includegraphics[width=0.32\textwidth]{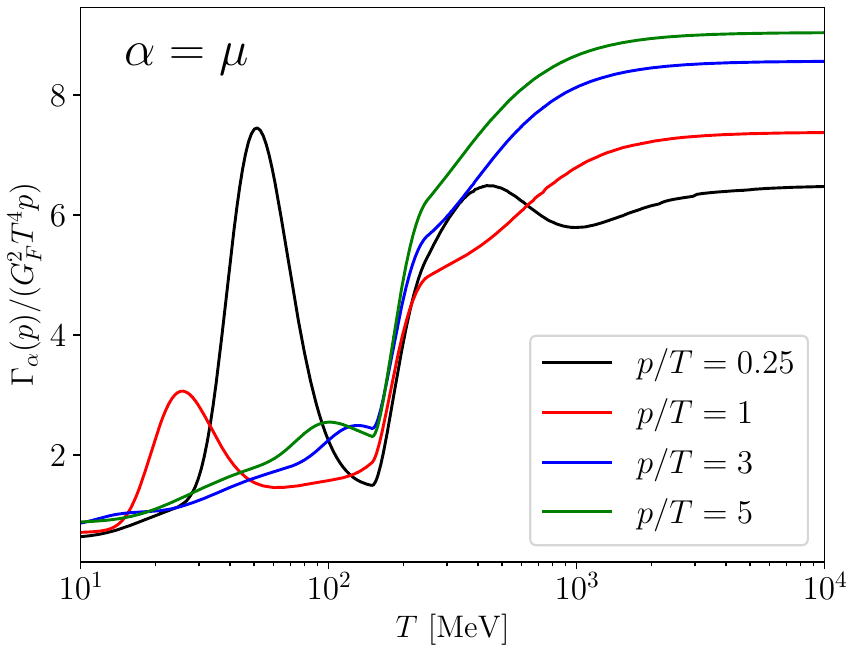}
    \includegraphics[width=0.32\textwidth]{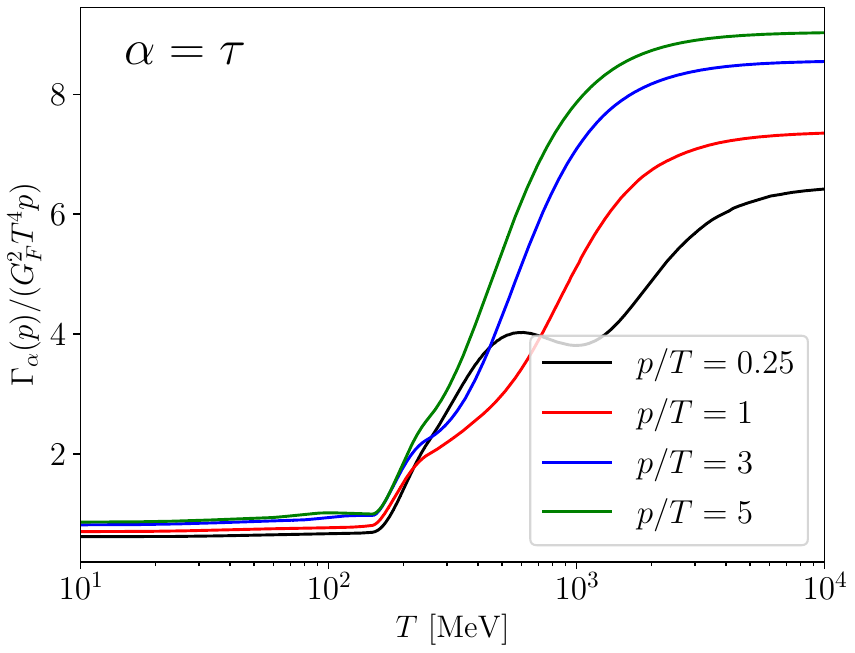}
    \vskip-12pt
    \caption{Neutrino interaction rate, Eq.~\eqref{Nu_rate}, with no lepton asymmetries for various temperature and momenta. The results are in excellent agreement with figure~9 in Ref.~\cite{Venumadhav:2015pla} and very good agreement with Ref.~\cite{Asaka:2006nq}.}
    \label{fig:Rates_Noasy}
\vskip 2pt
\includegraphics[width=0.5\textwidth]{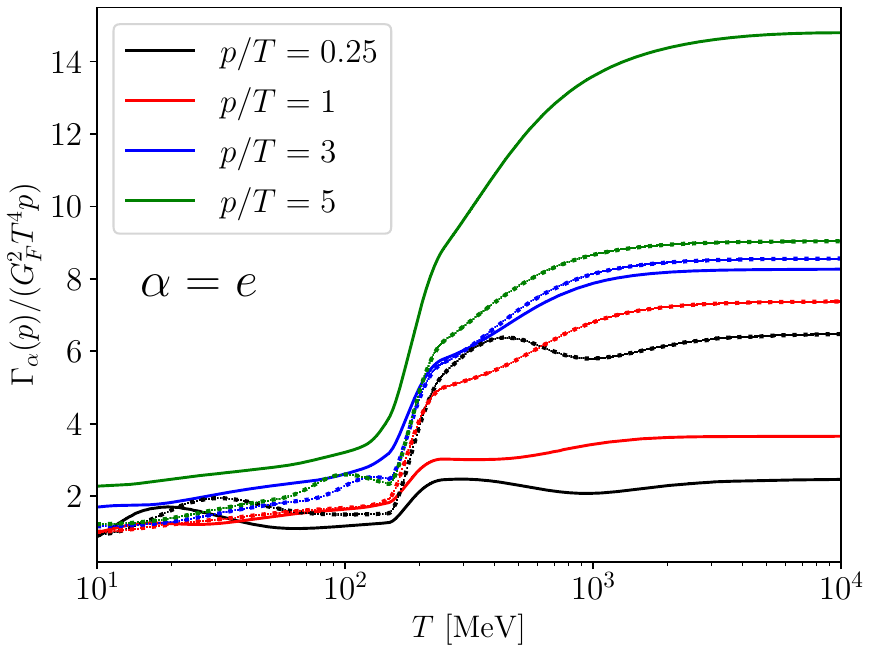}
\vskip-12pt
    \caption{Neutrino interaction rate, Eq.~ \eqref{Nu_rate}, for electron flavor with lepton flavor asymmetries of $L_e=-L_\mu=0.1$. Here we neglect a small reduction in lepton asymmetries due to the production of sterile neutrino DM (see Figure~\ref{fig:Lepton_evo} and Section~\ref{app:backreaction}.).
    Dotted lines denote the case for no lepton asymmetry for comparison.
    }
    \label{fig:Rate_01}
\end{figure*}


\subsection{Impact of QCD uncertainties on our results}
\label{app:QCD-uncertainties-impact}
Let us now summarize the impact of all the approximations and uncertainties in the description of strongly interacting particles discussed above on our results. These mainly concern the domain of the QCD transition, $120\text{ MeV}<T<280\text{ MeV}$, and are: validity of the perturbative expansion of the Taylor-expanded equation of state for the domain around the QCD transition; possibility of charged pion condensation; interpolation of the hadronic contribution to the active neutrino production rate $\Gamma_{\alpha}$ in the same domain; and uncertainties in the susceptibilities $\chi^{BB}$ and $\chi^{QQ}$.

Their impact on the robustness of the results presented in this study is small. This is because of the following reasons:
\begin{itemize}
\item[--] At large asymmetries $|L_{\mu}|\gtrsim 0.1$ for the scenario $L_{\mu} = -L_{\tau}, L_{e} = 0$, the Taylor expansion of the QCD equation of state utilized by Refs.~\cite{Vovchenko:2020crk,Middeldorf-Wygas:2020glx} completely breaks down; in addition, the pion condensation occurs. However, this range is out of the validity of the state-of-the-art studies of cosmological impact of lepton flavor asymmetries~\cite{Domcke:2025lzg,Domcke:2025jiy}, and so we do not consider them in Fig.~\ref{fig:parameterspace}.\footnote{For the direction $L_{\mu} = -L_{\tau}, \ L_{e} = 0$, the range of asymmetries allowed by BBN is $L_{\mu} < 0.06$. It includes the values $0.04<L_{\mu}<0.06$, which are outside the domain of the quantitative stability $L_{\mu}<0.04$ defined in Ref.~\cite{Middeldorf-Wygas:2020glx}. However, we believe that our results are still stable in this region. Apart from the point made above, this is motivated by the explicit form of the perturbativity criterion~\eqref{eq:EOS-perturbativity}.
Increasing $L_{\mu}$ from $0.04$ to $0.06$, we increase the left-hand side by a factor of $\approx (0.06/0.04)^{2} = 2.25$. Given the conservativity of Eq.~\eqref{eq:EOS-perturbativity}, we do not expect that such an increase would break perturbativity.} For the direction $L_{e} = -L_{\mu}$, there is no limitation from QCD, and so the domain $|L_e|\le 0.1$ allowed by BBN is treated fully robustly a priori.
\item[--] As shown in Sec.~\ref{app:behavior-abundances} and Fig.~\ref{fig:abundances}, at the boundary of the allowed parameter space (maximal lepton asymmetry $L\simeq 0.1$), the sterile neutrino production accumulates at $T\lesssim 100~\mathrm{MeV}$, well below the QCD transition, for all masses of interest. This keeps our main results -- the floor of the allowed sterile neutrino couplings in the presence of lepton flavor asymmetries in Fig.~\ref{fig:parameterspace} -- weakly sensitive to details of the transition. 
\item[--] The differences in susceptibilities $\chi^{BB}$ and $\chi^{QQ}$ propagate to the production of sterile neutrinos only for the range of masses and asymmetries where the production accumulates in the domain $120\text{ MeV}<T<280\text{ MeV}$; they only affect the redistribution of the asymmetry between leptons and baryons. Since both leptonic and hadronic asymmetries enter the neutrino potential~\eqref{eq:Valpha}, this redistribution does not lead to the disappearance or suppression of the resonant production and just slightly shifts the position of the resonant momenta (see Eq.~\eqref{eq:narrow-width-approximation}), which should not have a visible influence on the production. We have explicitly verified this by switching between the different lattice results and finding that the maximal impact on the sterile neutrinos' abundances is within 10\%.
\item[--] In the context of the redistribution of asymmetries across the transition by interpolating between the quark-gluon and hadronic phases, we ensure reliability across the domain $T=120$--$280~\mathrm{MeV}$ by enforcing smooth matching of the electric charge and baryon chemical potentials, $\mu_Q$ and $\mu_B$, at the boundaries between the lattice-based transition region and the adjoining quark gluon and hadronic regimes. The resulting evolution of the chemical potentials is regular and smooth (see, e.g., Fig.~\ref{fig:chemical-potentials}).
\item[--] The sterile neutrino production depends only on $\Gamma_{\alpha}$ when the production cannot be described by the narrow width approximation. This does not happen for the asymmetries $L_{\alpha}\gtrsim 10^{-3}$, which are required to open the sterile neutrino parameter space (see Sec.~\ref{app:simplified-approach}). Production for smaller asymmetries may be, in principle, sensitive to the details of the interpolation, and we attribute the uncertainty comparable to the variation of the rates from the interpolation scheme. We have explicitly checked that the dependence on the interpolation method modifies the rate within a factor of 2.
\end{itemize}

In principle, the same arguments may also allow considering the scenarios with large asymmetries $|L_{\alpha}|>0.1$. This is especially motivated by the fact that the potentially least constrained by BBN and CMB~\cite{Domcke:2025jiy,Domcke:2025lzg} direction in the flavor space $L_{e}\approx -L_{\mu}, L_{\tau} = 0$ does not lead to the pion condensation and induces negligible charge chemical potential. However, the investigation of such asymmetries in the context of sterile neutrino DM is tied to the question of their consistency with BBN and CMB, which is an open question nowadays. Therefore, we leave a detailed study of this domain for future work.

\subsection{Results for the evolution of sterile neutrinos}
\label{sec:results_sterile_nu}

In this section, we show some results for the evolution of sterile neutrinos and lepton asymmetries. Our own code reproduces the results for the evolution of sterile neutrinos in Ref.~\cite{Venumadhav:2015pla} very well. 

\begin{figure*}
 \centering
    \includegraphics[width=0.45\columnwidth]{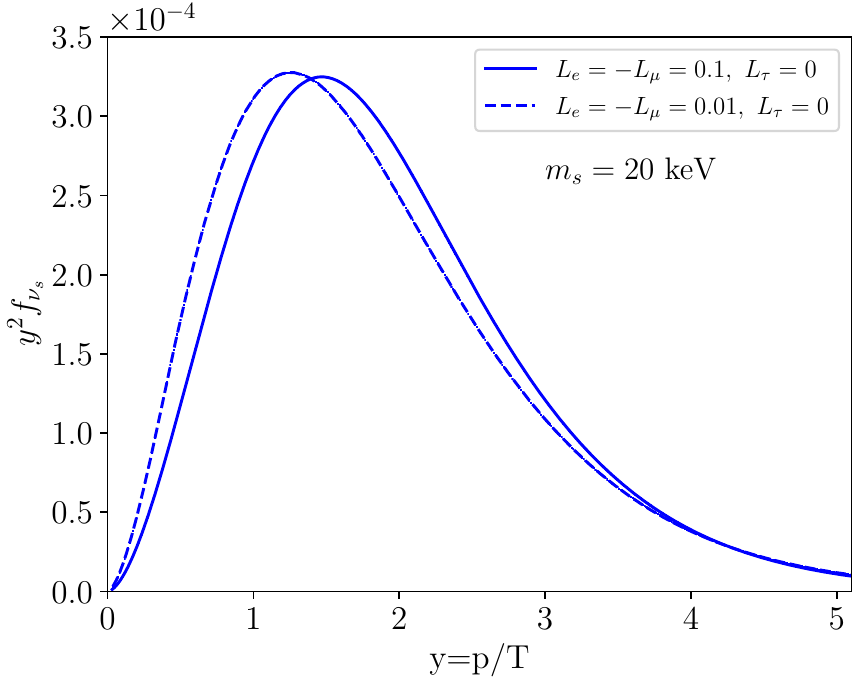}
    \includegraphics[width=0.45\columnwidth]{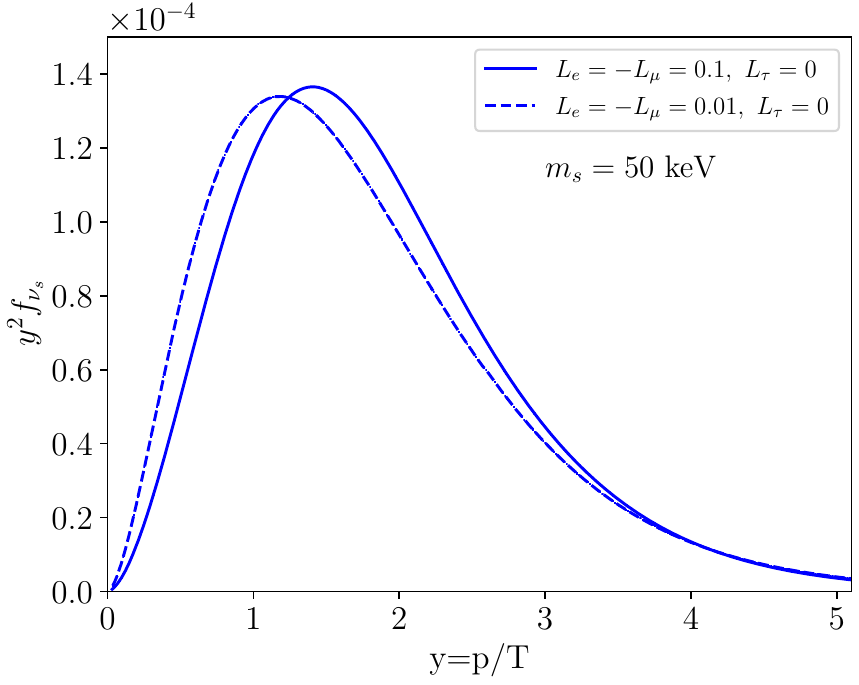}
    \vskip-6pt
    \caption{The momentum distributions of sterile neutrinos in the current Universe. The two panels show the cases of $m_s=20~{\rm keV}$ (left) and $m_s=50~{\rm keV}$ (right) with $L_e=-L_\mu=0.1,\ L_\tau=0$ (solid lines), $L_e=-L_\mu=0.01,\ L_\tau=0$ (dashed lines). The $\nu_s$ mixing with $\nu_e$ is considered, and mixing angles are fixed to explain the observed dark matter abundance with sterile neutrinos.}
    \label{fig:distribution}
\end{figure*}

Figure~\ref{fig:distribution} shows the momentum distributions of sterile neutrinos in the current Universe for some scenarios with large lepton asymmetries. The momentum distribution of sterile antineutrinos is negligibly small.

\begin{figure*}
 \centering
    \includegraphics[width=0.45\columnwidth]{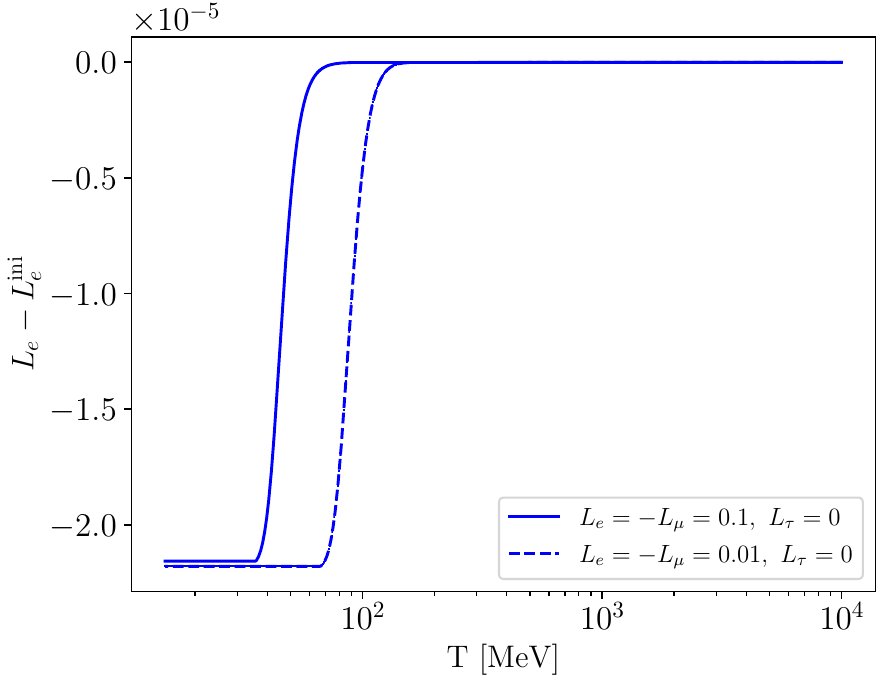}
    \includegraphics[width=0.45\columnwidth]{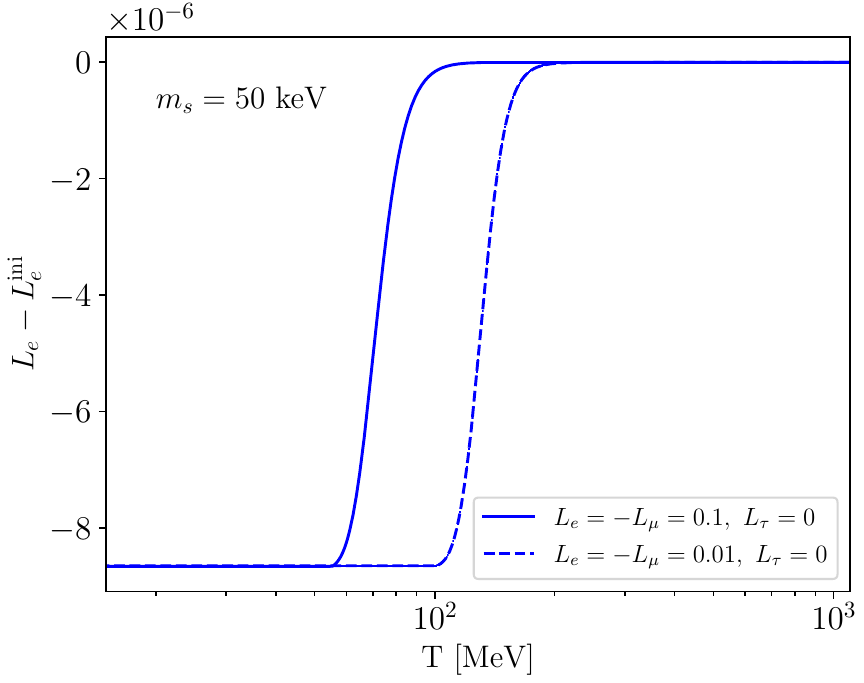}
    \vskip-6pt
    \caption{The temperature evolution of electron-flavor lepton asymmetry mixing with sterile neutrinos. The two panels show the cases of $m_s=20~{\rm keV}$ (left) and $m_s=50~{\rm keV}$ (right) with $L_e=-L_\mu=0.1,~L_\tau=0$ (solid lines), $L_e=-L_\mu=0.01,~L_\tau=0$ (dashed lines). Mixing angles are fixed to explain the observed dark matter abundance with sterile neutrinos.}
    \label{fig:Lepton_evo}
\end{figure*}

Figure~\ref{fig:Lepton_evo} shows some cases of the temperature evolution of lepton asymmetry mixing with sterile neutrinos. At the resonance production of sterile neutrinos, the lepton asymmetry slightly decreases. The resonance temperature is consistent with Eq.~\eqref{eq:Tres}. We confirm that this reduction of lepton asymmetry due to the sterile neutrino production is negligible for large lepton asymmetries of our interest.
For initial large lepton asymmetries, $L_\alpha^{\rm ini}$, the sterile neutrino abundance may be approximated as $\rho_{\nu_s}\simeq m_s|L_{\nu_s}|s\simeq m_s|\Delta L_\alpha|s$ with $|\Delta L_\alpha|=|L_\alpha-L_\alpha^{\rm ini}|$. If we fix $m_s$ and $\rho_{\nu_s}$, $|\Delta L_\alpha|$ is also fixed as shown in Fig.~\ref{fig:Lepton_evo}.

\subsection{Details of numerical calculations}
\label{sec:numerical_calculation}

We incorporate the system of Eqs.~\eqref{System_eq_1},~  \eqref{System_eq_1_2},~\eqref{System_eq_2},~\eqref{System_eq_3} and Eqs.~\eqref{eq_redist_1}--\eqref{eq_redist_3} in a \texttt{python} code with \texttt{scipy} and \texttt{numpy} libraries. Functions that are bottlenecks in computation time are compiled with the just-in-time compiler \texttt{numba}. To eliminate the inhomogeneous term $\partial/\partial p$ in Eq.~\eqref{System_eq_1} and simplify Eq.~\eqref{System_eq_2}, we introduce the following variables,
\begin{align}
    \tilde{y}=\left[\frac{s(T_{\rm ini},\xi_{\rm ini})/T_{\rm ini}^3}{s(T,\xi)/T^3}\right]^{1/3}\frac{p}{T},\ \ \ \ \ \ \xi=\frac{\mu}{T},
\end{align}
where $T_{\rm ini}$ is the initial temperature in the numerical calculation. It is convenient to use the plasma temperature as a clock and we numerically solve the following ordinary differential equations (ODEs), using Eq.~\eqref{System_eq_3},
\begin{align}
    \frac{df_{\nu_s}(\tilde{y},t)}{dT}=\frac{dt}{dT}\frac{df_{\nu_s}(\tilde{y},t)}{dt},\ \ \ \ \ \ \frac{dL_\alpha}{dT} = \frac{dt}{dT}\frac{dL_{\alpha}}{dt}.
\end{align}

To solve these ODEs, we use the \texttt{RK23} method in \texttt{solve\_ivp} distributed in \texttt{scipy}. The \texttt{RK45} method also works, but the \texttt{RK23} method is faster, and the results in both methods remain the same. In figure~\ref{fig:parameterspace}, we linearly discretize the momentum $\tilde{y}_i$ using $2\times 10^5$ grid points with $\tilde{y}_{\rm min}=0.1$ and $\tilde{y}_{\rm max}=16$. We estimate the evolution of sterile neutrinos in the plasma temperature range from $T_{\rm ini}=10\ {\rm GeV}$ to $T_{\rm fin}=15\ {\rm MeV}$, at which neutrino oscillations start.
We confirm that even if we take a smaller $\tilde{y}_{\rm min}$ and a larger $\tilde{y}_{\rm max}$, the sterile neutrino abundance converges within a few $\%$ level. 
We have also checked that the logarithmic momentum bins have worse numerical convergence than the linear ones.

\begin{figure}[t!]
\vskip-6pt
    \centering
    \includegraphics[width=0.45\columnwidth]{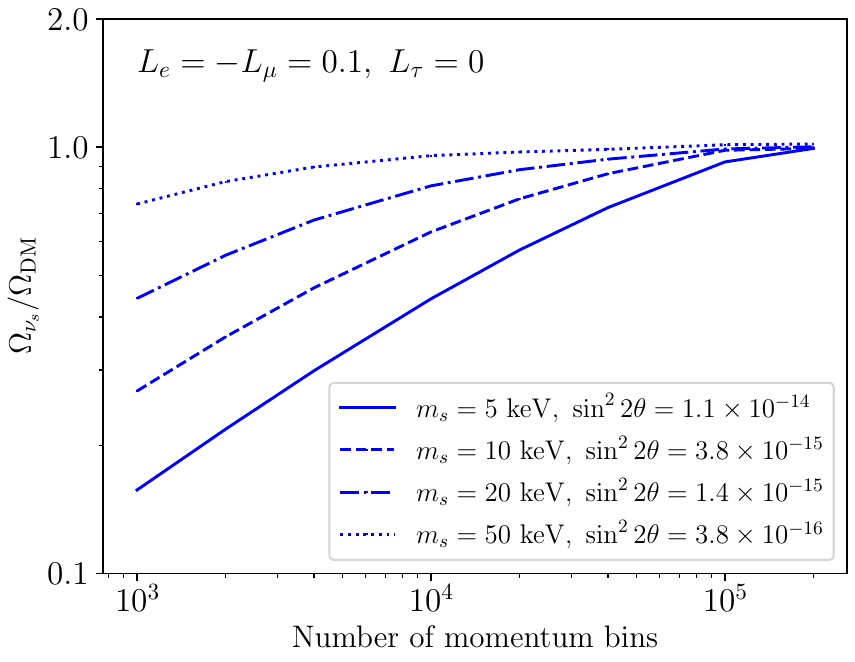}
    \includegraphics[width=0.45\columnwidth]{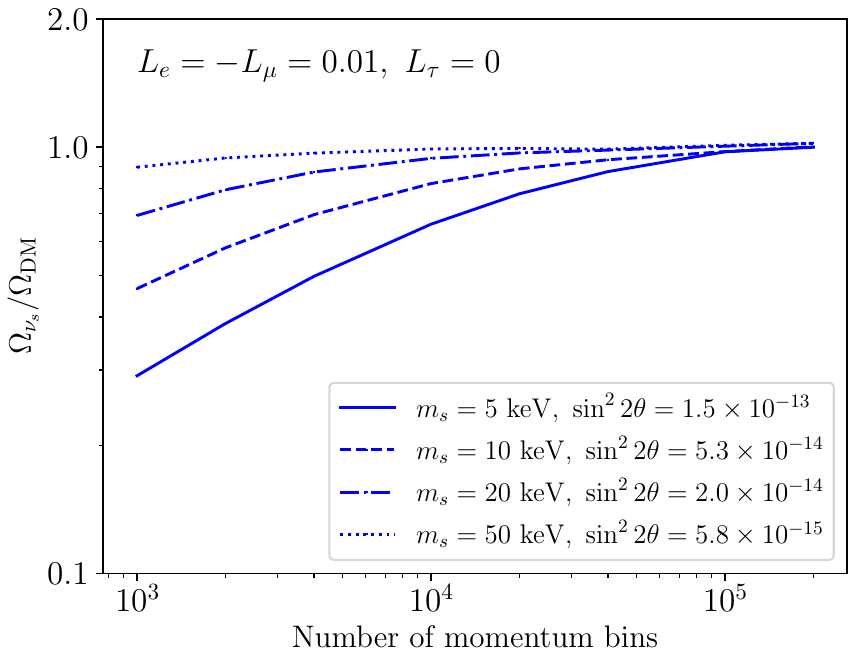}
    \vskip-6pt
    \caption{Numerical convergence of sterile neutrino abundance $\Omega_{\nu_s}$ (normalized to be the DM abundance $\Omega_{\rm DM}$) on the number of momentum bins. The two panels show the cases of $L_e=-L_\mu=0.1,\ L_\tau=0$ (left) and $L_e=-L_\mu=0.01,\ L_\tau=0$ (right) with $m_s=5~{\rm keV}$ (solid lines), $10~{\rm keV}$ (dashed lines), $20~{\rm keV}$ (dot-dashed lines) and $50~{\rm keV}$ (dotted lines). The $\nu_s$ mixing with $\nu_e$ is considered.}
    \label{fig:OmegasN}
\end{figure}

As reported in Ref.~\cite{Kasai:2024diy}, the numerical convergence of the sterile neutrino abundance with their momentum bins is rather poor.
Figure~\ref{fig:OmegasN} shows the dependence of the sterile neutrino abundance, $\Omega_{\nu_s}$, on the number of momentum bins in some setups. Here we consider the $\nu_s$ mixing with $\nu_e$, and nonzero $L_e=-L_\mu$ asymmetries with $L_\tau=0$. For lighter sterile neutrinos and larger asymmetries, the numerical convergence is worse. 
This is because the resonant width is narrower for lighter sterile neutrinos and larger asymmetries (see the next section~\ref{app:analytic}).
The small number of momentum bins underestimates $\Omega_{\nu_s}$ because they do not fully capture the narrow resonance. For $m_s\gtrsim 10~{\rm keV}$ with $|L_\alpha|\sim 0.1$, the numerical results for sterile neutrino abundance would converge well. On the other hand, for $m_s\lesssim 10~{\rm keV}$ with $|L_\alpha|\sim 0.1$, the abundance would still contain a few tens of percent numerical uncertainty.
We should note again that we use $2\times 10^5$ momentum bins with $\tilde{y}_{\rm min}=0.1$ and $\tilde{y}_{\rm max}=16$ in figure~\ref{fig:parameterspace}.

\clearpage


\section{A closer look at resonant production of sterile neutrinos}
\label{app:analytic}

In Section~\ref{app:analytic}, we will not write about the dependence of chemical potentials for simplicity unless they are necessary.

As discussed in Refs.~\cite{Shi:1998km,Abazajian:2001nj,Kishimoto:2008ic,Kasai:2024diy}, for extremely large lepton asymmetries, the resonance time scale is shorter than the neutrino oscillation length. In such a case, the semi-analytical kinetic equation with averaged oscillations, used in all previous literature, might not be appropriate to estimate the sterile neutrino abundance. 

Oscillations between active and sterile states may be suppressed by short resonance times and/or quantum Zeno damping. However, we find a compensating enhancement factor due to the fact that neutrinos produced cumulatively over the mean free path can experience the resonance. If a typical resonance scale is shorter than the neutrino mean free path, the resonance scale is effectively extended to the mean free path. Thus, even at very short resonance times, sterile neutrinos can be produced through sizable active-sterile neutrino oscillations.

The main purpose of this section is to construct the semi-analytical kinetic equations with non-averaged oscillations, which apply to any lepton asymmetries. To achieve this purpose, we analytically study the resonant production of sterile neutrinos with both averaged and non-averaged oscillations.

In Subsection~\ref{sec:nuosc}, we review neutrino oscillations in the Early Universe. In Subsection~\ref{sec:ave_osc}, we revisit the case of averaged neutrino oscillations and the validity of the averaged oscillations. In Subsection~\ref{sec:non_ave_osc}, we study the case of non-averaged neutrino oscillations.

We should note that the semi-classical kinetic equation with non-averaged neutrino oscillations is constructed using many analogies of quantum-mechanical-like neutrino oscillations and the Boltzmann equation. This is not derived by the more fundamental QKEs. We test the constructed effective kinetic equation by comparing the numerical results with those of QKEs in Section~\ref{app:QKEs}.


\subsection{Neutrino oscillations}
\label{sec:nuosc}

First, we review neutrino oscillations between active and sterile states in a thermal bath with lepton asymmetries to discuss the resonant production of sterile neutrinos.

We will assume that sterile neutrinos $\nu_s$ mix with only one flavor neutrinos $\nu_a$, characterized by the vacuum mixing angle $\theta$. When the oscillation length is much larger than the mean free path for $\nu_a$, the scattering event resets the phase of the active neutrino state to the initial state, suppressing the oscillation probability to the sterile state~\cite{Misra:1976by,Stodolsky:1986dx}. This is the so-called quantum Zeno effect. Incorporating this effect as an ansatz as in the previous studies, the oscillation probability is~\cite{Abazajian:2001nj}
\begin{align}
    &P_m(\nu_\alpha \rightarrow \nu_s; p,t)\approx\sin^22\theta_m\sin^2\left(\frac{m_m^2}{4p}t\right) \left[1+\left(\frac{\Gamma_\alpha (p) t}{2} \right)^2 \right]^{-1},
    \label{OsciProb}
\end{align}
where $[1+\left(\Gamma_\alpha t/2\right)^2]^{-1}$ is the quantum Zeno suppression factor.\footnote{\label{footnote:t_larger_than_lm} If the time of interest is longer than the oscillation length, $t>l_m$, the suppression factor would be replaced as $[1+\left(\Gamma_\alpha l_m/2\right)^2]^{-1}$.}
The factor of $1/2$ for $\Gamma_\alpha/2$ accounts for the fact that only active states (not sterile states) interact. 
$\Gamma_\alpha\sim G_F^2 T^4p$ is the interaction rate for active neutrinos, $\theta_m$, $m_m$, and $l_m$ are the effective mixing angle, mass, and the oscillation length, including the medium effects:
\begin{align}
    \sin^22\theta_m &= \frac{\Delta(p)^2\sin^22\theta}{\Delta(p)^2\sin^22\theta + \left[\Delta(p)\cos2\theta -V_\alpha(p)\right]^2}, \label{Mixing_eff} \\   
    m_m^2 &= 2p \sqrt{\Delta(p)^2\sin^22\theta+\left[\Delta(p)\cos2\theta-V_\alpha(p)\right]^2}, \\
    l_m &= \left\{\Delta (p)^2\sin^22\theta + \left[\Delta(p)\cos2\theta-V_\alpha(p)\right]^2 \right\}^{-1/2},
    \label{eq:lm}
\end{align}
where $\Delta(p)=\frac{m_s^2-m_\alpha^2}{2p}\simeq \frac{m_s^2}{2p}$ and $m_{s,\alpha}$ are sterile and active neutrino masses, respectively. $V_\alpha(p)$ is the matter potential for $\nu_\alpha$, which is schematically written as \cite{Notzold:1987ik}
\begin{align}
    V_\alpha(p) &\approx \sqrt{2}G_F Ls - \frac{8\sqrt{2}G_Fp}{3m_Z^2}\left(\rho_{\nu_\alpha} +\rho_{\bar{\nu}_\alpha} \right)-\frac{8\sqrt{2}G_Fp}{3m_W^2}\left(\rho_\alpha +\rho_{\bar{\alpha}} \right),
\end{align}
where $L\equiv (n_L-\bar{n}_L)/s$ is the lepton asymmetry, $n_L$ and $\bar{n}_L$ are the lepton and anti-lepton number densities, $s\approx2\pi^2/45g_{\ast}T^3$ is the total entropy density of the Universe with the effective number of relativistic species $g_{\ast}$, neglecting effects of chemical potentials, and $\rho_{\nu_\alpha},\  \rho_{\bar{\nu}_\alpha},\ \rho_\alpha,\ \rho_{\bar{\alpha}}$ are the energy densities for neutrino $\nu_\alpha$, charged-leptons $\alpha$ and their antiparticles.

The resonance condition in neutrino oscillations is 
\begin{align}
    \Delta(p)\cos2\theta = V_\alpha (p).
\end{align}
Two solutions satisfy the resonance condition: the first is the higher temperature satisfying $V_\alpha \simeq 0$ while the second is the lower temperature satisfying $\Delta\cos2\theta\simeq \sqrt{2}G_FLs$. Since, at the higher temperature, the oscillation probability \eqref{OsciProb} is significantly small, the resonance at the lower temperature is of interest. This resonance temperature is approximately, assuming $\cos2\theta\simeq1$,
\begin{align}
    &T_{\rm res}\sim 27\ {\rm MeV}\left(\frac{10.75}{g_{\ast}} \right)^{1/4}\left(\frac{3.15}{y} \right)^{1/4}\left(\frac{0.1}{L} \right)^{1/4}\left(\frac{m_s}{5\ {\rm keV}} \right)^{1/2},
    \label{Tres}
\end{align}
where $p=yT$ 
and we consider a fiducial value of $y=3.15$, which is the average energy for neutrinos in thermal equilibrium.


In the following subsections, we will estimate the resonant production of sterile neutrinos through averaged and non-averaged oscillations. Before closing this subsection, let us schematically discuss the regime for the validity of averaged neutrino oscillations.
The averaged description of the oscillations is valid when the resonance width $\delta t_{\rm res}^{\rm ave}$ is longer than the oscillation length at the resonance $l_m^{\rm res}$,
\begin{align}
    \gamma\equiv\frac{\delta t_{\rm res}^{\rm ave}}{l_m^{\rm res}}>1,
\end{align}
where $\gamma$ is the so-called adiabaticity parameter and if $\gamma<1$, the oscillations can no longer be averaged. $\delta t_{\rm res}^{\rm ave}$ is the resonance width, which is estimated when the averaged oscillation probability is maximized.
In the next subsection, we will see that for large lepton asymmetries of $|L_\alpha|\gtrsim 5\times 10^{-3}$, the adiabaticity parameter can be $\gamma<1$.
Thus, to estimate the sterile neutrino production with very large lepton asymmetries, it is necessary to formulate the resonant production with non-averaged oscillations.

\subsection{Resonant production with averaged
neutrino oscillations}
\label{sec:ave_osc}

The effective mixing angle in matter $\theta_m$ in Eq.~\eqref{Mixing_eff} is enhanced when $\Delta\cos2\theta \simeq V_\alpha$. Then, sterile neutrinos are resonantly produced through the enhanced neutrino oscillations.

First, we revisit the resonant production of sterile neutrinos with averaged neutrino oscillations. Even in this case, we find an enhancement factor by accumulating neutrinos during the resonance.
Then we study the resonant production with non-averaged neutrino oscillations in the next section~\ref{sec:enhancement_non_ave}.


\subsubsection{Resonance width and oscillation probability}

First, we review the oscillation probability and the resonant width for the averaged
neutrino oscillations in the previous work~\cite{Abazajian:2001nj,Kishimoto:2008ic,Venumadhav:2015pla}, where the oscillation is always averaged. The averaged oscillation probability is
\begin{align}
    \langle P_m(\nu_\alpha\rightarrow \nu_s;p) \rangle
    &\approx\frac{1}{2}\sin^22\theta_m\left[1 + \left(\frac{\Gamma_\alpha(p)l_m}{2} \right)^2
    \right]^{-1}, \nonumber \\
    &= \frac{1}{2}\frac{\Delta(p)^2\sin^22\theta}{\Delta(p)^2\sin^22\theta + \left[\Delta(p)\cos2\theta -V_\alpha(p)\right]^2+\left(\frac{\Gamma_\alpha}{2}\right)^2}.
    \label{Prob_ave_full}
\end{align}
The oscillation probability is maximized at 
\begin{align}
    \left|\Delta(p)\cos2\theta - V_\alpha (p) \right|\leq\max\left[\Delta(p)\sin2\theta, \frac{\Gamma_\alpha}{2}\right]
    \label{range_resonance_ave}
\end{align}
The corresponding resonance temperature width $\delta T_{\rm res}$ is
\begin{align}
    \frac{\delta T_{\rm res}}{T_{\rm res}}\sim\frac{1}{3V_\alpha}\max\left[\Delta(p)\sin2\theta, \frac{\Gamma_\alpha}{2}\right]
    \label{deltaT_ave}
\end{align}
The resonance time width $\delta t_{\rm res}^{\rm ave}$ is
\begin{align}
    \delta t_{\rm res}^{\rm ave}&=\frac{dt}{dT}\biggl|_{T_{\rm res}}\delta T_{\rm res} \nonumber \\
    &\sim \frac{1}{3HV_\alpha}\max\left[\Delta(p)\sin2\theta, \frac{\Gamma_\alpha}{2}\right], 
    \label{deltat_ave}
\end{align}
where we roughly approximate $dT/dt\sim HT$ for analytic estimations, where $H$ is the Hubble parameter. At this resonance width, the oscillation probability is
\begin{align}
    \langle P_m(\nu_\alpha\rightarrow \nu_s;p) \rangle_{\rm res} \sim \frac{\Delta(p)^2\sin^22\theta}{\Delta(p)^2\sin^22\theta+\left(\frac{\Gamma_\alpha}
    {2}\right)^2}.
    \label{Prob_ave}
\end{align}

Most of the sterile neutrinos would be produced during the resonance time width $\delta t_{\rm res}^{\rm ave}$ in Eq.~\eqref{deltat_ave}. This is because the oscillation probability \eqref{Prob_ave_full} is approximately proportional to $\left|\Delta\cos2\theta - V_\alpha \right|^{-2}$ while the resonance time scale is $\delta t\sim \frac{1}{3HV_{\alpha}}|\Delta\cos2\theta-V_\alpha|$. Thus, the production of sterile neutrinos would be maximized when the denominator of the oscillation probability \eqref{Prob_ave_full} is minimized.


\subsubsection{Semi-classical kinetic equations}

When the oscillation length is longer than the resonance (that is, the oscillations can be averaged), the quantum kinetic equation can be separated into the averaged oscillations and the classical kinetic equation \cite{Bell:1998ds,Volkas:2000ei,Lee:2000ej}.
This semi-classical Boltzmann equation for the sterile neutrino distribution function $f_{s}(p,t)$ at the resonance is~\cite{Abazajian:2001nj,Kishimoto:2008ic,Venumadhav:2015pla}
(see also Refs.~\cite{Asaka:2005pn,Asaka:2006rw,Asaka:2006nq})
\begin{align}
    \frac{\delta f_{s}(p,t)}{\delta t_{\rm res}^{\rm ave}} \approx\frac{\Gamma_\alpha(p)}{2}\langle P_m(\nu_\alpha\rightarrow \nu_s;p) \rangle_{\rm res}\left[f_\alpha(p,t)-f_s(p,t) \right],
    \label{Boltzmann_ave}
\end{align}
where $f_\alpha(p,t)$ is the active neutrino distribution function.
The factor of $1/2$ comes from the same reason as for the quantum Zeno suppression factor. The first term in Eq.~\eqref{Boltzmann_ave} denotes the production process for sterile neutrinos while the second term denotes their destruction process.

We should note that the derivations of the semi-classical kinetic equations are different for Refs.~\cite{Abazajian:2001nj,Kishimoto:2008ic,Venumadhav:2015pla} and Refs.~\cite{Asaka:2005pn,Asaka:2006rw,Asaka:2006nq}. In Refs.~\cite{Asaka:2005pn,Asaka:2006rw,Asaka:2006nq}, the semi-classical equations with averaged neutrino oscillations are derived from the QKEs under the assumption that the coherence of active and sterile neutrinos vanishes. 
In this study, we compare our formalism only with Refs.~\cite{Abazajian:2001nj,Kishimoto:2008ic,Venumadhav:2015pla} because the Boltzmann formalism is complicated and we have followed only Refs.~\cite{Abazajian:2001nj,Kishimoto:2008ic,Venumadhav:2015pla} carefully.


\subsubsection{Enhancement by accumulating neutrinos}
\label{sec:enhancement_ave}

Eq.~\eqref{Boltzmann_ave} would mean that this equation describes that active neutrinos ``produced during the oscillation length'' oscillates to sterile states,
\begin{align}
    \delta f_s\sim \frac{\Gamma_\alpha}{2}l_m^{\rm res}\times\langle P_m \rangle_{\rm res} \times \frac{\delta t_{\rm res}^{\rm ave}}{l_m^{\rm res}}\times\left[f_\alpha-f_s \right].
\end{align}
where $l_m^{\rm res}$ is the oscillation length at the resonance, 
\begin{align}
    l_m^{\rm res}\sim \max\left[\Delta(p)\sin2\theta, \frac{\Gamma_\alpha}{2}\right]^{-1}.
\end{align}
Here we substitute $\left|\Delta\cos2\theta - V_\alpha \right|\sim \max\left[\Delta\sin2\theta, \Gamma_\alpha/2\right]$ in Eq.~\eqref{eq:lm}. The first factor
$\frac{\Gamma_\alpha}{2}l_m^{\rm res}$ is the amount of neutrinos produced during one oscillation $l_m^{\rm res}$, the second factor $\langle P_m \rangle_{\rm res}$ is the averaged oscillation probability and the third factor $\frac{\delta t_{\rm res}^{\rm ave}}{l_m^{\rm res}}$ characterizes the number of oscillations.

However, active neutrinos are freely streaming during $\sim(\Gamma_\alpha/2)^{-1}$. If $(\Gamma_\alpha/2)^{-1}\gg  l_m$, such neutrinos would accumulate without initialization of their state by the quantum Zeno effects.
Since all accumulating neutrinos pass through the resonance, the amount of neutrinos produced during one oscillation should include an enhancement factor of $\sim (\Gamma_\alpha/2)^{-1}/l_m$,
\begin{align}
    \frac{\Gamma_\alpha}{2}l_m^{\rm res}\rightarrow \frac{\Gamma_\alpha}{2}l_m^{\rm res}\times \frac{(\Gamma_\alpha/2)^{-1}}{l_m^{\rm res}},
\end{align}
The resulting kinetic equation that includes this enhancement factor is
\begin{align}
    \frac{\delta f_s}{\delta t_{\rm res}^{\rm ave}}\sim \frac{\Gamma_\alpha}{2}\langle P_m \rangle_{\rm res} \left[f_\alpha-f_s \right]\times\frac{(\Gamma_\alpha/2)^{-1}}{l_m^{\rm res}}.
\end{align}
We will numerically confirm this enhancement factor is necessary by comparing the results of QKEs in section~\ref{app:QKEs}.

If $\Gamma_\alpha/2>\Delta\sin2\theta$ and $l_m^{\rm res}\sim(\Gamma_\alpha/2)^{-1}$
there is no enhancement factor. The kinetic equation~\eqref{Boltzmann_ave} is applicable to this case, which can be written as
\begin{align}
    \frac{\delta f_{s}(p,t)}{\delta t_{\rm res}^{\rm ave}} \approx\frac{\Gamma_\alpha(p)}{2}\frac{\Delta(p)^2\sin^22\theta}{\left(\frac{\Gamma_\alpha}{2}\right)^2}\left[f_\alpha(p)-f_s(p,t) \right],
    \label{Boltzmann_ave_Eff}
\end{align}

If $\Gamma_\alpha/2<\Delta\sin2\theta$, the oscillation length at the resonance is $l_m^{\rm res}\sim (\Delta\sin2\theta)^{-1} <(\Gamma_\alpha/2)^{-1}$. We should include an enhancement factor of $\sim (\Gamma_\alpha/2)^{-1}/l_m\sim(\Delta\sin2\theta)/(\Gamma_\alpha/2)$ in this case. 
Then we arrive at the same kinetic equation \eqref{Boltzmann_ave_Eff} after rescaling as
\begin{align}
&\frac{\delta f_s}{\delta t_{\rm \Gamma_\alpha/2<\Delta\sin2\theta}}= \frac{\delta f_s}{\delta t_{\Gamma_\alpha/2>\Delta\sin2\theta}}\frac{\delta t_{\Gamma_\alpha/2>\Delta\sin2\theta}}{\delta t_{\Gamma_\alpha/2<\Delta\sin2\theta}}, \\
&\delta t_{\Gamma_\alpha/2>\Delta\sin2\theta}=\frac{\Gamma_\alpha/2}{\Delta\sin2\theta}\delta_{\Gamma_\alpha/2<\Delta\sin2\theta},
\end{align}
where $\delta t_{\Gamma_\alpha/2>\Delta\sin2\theta}$ is the resonance width for $\Gamma_\alpha/2>\Delta\sin2\theta$ and $\delta t_{\Gamma_\alpha/2<\Delta\sin2\theta}$ is the width for $\Gamma_\alpha/2<\Delta\sin2\theta$.

Outside the resonance, the production of sterile neutrinos is negligible. As a result, we construct the following semi-classical kinetic equation for sterile neutrinos with averaged neutrino oscillations:
\begin{align}
    &\left(\frac{\partial}{\partial t}-Hp\frac{\partial}{\partial p} \right)f_{s}(p,t)\approx \frac{\Gamma_\alpha(p)}{2}P_{\rm eff}(\nu_\alpha\rightarrow \nu_s;p)\left[f_{\alpha}(p)-f_s(p,t) \right],
    \label{Boltzmann_Eff_fin_ave}
\end{align}
with the effective oscillation probability
\begin{align}
    P_{\rm eff}(\nu_\alpha\rightarrow \nu_s;p)=\frac{1}{2}\frac{\Delta(p)^2\sin^22\theta}{ \left[\Delta(p)\cos2\theta -V_\alpha(p)\right]^2+\left(\frac{\Gamma_\alpha}{2}\right)^2}.
    \label{OsciProb_Eff_ave}
\end{align}
The l.h.s of Eq.~\eqref{Boltzmann_Eff_fin_ave} takes into account the effect of the cosmic expansion. The effective oscillation probability in Eq.~\eqref{OsciProb_Eff_ave} has no term of $\Delta^2\sin^22\theta$ in the denominator, unlike the averaged oscillation probability in Eq.~\eqref{Prob_ave_full}.


\subsubsection{Validity of averaged neutrino oscillations}

So far, we have assumed that neutrino oscillations can be averaged.
Let us estimate the condition of this invalidity, i.e., when the estimated resonance width~\eqref{deltat_ave} and oscillation probability at the resonance~\eqref{Prob_ave} are not valid.

This condition is  $\delta t_{\rm res}^{\rm ave}< l_m^{\rm res} \sim \max\left[\Delta\sin2\theta, \frac{\Gamma_\alpha}{2}\right]^{-1}$,
which is translated as the so-called adiabaticity parameter $\gamma$
\begin{align}
    \gamma&\equiv \frac{\delta t_{\rm res}^{\rm ave}}{l_m^{\rm res}}, \nonumber \\
    &=\frac{1}{3HV_\alpha}\max\left[\Delta(p)\sin2\theta, \frac{\Gamma_\alpha}{2}\right]^2<1.
    \label{eq:adiabaticity_def}
\end{align}
Because of $V_\alpha\propto L$ at the resonance, we expect that Eqs.~\eqref{deltat_ave} and~\eqref{Prob_ave} are not valid for larger lepton asymmetries.
Eq.~\eqref{eq:adiabaticity_def} can be estimated, assuming $\Delta\sin2\theta<\Gamma_\alpha/2$,
\begin{align}
    \gamma \sim 0.05\left(\frac{10.75}{g_{\ast}} \right)^{3/4}\left(\frac{y}{3.15}\right)^{13/4}\left(\frac{10^{-2}}{L}\right)^{9/4}\left(\frac{m_s}{10\ {\rm keV}}\right)^{5/2},
    \label{eq:adiabaticity}
\end{align}
where we have used Eq.~\eqref{Tres} and consider the radiation-dominated Universe. Therefore, the adiabaticity condition is indeed violated for large lepton asymmetries.
We have confirmed that the adiabaticity parameter can be $\gamma<1$ for both the cases of $\Delta\sin2\theta<\Gamma_\alpha/2$ and $\Delta\sin2\theta>\Gamma_\alpha/2$ in the parameter space of sterile neutrino DM.

\subsection{Resonant production with non-averaged neutrino oscillations}
\label{sec:non_ave_osc}

For very large lepton asymmetries, sterile neutrinos would not be produced fully incoherently at the resonance as can be seen in Eq.~\eqref{eq:adiabaticity}. Let us now estimate the resonant width and the oscillation probability in such a case without the averaging procedure as in Eq.~\eqref{Prob_ave_full} and construct the semi-classical kinetic equation with non-averaged neutrino oscillations for sterile neutrinos.



\subsubsection{Oscillation probability and resonance width}
\label{sec:width_non_ave}

First, we look for the maximum value of the oscillation probability \eqref{OsciProb} in the case of non-averaged oscillation ($\gamma<1$) and the corresponding resonance width. We expect most of the sterile neutrinos to be produced during this resonance width. We will confirm this later.

$\delta t_{\rm res}$ is the width centered at the cosmic time $t_{\rm res}$ corresponding $T_{\rm res}$ that satisfies $\Delta\cos2\theta-V_\alpha=0$.
The corresponding range in the cosmic time at the resonance is
\begin{align}
t\in [t_{\rm res}-\delta t_{\rm res}/2,\  t_{\rm res}+\delta t_{\rm res}/2].
\end{align}
The oscillation probability is
\begin{align}
    P_m(\nu_\alpha \rightarrow \nu_s; p,\delta t_{\rm res})&\approx\sin^22\theta_m\sin^2\left(\frac{m_m^2}{4p}\delta t_{\rm res}\right) \left[1+\left(\frac{\Gamma_\alpha \delta t_{\rm res}}{2} \right)^2 \right]^{-1}.
\end{align}
The smaller $\delta t_{\rm res}$ (i.e., $\Delta\cos2\theta-V_\alpha\rightarrow 0$) corresponds to the larger $\sin^22\theta_m$ (i.e., $\sin2\theta_m\rightarrow 1$). For large $\delta t_{\rm res}$ such as  $\sin^2\left(\frac{m_m^2}{4p}\delta t_{\rm res} \right)\sim 1/2$, the oscillation probability increases as $\delta t_{\rm res}$ decreases.
On the other hand, the oscillation probability can be written as,
for $\delta t_{\rm res}$ small enough to approximate $\sin\left(\frac{m_m^2}{4p}\delta t_{\rm res} \right)\sim\frac{m_m^2}{4p}\delta t_{\rm res}$,
\begin{align}
    P_m(\nu_\alpha \rightarrow \nu_s; p,\delta t_{\rm res})&\approx\sin^22\theta_m\sin^2\left(\frac{m_m^2}{4p}\delta t_{\rm res}\right) \left[1+\left(\frac{\Gamma_\alpha \delta t_{\rm res}}{2} \right)^2 \right]^{-1}, \nonumber \\
    &\sim\sin^22\theta_m\left(\frac{m_m^2}{4p}\delta t_{\rm res}\right)^2 \left[1+\left(\frac{\Gamma_\alpha \delta t_{\rm res}}{2} \right)^2 \right]^{-1}, \nonumber \\
    &\sim \frac{1}{4}\Delta^2\sin^22\theta\delta t_{\rm res}^2\left[1+\left(\frac{\Gamma_\alpha \delta t_{\rm res}}{2} \right)^2 \right]^{-1}, \nonumber \\
    &\sim \frac{1}{4}\Delta^2\sin^22\theta \frac{1}{(\delta t_{\rm res})^{-2}+\left(\frac{\Gamma_\alpha}{2}\right)^2}.
    \label{Osc_full_another}
\end{align}
In this case, the oscillation probability decreases as $\delta t_{\rm res}$ decreases. 
Thus, the oscillation probability \eqref{Osc_full_another} is maximized at
\begin{align}
\frac{m_m^2}{4p}\delta t_{\rm res}\sim \frac{1}{2}.
\end{align}

Let us estimate the values of the resonant width and the corresponding oscillation probability. We parametrize the resonance width as
\begin{align}
    |V_\alpha-\Delta\cos2\theta|=\epsilon V_\alpha,
\end{align}
where $\max\left[\Delta\sin2\theta, \frac{\Gamma_\alpha}{2}\right]/V_\alpha \leq \epsilon \leq 1$. Then the effective mass is $m_m^2\sim 2p\epsilon V_\alpha$. The resonance width $\delta t_{\rm res}^{\rm non\text{-}ave}$ is, following the same procedure as
Eqs.~\eqref{range_resonance_ave}--\eqref{deltat_ave},
\begin{align}
    \delta t_{\rm res}^{\rm non\text{-}ave}\sim \frac{\epsilon}{3H} \geq \delta_{\rm res}^{\rm ave}
\end{align}
In addition, following the condition of $\frac{m_m^2}{4p}\delta t_{\rm res}\sim \frac{1}{2}$, we find
\begin{align}
    \epsilon\sim \left(\frac{3H}{V_\alpha}\right)^{1/2}.
\end{align}
$\delta t_{\rm res}^{\rm non\text{-}ave}$ can be rewritten as
\begin{align}
    \delta t_{\rm res}^{\rm non\text{-}ave} \sim \left(\epsilon V_\alpha \right)^{-1} < \max\left[\Delta\sin2\theta, \frac{\Gamma_\alpha}{2}\right]^{-1}.
    \label{deltat_nonave}
\end{align}
The corresponding oscillation probability is 
\begin{align}
    &P_m(\nu_\alpha \rightarrow \nu_s; p,\delta t_{\rm res}^{\rm non\text{-}ave}) 
    \sim \frac{1}{4}\frac{\Delta(p)^2\sin^22\theta}{\epsilon^2 V_\alpha^2}.
    \label{Prob_nonave}
\end{align}
Eq.~\eqref{Prob_nonave} is smaller than Eq.~\eqref{Prob_ave}. 

\subsubsection{Semi-classical kinetic equations}

Let us construct the semi-classical Boltzmann equations for sterile neutrinos in the case of non-averaged oscillation ($\gamma<1$), using the analogy of Eq.~\eqref{Boltzmann_ave}, the resonant width~\eqref{deltat_nonave} and the oscillation probability~\eqref{Prob_nonave}. We expect that this equation is at the resonance
\begin{align}
    \frac{\delta f_{s}(p,t)}{\delta t_{\rm res}^{\rm non\text{-}ave}}&\approx\frac{\Gamma_\alpha(p)}{2} P_m(\nu_\alpha\rightarrow \nu_s;p, \delta t_{\rm res}^{\rm non\text{-}ave}) \left[f_\alpha(p,t)-f_s(p,t) \right].
    \label{Boltzman_Eff_incomplete}
\end{align}
The oscillation probability \eqref{Prob_nonave} is suppressed compared to the average probability \eqref{Prob_ave}. The abundance of the produced sterile neutrinos may also be suppressed.
However, an enhancement factor as discussed in the previous section~\ref{sec:enhancement_ave} would exist in the semi-classical kinetic equation with 
non-averaged oscillations, which will be discussed in the next section.


\subsubsection{Enhancement by accumulating neutrinos}
\label{sec:enhancement_non_ave}

As in section~\ref{sec:enhancement_ave}, 
Eq.~\eqref{Boltzman_Eff_incomplete} means that these equations describe that active neutrinos ``produced during the resonance width $\delta t_{\rm res}$'' oscillates to sterile states,
\begin{align}
    \delta f_{s}&\sim\frac{\Gamma_\alpha}{2}\delta t_{\rm res}^{\rm non\text{-}ave}\times P_m(\nu_\alpha\rightarrow \nu_s;p,\delta t_{\rm res}^{\rm non\text{-}ave})\times\left[f_\alpha-f_s \right],
\end{align}
where $(\Gamma_\alpha/2)\delta t_{\rm res}^{\rm non\text{-}ave}$ is the amount of active neutrinos produced during the resonance width.

Similarly, active neutrinos are freely streaming during $\sim(\Gamma_\alpha/2)^{-1}$. If $(\Gamma_\alpha/2)^{-1}\gg \delta t_{\rm res}$, such neutrinos would accumulate without initialization of their state by the quantum Zeno effects.
Since all accumulating neutrinos pass through the resonance, the kinetic equation should include an enhancement factor of $\sim (\Gamma_\alpha/2)^{-1}/\delta t_{\rm res}$,
\begin{align}
     \delta f_{s}&\sim\frac{\Gamma_\alpha}{2}\delta t_{\rm res}^{\rm non\text{-}ave}\times P_m(\nu_\alpha\rightarrow \nu_s;p,\delta t_{\rm res}^{\rm non\text{-}ave})\times\left[f_\alpha-f_s \right] \times \frac{(\Gamma_\alpha/2)^{-1}}{\delta t_{\rm res}^{\rm non\text{--}ave}}, \nonumber \\
     &\sim P_m(\nu_\alpha\rightarrow \nu_s;p,\delta t_{\rm res}^{\rm non\text{-}ave})\left[f_\alpha-f_s \right]
     \label{deltaf_non_ave}
\end{align}
The resonance width is included only in the oscillation probability. At the resonance width that maximizes the oscillation probability, most of the sterile neutrinos are produced. We expected this fact in the previous section, and this has now been confirmed by Eq.~\eqref{deltaf_non_ave}.

Let us construct the effective semi-classical kinetic equation for non-averaged oscillations, including this enhancement factor.
Eq.~\eqref{Boltzman_Eff_incomplete} should always include this enhancement factor because of $\delta t_{\rm res}^{\rm non\text{-}ave}\sim (\epsilon V_{\alpha})^{-1} \ll (\Gamma_\alpha/2)^{-1}$,
\begin{align}
    \frac{\delta f_{s}(p,t)}{\delta t_{\rm res}^{\rm non\text{-}ave}}&\approx\frac{\Gamma_\alpha(p)}{2} P_m(\nu_\alpha\rightarrow \nu_s;p, \delta t_{\rm res}^{\rm non\text{-}ave}) \left[f_\alpha(p)-f_s(p,t) \right] \times\frac{(\Gamma_\alpha/2)^{-1}}{\delta t_{\rm res}^{\rm non\text{-}ave}}.
\end{align}
After some calculations, we arrive at an effective semi-classical kinetic equation for the case of non-averaged oscillations,
\begin{align}
    &\frac{\delta f_{s}(p,t)}{\delta t_{\rm res}^{\rm eff}} \approx \frac{\Gamma_\alpha(p)}{2}P_{\rm eff}(\nu_\alpha\rightarrow \nu_s;p)\left[f_{\alpha}(p)-f_s(p,t) \right],
    \label{Boltzmann_Eff}
\end{align}
with the effective oscillation probability
\begin{align}
    P_{\rm eff}(\nu_\alpha\rightarrow \nu_s)=\frac{1}{2}\frac{\Delta(p)^2\sin^22\theta}{ \left[\Delta(p)\cos2\theta -V_\alpha(p)\right]^2+\left(\frac{\Gamma_\alpha}{2}\right)^2}.
    \label{OsciProb_Eff}
\end{align}
We should note that $\Delta \cos2\theta-V_\alpha\ll \Gamma_\alpha/2$ during the ``effective'' resonance and we have rescaled the resonance width as $\delta f_s/\delta t^{\rm non\text{-}ave}_{\rm res}=\delta t^{\rm eff}_{\rm res}/\delta t^{\rm non\text{-}ave}_{\rm res}\times\delta f_s/\delta t^{\rm eff}_{\rm res}$, where $\delta t^{\rm eff}_{\rm res}$ is the resonance width for the effective oscillation probability~\eqref{OsciProb_Eff},
\begin{align}
    \delta t^{\rm eff}_{\rm res}\sim \frac{1}{3HV_\alpha}\frac{\Gamma_\alpha}{2}.
\end{align}

Finally, we conclude that the following semi-classical kinetic equation with non-averaged neutrino oscillations applies to any lepton asymmetries:
\begin{align}
    &\left(\frac{\partial}{\partial t}-Hp\frac{\partial}{\partial p} \right)f_{s}(p,t)\approx \frac{\Gamma_\alpha(p)}{2}P_{\rm eff}(\nu_\alpha\rightarrow \nu_s;p)\left[f_{\alpha}(p,t)-f_s(p,t) \right].
    \label{Boltzmann_Eff_fin}
\end{align}
This equation with non-averaged oscillations applies to the case of averaged oscillations because non-averaged oscillation is a generalization of averaged oscillations.
In fact, this equation is the same as the equation with averaged oscillations, but including the enhancement factor by accumulating neutrinos~\eqref{Boltzmann_Eff_fin_ave}.

\clearpage


\section{Back-reaction on lepton asymmetries}
\label{app:backreaction}

Let us introduce the sterile neutrino number-to-entropy ratio, 
\begin{equation}
L_{\nu_{s}} \equiv \frac{n_{\nu_{s}}-n_{\bar{\nu}_{s}}}{s}
\end{equation}
Its value determines the back-reaction on the lepton asymmetry $L_{\alpha}$, because of the conservation law
\begin{equation}
    L_{\alpha}(T)+L_{\nu_{s}}(T) = \text{const}
\end{equation}
In this section, we estimate the upper bound on $|L_{\nu_{s}}|$ and discuss its
impact on the back-reaction. 

Typically, only $n_{\nu_{s}}$ or $n_{\bar{\nu}_{s}}$ is accumulated throughout the evolution. Because of this, assuming that $\nu_{s}+\bar{\nu}_{s}$ populate the whole dark matter of the Universe, we may easily relate $L_{s}$ to the dark matter abundance:
\begin{equation}
    \Omega_{\nu_{s}} = \frac{\big(n_{\nu_{s}}(T_{\text{today}})+n_{\bar{\nu}_{s}}(T_{\text{today}})  \big) \cdot m_{\nu_{s}}}{\rho_{\text{critical}}} \approx \frac{|L_{\nu_{s}}(15\text{ MeV})|\cdot s_{\text{today}}\cdot m_{s}}{\rho_{\text{critical}}} = \Omega_{\text{DM}},
\end{equation}
where today's entropy density is
\begin{equation}
s_{\text{today}} = \frac{2\pi^{2}}{45}g_{*,\text{today}}T^{3}_{\text{today}} = 2.23\cdot 10^{-29}\text{ MeV}^{3}, \quad g_{*,\text{today}}\approx 3.91, \quad T_{\text{today}} = T_{\text{CMB}},
\end{equation}
\begin{equation}
    \rho_{\text{critical}} = 3.66\cdot 10^{-35}\text{ MeV}^{4}, \quad \Omega_{\text{DM}} = 0.265
\end{equation}
Now, we may get the maximal value of $L_{s}$:
\begin{equation}
    |L_{\nu_{s}}(T)|<|L_{\nu_{s}}(15\text{ MeV})| = \frac{4.4\cdot 10^{-4}}{m_{s}/1\text{ keV}}
\end{equation}
This means that, for $m_{\nu_{s}} >5\text{ keV}$, the maximal correction to the lepton asymmetry $L_{\alpha}$ throughout the evolution is
\begin{equation}
    |\Delta L_{\alpha}| <  8.7\cdot 10^{-5}\frac{5\text{ keV}}{m_{\nu_{s}}}
\end{equation}
As far as $|\Delta L_{\alpha}/L_{\alpha}|\ll 1$, we may safely neglect the effect of back-reaction on the evolution of the lepton asymmetries.

\clearpage

\section{Cross-checks}
\label{app:cross-checks}

In this section, we validate our approach to describe the production of sterile neutrinos in the Early Universe. To this end, we perform two independent cross-checks. The first one (Sec.~\ref{app:QKEs}) concerns the correctness of the treatment of the semi-classical Boltzmann equation with averaged and non-averaged oscillations; we compare their solutions with the quantum kinetic equations. The second one (Sec.~\ref{app:cross-check-simple-thermo}) is devoted to reproducing thermodynamic identities and checking the numerical stability of the code; we compare the results of the full Boltzmann code with the very simple but accurate code that uses narrow width approximation and neglects back-reaction on the lepton asymmetries (see Sec.~\ref{app:simplified-approach}).

\subsection{Comparison with QKEs}
\label{app:QKEs}

We have constructed the semi-classical kinetic equation with non-averaged neutrino oscillations \eqref{Boltzmann_Eff_fin}, which applies to any lepton asymmetries, using many analogies of quantum-mechanical-like neutrino oscillations and the classical Boltzmann equations. However, Eq.~\eqref{Boltzmann_Eff_fin} is not derived from QKEs. To test this effective equation more rigorously, we compare the results of this equation with those of QKEs.

The QKEs for active and sterile neutrinos are~\cite{Sigl:1993ctk}
\begin{align}
    &i\left(\frac{\partial}{\partial t}-Hp\frac{\partial}{\partial p} \right)\rho(p,t) =\left[\mathcal{H} ,\ \rho \right]- i\left\{\Gamma ,\rho \right\}+i\left\{\Gamma^p,\ 1-\rho \right\},
    \label{QKEs}
\end{align}
where $\rho$ is the density matrix for active and sterile neutrinos,
\begin{align}
    \rho=\begin{pmatrix}
   \langle a_\alpha^\dagger a_\alpha \rangle & \langle a_s^\dagger a_\alpha \rangle \\
   \langle a_\alpha^\dagger a_s \rangle & \langle a_s^\dagger a_s \rangle
\end{pmatrix},
\end{align}
$a_i(p,t)$ and $a_i^\dagger(p,t)$ $(i=\alpha,~s)$ denote the creation and annihilation operators for active and sterile neutrinos and 
\begin{align}
    \mathcal{H}&=\begin{pmatrix}
   V_\alpha - \Delta\cos2\theta & \Delta\sin2\theta \\
   \Delta\sin2\theta & \Delta\cos2\theta
\end{pmatrix}, \\ 
\Gamma &= 
\begin{pmatrix}
   \Gamma_\alpha/2 & 0 \\
   0 & 0
\end{pmatrix},\ \ \ \ \ 
\Gamma^{p} = 
\begin{pmatrix}
   \Gamma_\alpha^p/2 & 0 \\
   0 & 0
\end{pmatrix}.
\end{align}
Here, $f_{\alpha}(p,t)\equiv \langle a_\alpha^\dagger a_\alpha \rangle$ and $f_{s}(p,t)\equiv \langle s_\alpha^\dagger s_\alpha \rangle$. The off-diagonal parts of $\rho$ characterize the coherence between active and sterile neutrinos. Using the detailed balance to equate the forward and backward reaction rates and assuming active neutrinos are in thermal equilibrium, we have 
\begin{align}
\Gamma_\alpha^p=\Gamma_\alpha\exp[-(p-\mu)/T].
\label{Gamma_ap}
\end{align}
The QKEs are computationally expensive, but if we only consider $\rho(y)$ with a fixed $y=p/T=3$, which is the average momentum for thermal active neutrinos, and a narrow temperature range around the resonance, they might be easily solvable. This setup would be sufficient for our purposes to compare the effective semi-classical kinetic equation with the QKEs.

\begin{figure*}
    \centering
    \includegraphics[width=0.32\textwidth]{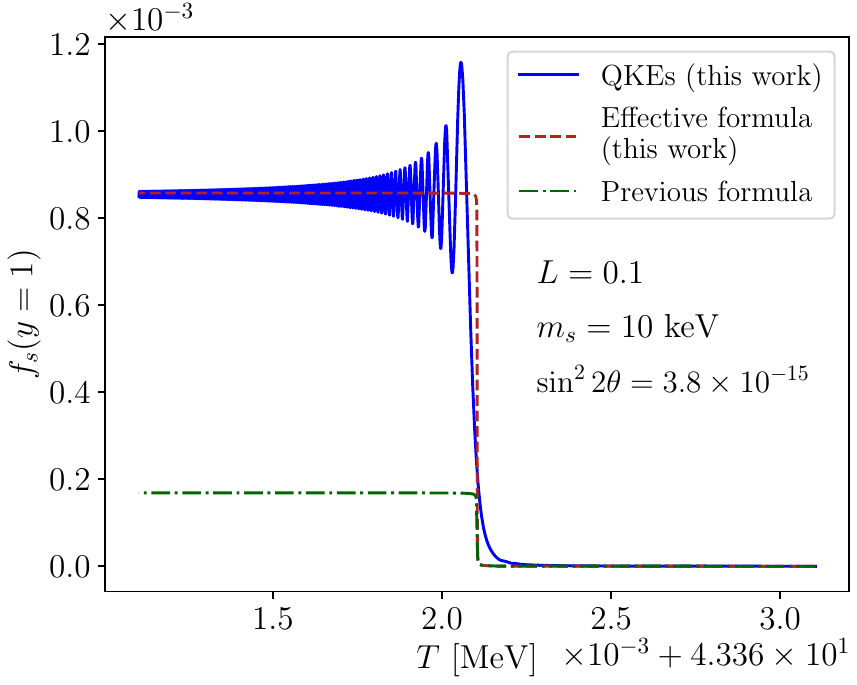}
    \includegraphics[width=0.32\textwidth]{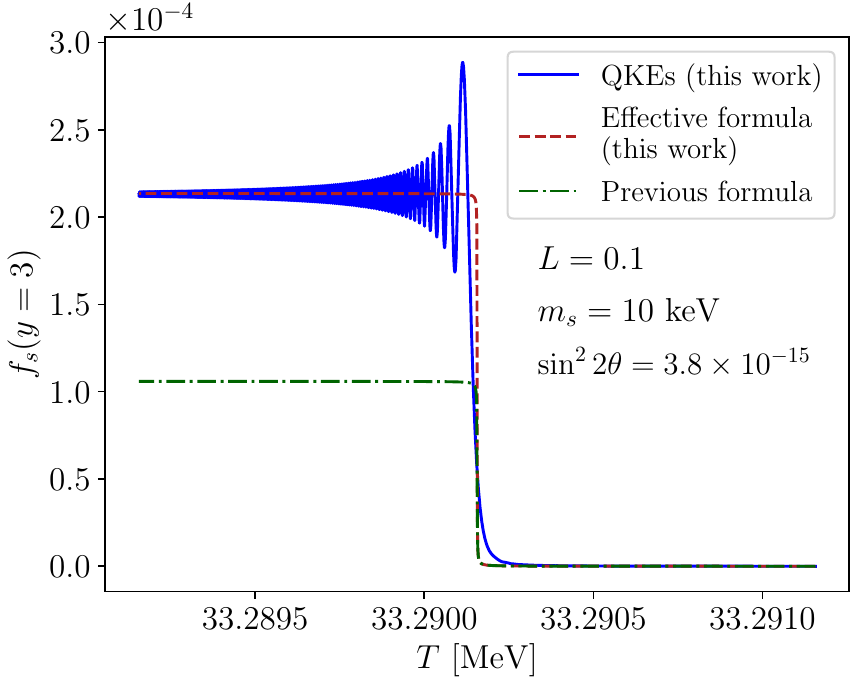}
    \includegraphics[width=0.32\textwidth]{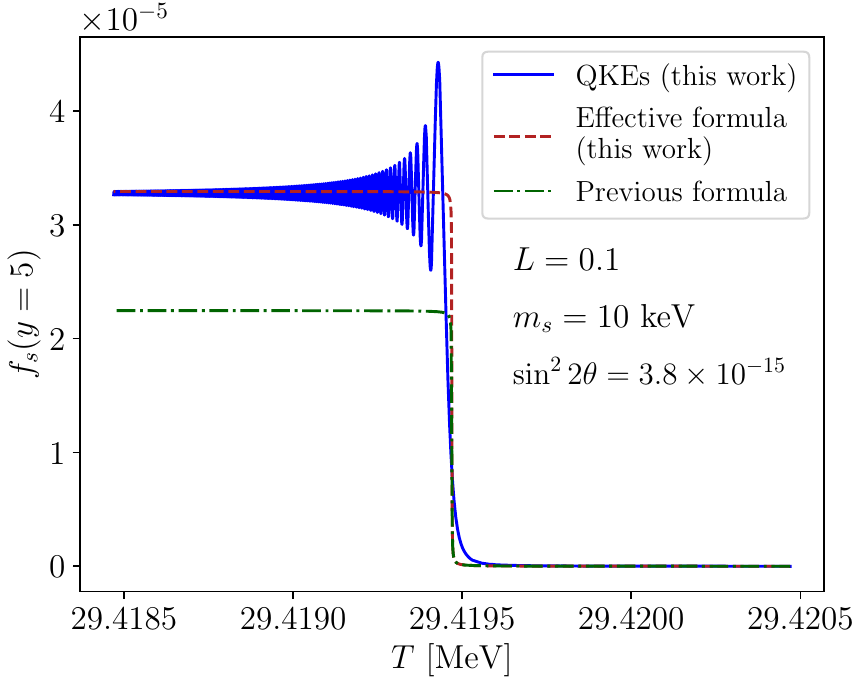}
    \includegraphics[width=0.32\textwidth]{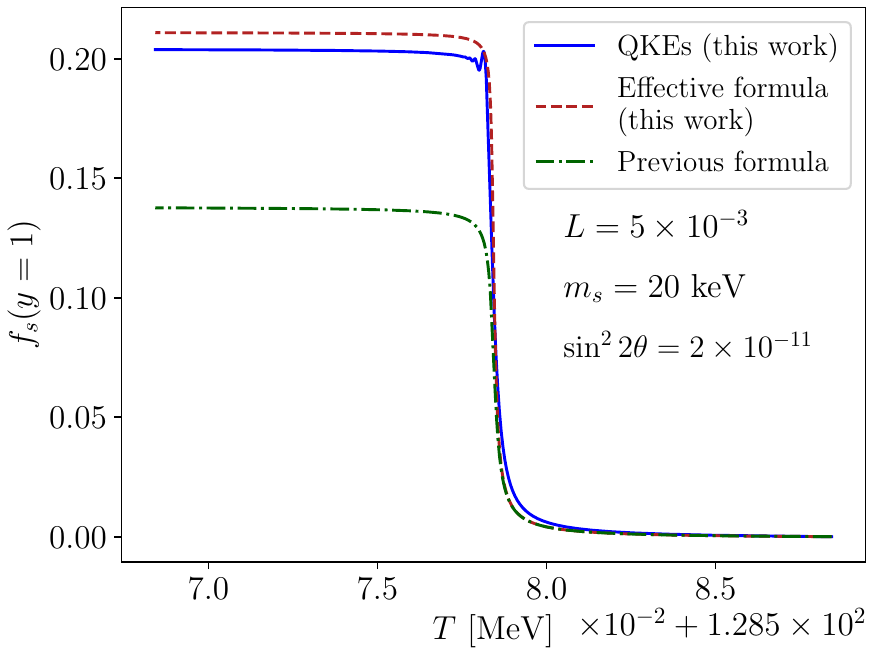}
    \includegraphics[width=0.32\textwidth]{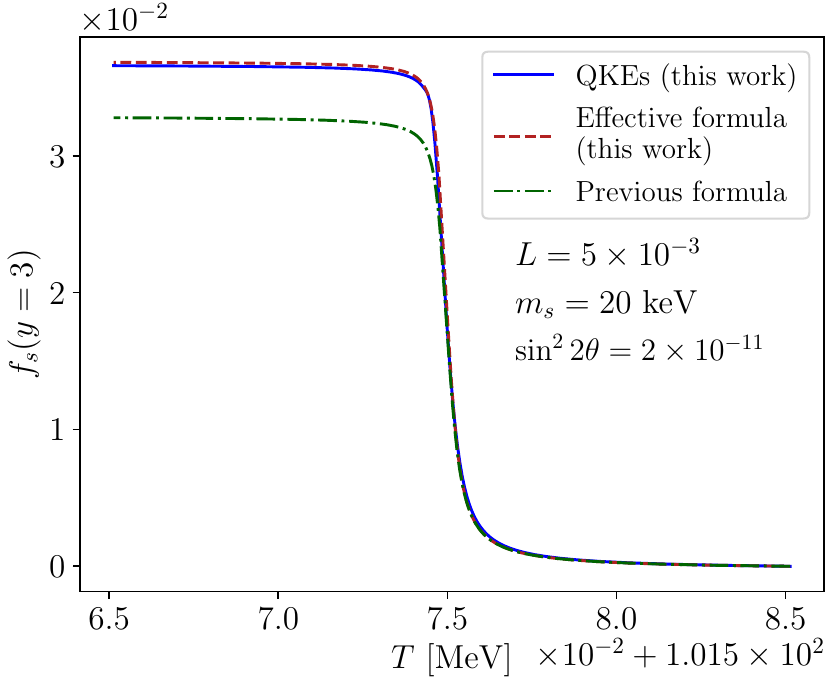}
    \includegraphics[width=0.32\textwidth]{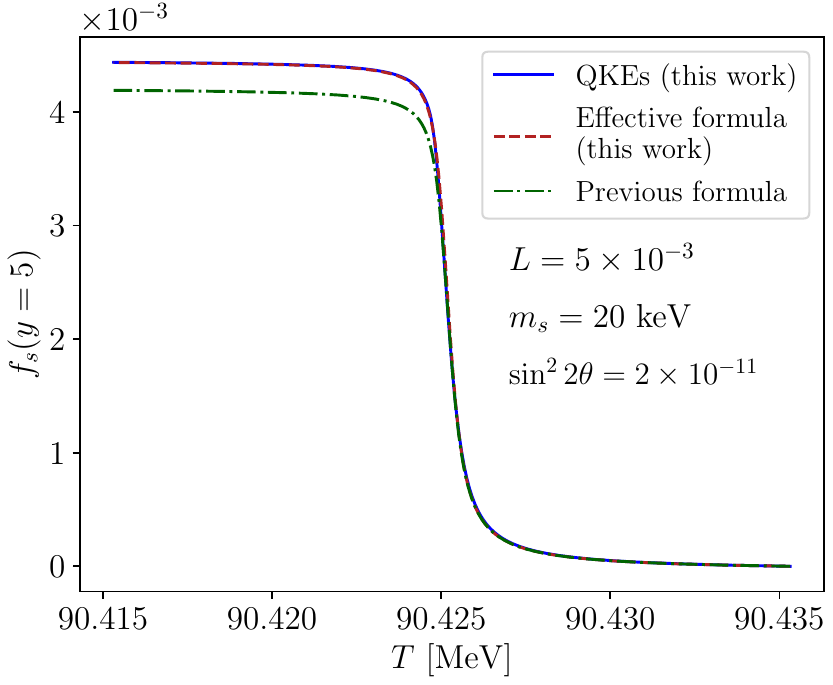}
    \caption{Evolution of sterile neutrino distribution function with $y=p/T=1$ (left), $y=3$ (middle), and $y=5$ (right) around the resonance.
    The top panels denote the case of $\nu_s$ production with 
    non-averaged
    neutrino oscillations ($\gamma<1$) while the bottom panels denote the case of $\nu_s$ production with 
    averaged
    neutrino oscillations $(\gamma>1)$. We consider asymmetries of $L_e=-L_\mu=L,\ L_\tau=0$ and the mixing between $\nu_s$ and $\nu_e$. We compare three kinetic equations, QKEs \eqref{QKEs} (blue solid line), the effective semi-classical equation~\eqref{Boltzmann_Eff_fin} constructed in this work (red dashed line), and Eq.~\eqref{Boltzmann_ave} used in the previous work~\cite{Abazajian:2001nj,Kishimoto:2008ic,Venumadhav:2015pla} (green dot-dashed line). The top panels are the case that sterile neutrinos constitute all dark matter.}
    \label{fig:QKEs}
\end{figure*}

However, to close the system for sterile neutrinos and thermal plasma, we additionally have to solve the evolution equations for lepton asymmetries and the plasma temperature,
\begin{align}
    \frac{d}{dt}L&=-\frac{1}{s}\int dp\  p^2\frac{d}{dt}\left[f_{s}(p,t)-f_{\bar{s}}(p,t) \right], \label{eq:Levo_QKEs} \\
    \frac{dT}{dt}&=-\frac{3H(\rho_{\rm SM}+P_{\rm SM})+\delta \rho_s/\delta t}{d\rho_{\rm SM}/dT},
    \label{eq:Tevo_QKEs}
\end{align}
where $\rho_{\rm SM}$ and $P_{\rm SM}$ are the energy density and pressure for the SM particles. $f_{\bar{s}}$ is the distribution for anti sterile neutrinos and 
\begin{align}
\frac{\delta \rho_{s}}{\delta t}\equiv \frac{1}{2\pi^2}\int dp\ p^2\sqrt{p^2+m_s^2}\frac{d}{dt}\left[f_s(p,t)+f_{\bar{s}}(p,t) \right].
\end{align}
These equations include the integrals of $df_{s}(y,t)/dt$. Therefore, $df_{s}(y,t)/dt$ with different $y$ are correlated.

However, at the resonance of $y=3$, the production of sterile neutrinos with $y\ll 3$ and $y \gg3$ would be negligible. At this resonance, we may approximate such integrals, for example, as follows:
\begin{align}
    \int dp\ p^{n-1} \frac{d}{dt}f_s(p) \approx 3^n T_{\rm res}^n\frac{\delta T_{\rm res}}{T_{\rm res}}\frac{d}{dt}f_s(y)\biggl|_{y=3},
\end{align}
where $\delta T_{\rm res}$ is the resonance width for temperature given by Eq.~\eqref{deltat_ave}. Using this approximation, we can close the system only for $y=3$. We have also performed a consistency check that the contributions of sterile neutrinos in Eqs.~\eqref{eq:Levo_QKEs} and \eqref{eq:Tevo_QKEs} are negligible, using our Boltzmann code. 
We will solve this system around the resonance and compare the results between the effective semi-classical kinetic equation and the QKEs.

Figure~\ref{fig:QKEs} shows the evolution of the sterile neutrino distribution with $y=1$ (left), $y=3$ (middle), and $y=5$ (right) around the resonance. The resonant productions have actually been observed. We compare three kinetic equations for $\nu_s$, QKEs \eqref{QKEs} (blue solid line), effective semi-classical equation \eqref{Boltzmann_Eff_fin} constructed in this work (red dashed line), eq~\eqref{Boltzmann_ave} used in the previous literature~\cite{Abazajian:2001nj,Kishimoto:2008ic,Venumadhav:2015pla} (green dot-dashed line). The top panels correspond to the case of non-averaged neutrino oscillations ($\gamma<1$) (and all dark matter with sterile neutrinos), while the bottom panels correspond to the case of the averaged oscillations ($\gamma>1$). In both panels, we consider $\delta t_{\rm res} <(\Gamma_\alpha/2)^{-1}$ or $l_m <(\Gamma_\alpha/2)^{-1}$, 
where the enhancement of accumulating neutrinos discussed in Section~\ref{app:analytic} would be crucial. The results of Eq.~\eqref{Boltzmann_Eff_fin} constructed in this work agree excellently with those of the QKEs, significantly improving Eq.~\eqref{Boltzmann_ave} utilized in the previous studies.

In the top panels, sterile neutrinos are produced partly coherently, and the QKE results do not match the results of the semi-classical equations microscopically. Macroscopically, Eq.~\eqref{Boltzmann_Eff_fin} still describes the QKEs very well.

When we solve the QKEs, we track the evolution of the active neutrino distribution function $f_\alpha(p,t)$. On the other hand, when we solve the semi-classical kinetic equations \eqref{Boltzmann_Eff_fin} and \eqref{Boltzmann_ave}, we assume active neutrinos are in thermal equilibrium.
However, we should note that, even for the case of the QKE, we use this assumption in Eq.~\eqref{Gamma_ap}. 
The excellent agreement of the effective semi-classical equation~\eqref{Boltzmann_Eff_fin} with the QKEs implies that the assumption that active neutrinos are in thermal equilibrium is valid. We confirm that $f_\alpha(y,t)$ deviates from the Fermi-Dirac distribution only by $0.3\%$ at most for the top panels in Figure~\ref{fig:QKEs}, which are the case that sterile neutrinos constitute all dark matter.

\subsection{Comparison with simplified approach and checking thermodynamics}
\label{app:cross-check-simple-thermo}

Let us start with checking the implementation of the evolution of particle-antiparticle asymmetries and thermodynamics. The list of cross-checks is summarized below.

\begin{figure}[h!]
    \centering
    \includegraphics[width=0.5\linewidth]{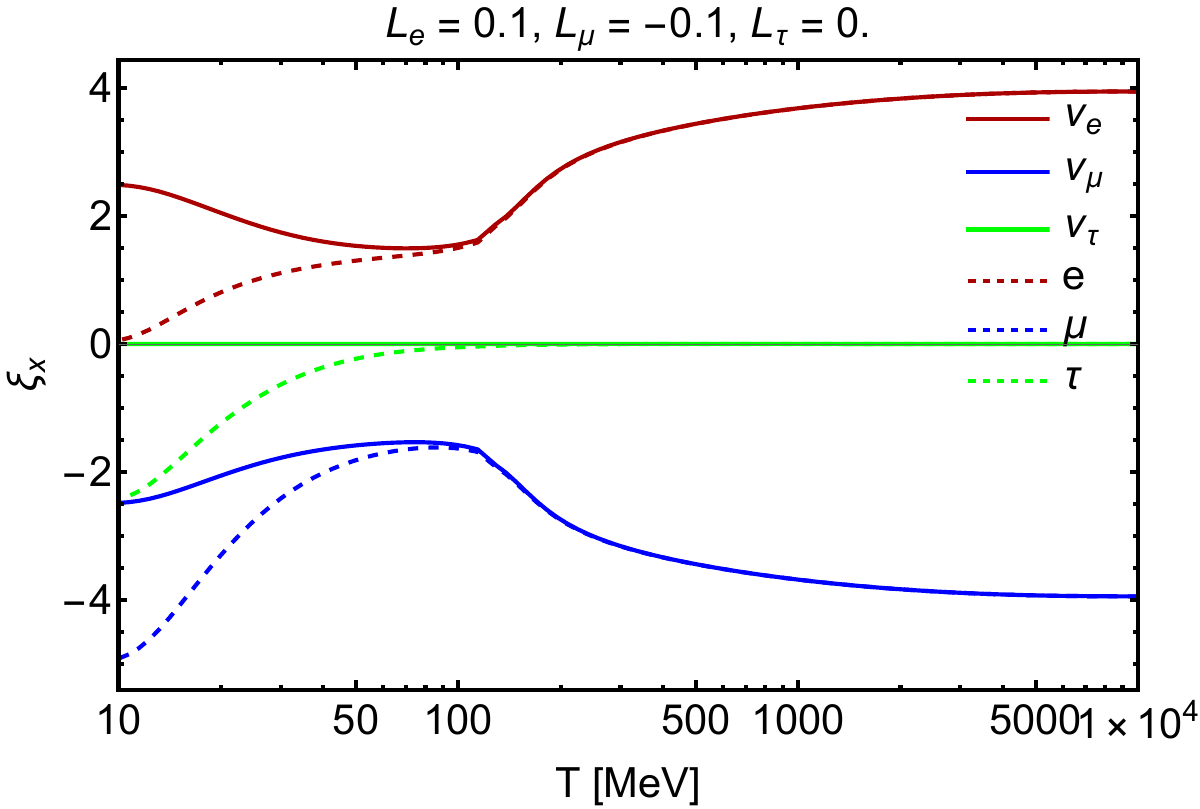}~\includegraphics[width=0.5\linewidth]{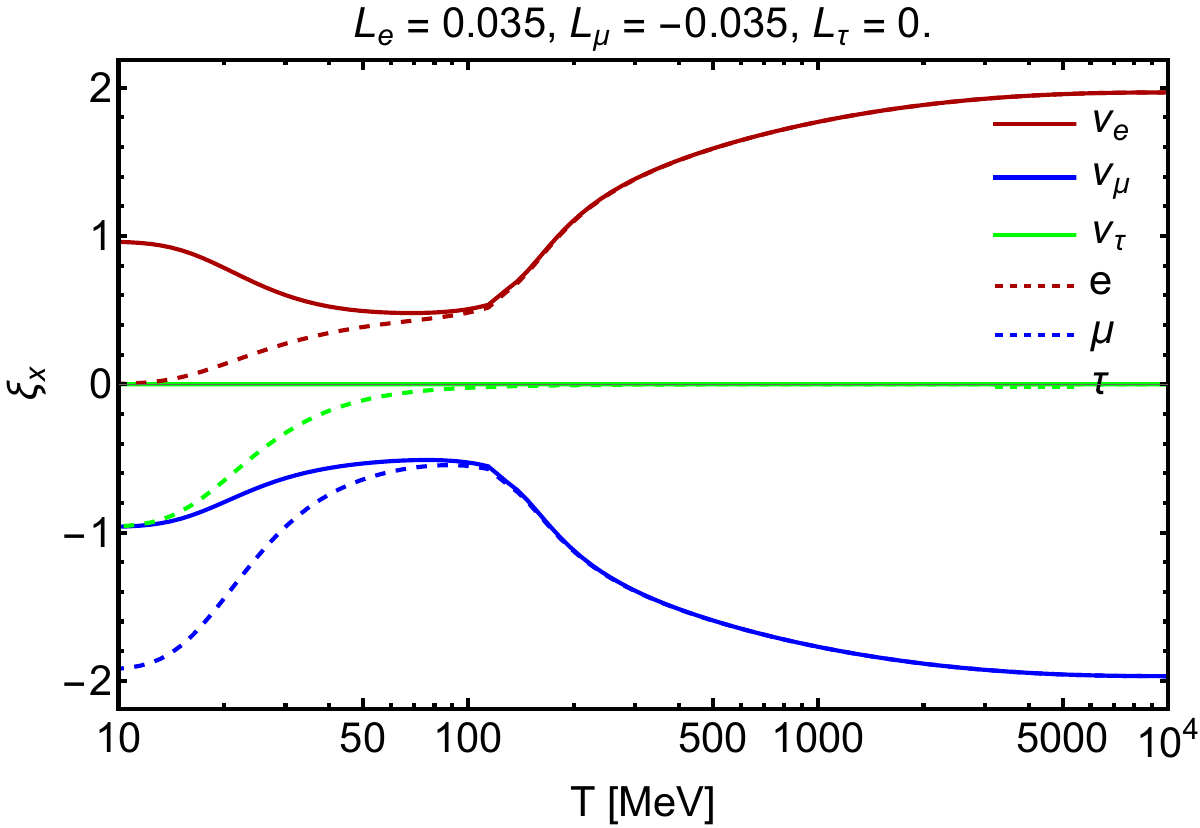}
    \caption{The evolution of the leptons' chemical potentials for two cases: $L_{e} = -L_{\mu} = 0.1, L_{\tau} = 0$  (the left panel) and $L_{e} = -L_{\mu} = 0.035, L_{\tau} = 0$ (the right panel). The non-zero tau chemical potential is a numerical error and negligible because of $m_\tau/T\gg \xi_\tau$.}
    \label{fig:chemical-potentials}
\end{figure}
\begin{itemize}
    \item Redistribution of asymmetries (see Fig.~\ref{fig:chemical-potentials} as an example). Since neutrinos, charged leptons $l_{\alpha}$, and hadrons are in equilibrium, we have two effects: the asymmetry $L_{\alpha}$ is redistributed between $\nu_{\alpha}$ and $l_{\alpha}$, generating the electric charge potential, and hadronic sector also acquires asymmetries. At large $T$, $\mu_{\nu_{\alpha}} = \mu_{l_{\alpha}}$. For the setup with $L_{e} = -L_{\mu}$, at low $T$, the value of $\mu_{e}$ tends to zero, whereas the values of $\mu_{\nu_{\alpha}}$ are fixed in a way such that the neutrino-antineutrino asymmetry satisfies the analytic relation
\begin{equation}
\mu_{\nu_{\alpha}} = \frac{\pi ^{2/3} \left(\sqrt[3]{3} \left(\sqrt{729 s^2 L_{\alpha}^2+3 \pi ^2 T^6}+27 s L_{\alpha}\right)^{2/3}-(3 \pi )^{2/3}
   T^2\right)}{3 (\sqrt{729 s^2 L_{\alpha}^2+3 \pi ^2 T^6}+27 s L_{\alpha})^{\frac{1}{3}}}
\label{eq:chemical-potential-asymmetry}    
\end{equation}
which follows from inverting the relation $L_{\alpha} =\Delta n(\mu_{\nu_{\alpha}})/s$.
\item We have checked that the Gibbs identity
    \begin{equation}
        s\cdot T = p+\rho - \sum_{i}\mu_{i}\Delta n_{i}
        \label{eq:Gibbs-identity}
    \end{equation} 
    is satisfied within less than $1\%$.
    \item Using the scale factor, $\dot{a}/a = H \Rightarrow da/dT = dt/dT \cdot H a$, we have checked that the entropy conservation law $a^{3}\cdot s = \text{const}$ holds up to 4\%. The slight deviation from the constant behavior is caused by the interpolations of $g_{*,s},g_{*,\rho}$ we use from Ref.~\cite{Saikawa:2018rcs}. Adding the effects of particle-antiparticle asymmetries is performed in a fully consistent way and only dilutes the non-constant behavior. 
\end{itemize}

\begin{figure}[h!]
    \centering
    \includegraphics[width=0.5\linewidth]{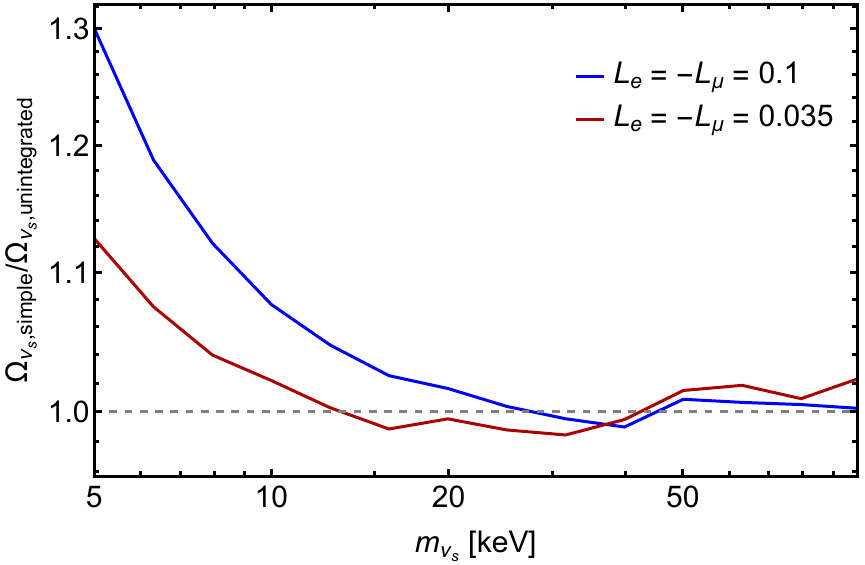}~\includegraphics[width=0.5\linewidth]{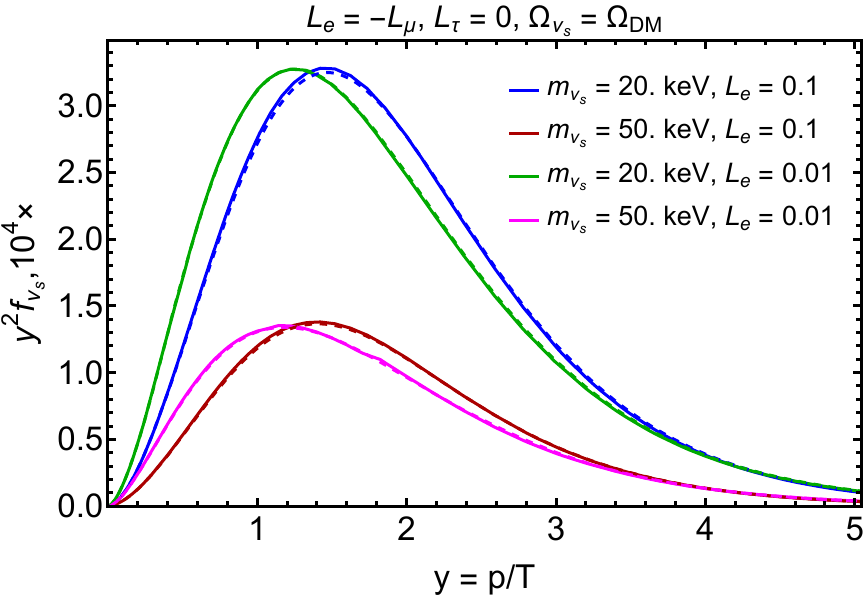}
    \caption{Comparison between the quasi-classical full Boltzmann equation~\eqref{eq:boltzmann-main} and the simplified approach discussed in Sec.~\ref{app:simplified-approach}. Left panel: the ratio $\Omega_{s}$ obtained by the simplified approach and using the full Boltzmann solver for the parameter space corresponding to the Fig.~\ref{fig:parameterspace} of the draft for $L_{e} = -L_{\mu} = 0.1$ and $L_{e} = -L_{\mu} = 0.035$. Right panel: the behavior of the sterile neutrino DM distribution function $f_{\nu_{s}}(y,T_{\text{today}})$, obtained by the full Boltzmann solver (solid lines) and the simplified approach (dashed lines), for the masses and lepton asymmetries considered in Fig.~\ref{fig:distribution}.}
    \label{fig:deviation}
\end{figure}

Now, let us proceed with comparing the results on the sterile neutrino DM abundance from the full Boltzmann and the simplified code from Sec.~\ref{app:simplified-approach}. For the values $\{m_{\nu_{s}},\sin^{2}(2\theta)\}$ from the lower boundary of Fig.~\ref{fig:parameterspace}, the ratio between the sterile neutrino abundance from the semi-classical full Boltzmann equation~\eqref{eq:boltzmann-main}, $\Omega_{\nu_s,{\rm unintegrated}}$, and from the simplified equation 
Eq.~\eqref{eq:abundance}, $\Omega_{\nu_s,{\rm simple}}$, is given by (see the left panel in Fig.~\ref{fig:deviation})
\begin{equation}
 \Omega_{\nu_s,{\rm simple}}/\Omega_{\nu_s,{\rm unintegrated}} = \begin{cases} 1-1.4, \quad L_{e} = -L_{\mu} = 0.1, \\ 1-1.1, \quad L_{e} = -L_{\mu} = 0.035.
 \end{cases}
\end{equation}

For the considered values of the lepton asymmetries, the discrepancy is within $30\%$, being maximal at small masses $m_{\nu_{s}}\simeq 5\text{ keV}$ and decreasing down to a ten percent level for the masses $m_{\nu_{s}}\simeq 10\text{ keV}$. The discrepancy is also significantly smaller for the smaller $L_{e}$. It may be due to the numeric instability of the full Boltzmann solver in the case of narrowing resonance $T$ (cf. Fig.~\ref{fig:OmegasN} and Sec.~\ref{sec:numerical_calculation}); to fix it, one would need to significantly increase the number of momentum bins, which heavily impacts the computation time. Nevertheless, we do believe that the agreement is quite good.

The right panel of Fig.~\ref{fig:deviation} shows the comparison of the sterile neutrino distributions obtained using the two approaches for a reference sterile neutrino mass. The results are in excellent agreement.

\clearpage

\section{Simplified approach to solve the Boltzmann equation}
\label{app:simplified-approach}
In this section, we discuss the simplified approach to solving the Boltzmann equation for sterile neutrinos. The code \texttt{sterile-dm-lfa} is available on \gitlink~\cite{sterile-dm-lfa} and is based on the following approximations:
\begin{enumerate}
\item $L_{\alpha}$ does not have back-reaction from accumulating sterile neutrinos. For the sterile DM, this approximation imposes the requirement 
\begin{equation}
4.4\cdot 10^{-4} \ \frac{1\text{ keV}}{m_{s}} \ll |L_{\alpha}|,
\end{equation}
see Section~\ref{app:backreaction}. In particular, for large lepton asymmetries $|L_{\alpha}|\gtrsim 0.01$ and masses $m_{s}>5\text{ keV}$, this condition is well-satisfied.
\item Narrow width approximation. In terms of the momentum, it is
\begin{equation}
    P_{\rm eff}(\nu_\alpha\rightarrow \nu_s,p,T) \approx \frac{1}{2}\sin^{2}(2\theta)\frac{2\pi}{\Gamma_{\alpha}}\sum_{p_{\text{res}}}h(p_{\text{res}})\delta(p-p_{\text{res}})\frac{\Delta^{2}(p_{\text{res}})}{\left|\frac{\partial}{\partial p}(\Delta(p)-V_{\alpha})\right|_{p_{\text{res}}}},
    \label{eq:narrow-width-approximation}
\end{equation}
with $h(p)$ being the Heaviside function; equivalently, it may be formulated in terms of temperature. The validity of the approximation is discussed in Section~\ref{app:narrow-width}.
\end{enumerate}

\subsection{For the number density}
After integrating over momenta, the Boltzmann equations of the evolution for sterile neutrinos and antineutrinos~\eqref{System_eq_1},~\eqref{System_eq_1_2} become the equations on the sterile neutrinos' number densities $n_{\nu_{s}}, n_{\bar{\nu}_{s}}$:
\begin{align}
    \frac{dn_{\nu_{s}}}{dt} +3H(t)n_{\nu_{s}} &= \frac{4\pi}{(2\pi)^{3}}\int p^{2}dp
    \ \frac{\Gamma_{\alpha}}{2} P_{\rm eff}(\nu_\alpha\rightarrow \nu_s,p,T)f_{\nu_{\alpha}}(p,T,\mu_{\nu_{\alpha}})
    \label{eq:boltzmann}
    \\ \frac{dn_{\bar{\nu}_{s}}}{dt} +3H(t)n_{\bar{\nu}_{s}} &= \frac{4\pi}{(2\pi)^{3}}\int p^{2}dp
    \ \frac{\Gamma_{\alpha}}{2} P_{\rm eff}(\bar{\nu}_\alpha\rightarrow \bar{\nu}_s,p,T)f_{\bar{\nu}_{\alpha}}(p,T,\mu_{\nu_{\alpha}})
\end{align}
Plugging Eq.~\eqref{eq:narrow-width-approximation} in Eq.~\eqref{eq:boltzmann}, we get
\begin{equation}
    \frac{dn_{\nu_{s}}}{dt} +3H(t)n_{\nu_{s}} = \frac{\sin^{2}(2\theta)}{4\pi}\sum_{p_{\text{res}}}h(p_{\text{res}})\frac{p_{\text{res}}^{2}f_{\nu_{\alpha}}(p_{\text{res}},T,\mu_{\nu_{\alpha}})
    \Delta^{2}(p_{\text{res}})}{{\left|\frac{\partial}{\partial p}(\Delta(p)-V_{\alpha})\right|_{p_{\text{res}}}}}\,.
    \label{eq:nwa-sn}
\end{equation}

\begin{figure}[h!]
    \centering
\includegraphics[width=0.5\linewidth]{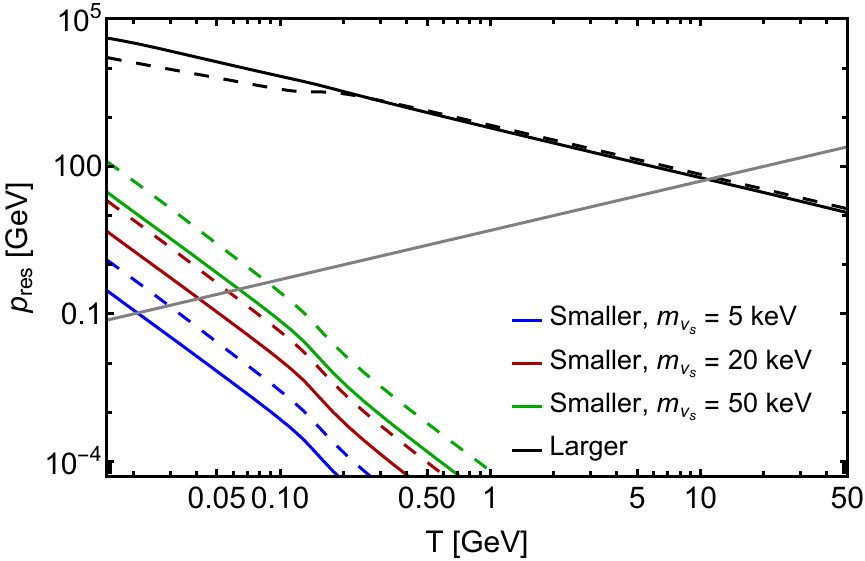}~\includegraphics[width=0.5\linewidth]{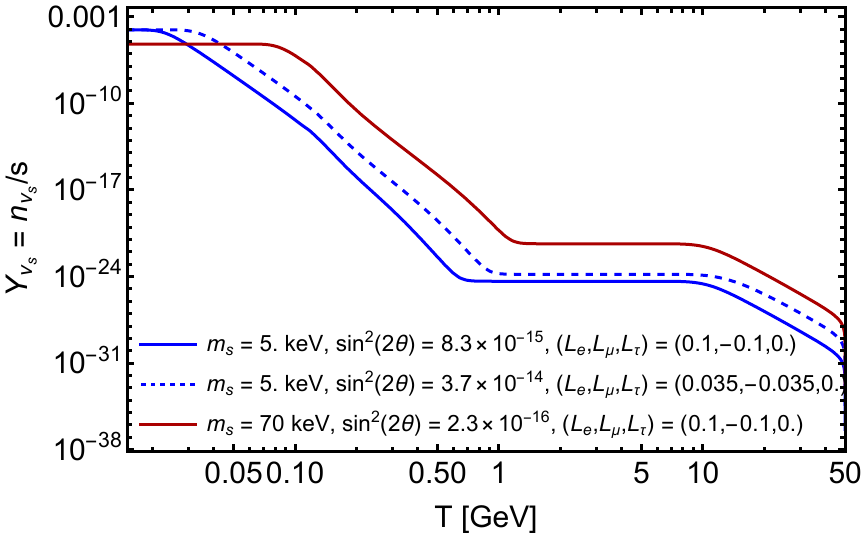}
    \caption{\textit{Left panel}: behavior of the resonant momenta $p_{\text{res}}$ for various sterile neutrino masses and the lepton asymmetries. The solid lines denote $L_{e} = -L_{\mu} = 0.1, L_{\tau} = 0$, whereas the dashed lines show $L_{e} = -L_{\mu} = 0.035, L_{\tau} = 0$. The ``Smaller'' branch significantly depends on the sterile neutrino mass, whereas the ``Larger'' branch is practically independent of it. The gray line shows the domain $p=5T$, above which the neutrino distribution function gets exponentially suppressed. \textit{Right panel}: the evolution of the sterile neutrino number-density-to-entropy ratio $Y_{\nu_{s}}(T) = (n_{\nu_{s}}+n_{\bar{\nu}_{s}})/s$ for the mass $m_{\nu_{s}} = 5\text{ keV}$ and two combinations of the lepton asymmetries: $L_{e} = -L_{\mu} = 0.1$ and $L_{e} = -L_{\mu} = 0.035$, and for the mass $m_{s} = 70\text{ keV}$ with $L_{e} = -L_{\mu} = 0.1$. $L_{\tau} = 0$ for all setups. There are two domains where the abundance grows, corresponding to the larger (temperatures $T\gtrsim 10\text{ GeV}$) and smaller branches of the resonant momentum of $p_{\text{res}}$.}
    \label{fig:pres}
\end{figure}

In particular, the $\Gamma_{\alpha}$-dependence cancels out. Finally, introducing the scale factor $H = \dot{a}/a$ and the derivative $dt/dT$, we obtain
\begin{equation}
    n_{\nu_{s}}(T_{\text{fin}}) = \left(\frac{a(T_{\text{ini}})}{a(T_{\text{fin}})}\right)^{3}\frac{\sin^{2}(2\theta)}{4\pi} \times \int \limits_{T_{\text{fin}}}^{T_{\text{ini}}} dT \  \frac{dt}{dT} \cdot\left(\frac{a(T)}{a(T_{\text{ini}})}\right)^{3} \sum_{p_{\text{res}}}h(p_{\text{res}})\frac{p_{\text{res}}^{2}f_{\nu_{\alpha}}(p_{\text{res}},T,\mu_{\nu_{\alpha}})
    \Delta^{2}(p_{\text{res}})}{{\left|\frac{\partial}{\partial p}(\Delta(p)-V_{\alpha})\right|_{p_{\text{res}}}}},
\end{equation}
where we consider $T_{\text{ini}} = 10\text{ GeV}$ and, similar to the full Boltzmann solver, $T_{\text{fin}} = 15\text{ MeV}$ (below which our approximation of neglecting neutrino oscillations breaks down). The ratio of the scale factors can be calculated using the entropy conservation, $\big(a(T)/a(T_{\text{ini}})\big)^{3} = s(T_{\text{ini}})/s(T)$.

A similar equation for sterile antineutrinos is obtained by replacing the neutrino distribution function and effective potential with the corresponding quantities for antineutrinos.

The sterile neutrino abundance is calculated using the following formula:
\begin{equation}
\Omega_{\nu_{s}} = \frac{1}{\rho_{\text{cr}}}\big(n_{\nu_{s}}(T_{\text{fin}})+n_{\bar{\nu}_{s}}(T_{\text{fin}})\big)\frac{s(T_{\text{today}})}{s(T_{\text{fin}})}\cdot m_{s}
    \label{eq:abundance}
    \end{equation}
Here, $T_{\text{today}} = T_{\text{CMB}}=2.7254\text{ K}$ is the today's temperature of the Universe, and $g_{*,s}(T_{\text{CMB}})\approx 2+6\cdot \frac{7}{8}\left(\frac{T_{\nu}}{T_{\gamma}}\right)^{3} \approx 3.91$, and $\rho_{\text{cr}}\approx 3.67\cdot 10^{-47}\text{ GeV}^{4}$.

\subsection{For the distribution function}
\label{app:simplified-approach-distribution}
Proceeding completely analogously, it is possible to derive the neutrino distribution function in the momentum space at the moment $T>T_{\text{fin}}$
The expression has the form
\begin{equation}
    \frac{df_{\nu_{s}}(\bar{y},T)}{dT} = -\frac{dt}{dT}\ h(T-T_{\text{fin}})
    \frac{\pi}{2}
    \frac{\Delta^{2}(\bar{y},T)f_{\nu_{\alpha}}(\bar{y},T,\mu_{\nu_{\alpha}})}{\left|\frac{\partial}{\partial T}\left(\Delta - V_{\alpha} \right)\right|}\delta(T-T_{\text{res}}),
    \label{eq:distribution-function-integrated}
\end{equation}
with $\bar{y} = (a/a(T_{\text{ini}}))\cdot p$ being comoving momenta ($a(T_{\text{ini}}) \equiv 1$) and $T_{\text{res}}(\bar{y})$ the solution of $V_{\alpha}-\Delta = 0$. The momentum argument in all the quantities entering Eq.~\eqref{eq:distribution-function-integrated} is replaced with $p = \bar{y}/a$. Integrated over $T$ from $T_{\text{max}}$ to $T_{\text{fin}}$, we get
\begin{equation}
    \frac{df_{\nu_{s}}(\bar{y},T)}{dT} = -\frac{dt}{dT}\ h(T-T_{\text{fin}})\frac{\Delta^{2}(\bar{y},T)f_{\nu_{\alpha}}(\bar{y},T,\mu_{\nu_{\alpha}})}{\left|\frac{\partial}{\partial T}\left(\Delta - V_{\alpha} \right)\right|}\bigg|_{T = T_{\text{res}}}
\end{equation}

In terms of the physical momenta, the distribution function today is
\begin{equation}
    f_{\nu_{s}}(p,T_{\text{today}}) =  f_{\nu_{s}}(\bar{y} \to p\cdot a(T_{\text{today}}),T_{\text{fin}})
    \label{eq:final-distribution}
\end{equation}
We have checked that the integral
\begin{equation}
    \Omega_{\nu_{s}} = \frac{m_s}{\rho_{\text{cr}}}\int \frac{d^{3}\mathbf{p}}{(2\pi)^{3}}f_{\nu_{s}}(p,T_{\text{today}})
\end{equation}
matches Eq.~\eqref{eq:abundance} with better than 5\% accuracy. The distribution functions for a few choices of the sterile neutrino masses and asymmetries are shown in Fig.~\ref{fig:pres}.

\subsection{Behavior of the sterile abundances}
\label{app:behavior-abundances}
For each temperature $T$, there are two branches of the resonant momenta $p_{\text{res}}$, see Fig.~\ref{fig:pres} (left panel). The larger branch is practically mass-independent and weakly depends on the asymmetry. It is typically irrelevant as it lies in the domain of momenta for which the neutrino distribution function entering Eq.~\eqref{eq:boltzmann} gets exponentially suppressed. As for the smaller branch, it substantially depends on $m_{\nu_{s}}$ and increases with $1/|L_{\alpha}|$ in the asymmetry. Overall, it causes a drop in the sterile abundance for the fixed mass and mixing angle as a function of $1/|L_{\alpha}|$.

\begin{figure}[h!]
    \centering
    \includegraphics[width=0.5\linewidth]{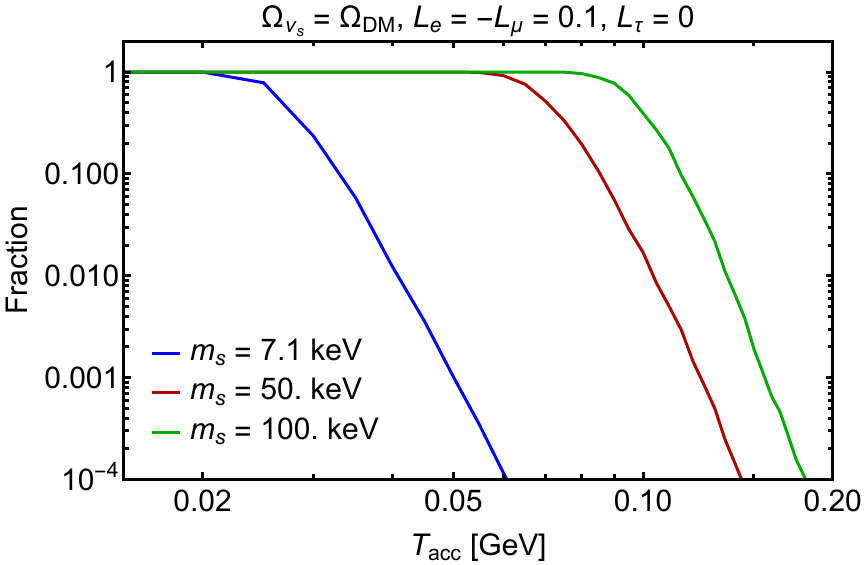}~\includegraphics[width=0.5\linewidth]{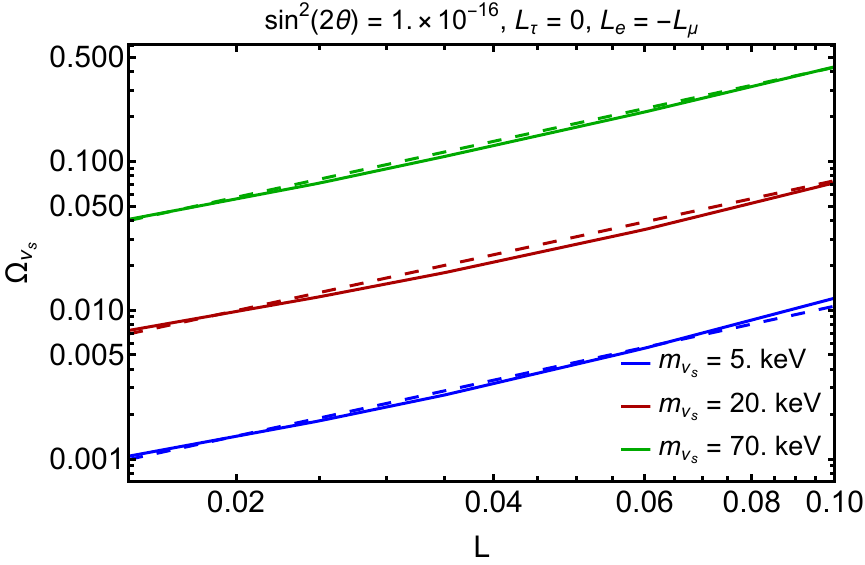}
    \caption{Left panel: the fraction of the number density of sterile neutrinos produced at temperatures $T>T_{\text{acc}}$, as a function of $T_{\text{acc}}$, for various masses $m_{s} = 7.1, 50, 100\text{ keV}$. We consider the scenario $L_{e}-L_{\mu} = 0.1, L_{\tau} = 0$, corresponding to the floor of sterile neutrino couplings in Fig.~\ref{fig:parameterspace}. Right panel: the scaling of the sterile abundances $\Omega_{\nu_{s}}$ with the sterile neutrino mass and the modulus of the electron flavor asymmetry $L_{e}$, for the setup $L_{e} = -L_{\mu},L_{\tau} = 0$. The dashed lines show the approximation with the fit~\eqref{eq:fit}.}
    \label{fig:abundances}
\end{figure}

To illustrate the role of the $p_{\text{res}}$ branches in accumulating the sterile neutrino abundance, in Fig.~\ref{fig:pres} (right panel), we show the behavior of the sterile number-density-to-entropy ratio $Y_{\nu_{s}} = (n_{\nu_{s}}+n_{\bar{\nu}_{s}})/s$. There are two domains where it increases -- one at large temperatures and another one at smaller temperatures, due to, correspondingly, the larger and the smaller branches $p_{\text{res}}$.

It also explains the range of temperatures where sterile neutrinos are mostly produced. It happens in the domain $p_{\text{res}}(L,T,m_{s})/T = 1-10$; otherwise, the population of neutrinos (producing $\nu_{s}$ via oscillations) is suppressed either by the phase space $p^{2}$ factor or by the Boltzmann exponent. If increasing $L_{\alpha}$, we decrease the resonant momentum $p_{\text{res}}$ and shift the domain $p_{\text{res}}/T = 1-10$ to lower temperatures. For the asymmetries $L_{\alpha} \gtrsim 10^{-4}$, opening up the parameter space of sterile neutrino DM, the production accumulates at temperatures $T\lesssim 1\text{ GeV}$. 

To further investigate this point, in the left panel of Fig.~\ref{fig:abundances}, we show the accumulation of the sterile neutrino abundance as a function of temperature for the floor of the parameter space in Fig.~\ref{fig:parameterspace}, corresponding to $L_{e} = -L_{\mu} = 0.1,\ L_{\tau}= 0$. We see two important points. First, the production accumulates quite fast: 95\% of sterile neutrinos are produced within 15-30 MeV temperature windows. Second, the sterile neutrino production happens at $T\lesssim 120\text{ MeV}$, i.e., below the domain of the QCD transition. This means that the floor we derived is weakly sensitive to the description of the QCD transition, as was indicated in Sec.~\ref{app:QCD-transition}.

It is also interesting to analyze the behavior of the abundances $Y_{\nu_{s}}(T_{\text{today}})$ as a function of mass and the modulus of the asymmetry $L_{\alpha}$, see Fig.~\ref{fig:abundances}. For the setup $L_{e} = -L_{\mu}, L_{\tau} = 0$, the scaling is 
\begin{equation}
\Omega_{\nu_{s}} \approx 0.04\ \frac{\sin^{2}(2\theta)}{10^{-16}}\cdot \left(\frac{L_{e}}{0.015}\right)^{1.25}\cdot\left(\frac{m_{\nu_{s}}}{70\text{ keV}}\right)^{1.4}
\label{eq:fit}
\end{equation}

\subsection{Checking the applicability of the narrow width approximation}
\label{app:narrow-width}
To cross-check the applicability of the narrow width approximation, we have considered the full integral~\eqref{eq:boltzmann} for the particular point 
\begin{equation}
m_{s} = 5\text{ keV}, \quad \sin^{2}(2\theta) = 1.7\cdot 10^{-14}
\end{equation}
and the lepton asymmetries
\begin{equation}
    L_{e} = -L_{\mu} = 0.1, \quad L_{\tau} = 0
\end{equation}
For this setup, the resonance is present for $\nu_{s}$ but absent for $\bar{\nu}_{s}$.

Then, we have represented the right-hand-side of Eq.~\eqref{eq:boltzmann} by
\begin{equation}
    \mathcal{I} = \frac{1}{2\pi^{2}} \times \begin{cases}
       \sum_{p_{\text{res}}} \int \limits_{p_{\text{res}}(1-\delta)}^{p_{\text{res}}(1+\delta)}dp \dots, \quad p_{\text{res}}\in \mathcal{P}, \\ \int \limits_{\mathcal{P}} dp \dots, \quad p_{\text{res}}\notin \mathcal{P}  \end{cases}
\end{equation}
Here, $\mathcal{P}$ is the integration domain defined by the comoving grid $\{y\}$ generated by the unintegrated code. Namely, if at least one of the $p_{\text{res}}$\!s lies inside $\mathcal{P}$, the integral is evaluated only in a close vicinity of $p_{\text{res}}$. Otherwise, it is integrated over the whole $\mathcal{P}$.

Using \texttt{Mathematica} and method \texttt{"InterpolationPointsSubdivision"}, we have found that $\mathcal{I}$ converges to Eq.~\eqref{eq:nwa-sn} \textbf{from below} once $\delta$ \textbf{decreases}. For $\delta \to 5\cdot 10^{-5}$, the results match within $\mathcal{O}(0.5\%)$.

If instead integrating over the whole domain outside the resonance domain, to check if the non-resonant contribution may sizeably increase the right-hand side, we have found that it is typically 2-3 orders of magnitude smaller, except for at the boundary of $\mathcal{P}$.

\clearpage

\section{Qualitative features of sterile neutrino spectrum in presence of lepton flavor asymmetries}
\label{app:sterile-spectrum}
In this section, we analyze the dependence of the sterile-neutrino momentum distribution on the lepton flavor asymmetry $|L_\alpha|$. Our goal is to clarify the qualitative trend of the final spectrum—most notably whether increasing $|L_\alpha|$ leads to an effectively warmer population once production and post-production dilution are taken into account. We then use the resulting spectra to define an illustrative Lyman-$\alpha$-motivated reference, indicating where structure-formation information may further restrict the low-mass region.

\subsection{Warming sterile neutrino spectrum}

\begin{figure}[h!]
    \centering
    \includegraphics[width=0.5\linewidth]{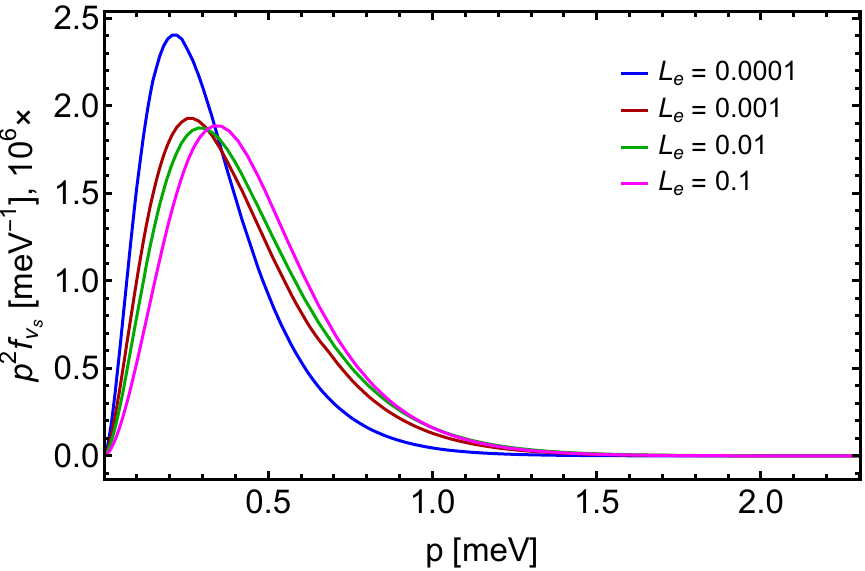}~\includegraphics[width=0.5\linewidth]{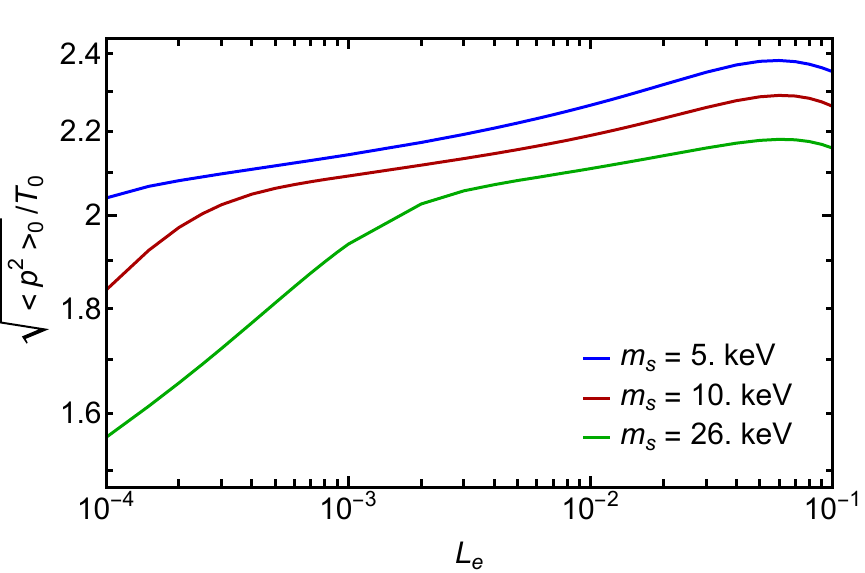}
    \caption{Spectrum of sterile neutrinos as a function of the lepton flavor asymmetry $L_{e}$, in the scenario $L_{e} = -L_{\mu}$, $L_{\tau} = 0$. Left panel: the sterile neutrino distribution function for mass $m_{s} = 20\text{ keV}$, normalized by the number density. Right panel: the ratio $\sqrt{\langle p^{2}\rangle_{0}}/T_{0}$~\eqref{eq:ratio-Ly-a} entering~\eqref{eq:ly-alpha-bound}, computed for several sterile neutrino masses. See text for details.}
    \label{fig:p2-to-T}
\end{figure}

Let us first assume that the asymmetries $L$ are conserved throughout the evolution of the Universe before the onset of neutrino oscillations, i.e., at $T \gtrsim 20\text{ MeV}$. We compute the quantity $\sqrt{\langle p^{2}\rangle_{0}}/T_{0}$ using both the unintegrated Boltzmann and semi-analytic simplified approaches, for cross-checking. Its behavior for various masses $m_{s}$, as well as the behavior of the overall sterile neutrino distribution as a function of $L_{\alpha}=10^{-4}$-$10^{-1}$, are shown in Fig.~\ref{fig:p2-to-T}. We find that $\sqrt{\langle p^{2}\rangle_{0}}/T_{0}$ generally increases with $L_{\alpha}$. For $m_{s}\simeq 20\text{ keV}$, the relative increase can reach a factor of $\sim1.5$, whereas for smaller masses, $m_{s}\simeq 5\text{ keV}$, the growth is less pronounced.

The origin of this increase is as follows. As discussed in Sec.~\ref{app:behavior-abundances}, increasing the asymmetry $L$ shifts sterile-neutrino production to lower temperatures. Production at lower temperatures entails weaker dilution from the annihilation of plasma species, leading to a larger $\sqrt{\langle p^{2}\rangle_{0}}/T_{0}$. In particular, as $L_{\alpha}$ varies from $10^{-4}$ to $0.1$, the temperature at which production saturates decreases from $T=500\text{ MeV}$ to $T\simeq 20\text{ MeV}$ (depending on $m_{s}$). Since $g_{*,s}(500\text{ MeV})\approx 60$ and $g_{*,s}(20\text{ MeV})\approx 11$, the dilution of sterile-neutrino momenta can be weaker by $(60/11)^{1/3}\approx 1.5$ relative to the $L=10^{-4}$ case. The much smaller impact of $L$ on $\sqrt{\langle p^{2}\rangle_{0}}/T_{0}$ for $m_{s}\simeq 5\text{ keV}$ arises because, for such small masses, production saturates at $T\lesssim 100\text{ MeV}$ throughout the considered $L$ range, so the dilution weakens only by $(g_{*,s}(100\text{ MeV})/g_{*,s}(20\text{ MeV}))^{1/3}\approx 1.15$.

\subsubsection{Case of varying $L$}
\label{app:ly-alpha-varying-L}

As a final remark, consider the scenario discussed in Ref.~\cite{Gorbunov:2025nqs}, in which a large initial asymmetry $L_{\text{ini}}$ vanishes around a temperature $T_{\text{trunc}}$.

Since the resonant momentum $p_{\text{res}}(T)$ increases as the temperature decreases (see Fig.~\ref{fig:pres}, left panel), lowering $T_{\text{trunc}}$ removes the high-momentum portion of the spectrum, i.e., truncates it from above. This effect is appreciable only if $L$ drops rapidly: most of the sterile population is produced within a narrow temperature window $\Delta T \lesssim 20~\text{MeV}$, see Figs.~\ref{fig:pres} (right panel) and~\ref{fig:abundances} (left panel). Consequently, the disappearance of $L$ must occur within this window and on a timescale much shorter than $\Delta T$.

To assess the impact, we compare the iso-abundance families
\begin{equation}
\{\sin^{2}2\theta,\; L(T;T_{\text{trunc}})\} \quad \text{vs.} \quad \{\sin^{2}2\theta,\; \tilde{L}=\text{const}\},
\label{eq:iso-families}
\end{equation}
where $\{\sin^{2}2\theta,\; L(T;T_{\text{trunc}})\}$ yields the observed dark matter abundance with a disappearing asymmetry starting from $L_{\text{ini}}$, while $\{\sin^{2}2\theta,\; \tilde{L}=\text{const}\}$ achieves the same abundance with a constant asymmetry $\tilde{L}\simeq L_{\text{ini}}$. Because the truncated case removes the high-momentum tail, it can produce a colder spectrum, reflected in a smaller ratio $\sqrt{\langle p^{2}\rangle_{0}}/T_{0}$, thereby relaxing the structure formation constraints.

For a quantitative illustration, we implement the minimal step-function model
\begin{equation}
L(T;T_{\text{trunc}})=
\begin{cases}
L_{\text{ini}}, & T>T_{\text{trunc}},\\
0, & T<T_{\text{trunc}},
\end{cases}
\label{eq:truncation}
\end{equation}
and find that the quantity in Eq.~\eqref{eq:ratio-Ly-a} can differ between the two families in Eq.~\eqref{eq:iso-families} by as much as a factor of 10.

Ref.~\cite{Gorbunov:2025nqs} proposed neutrino oscillations as a realistic mechanism of the $L$-truncation. However, within the mass range we consider, $5\text{ keV} < m_{s}<100\text{ keV}$, the production of sterile neutrinos is saturated at temperatures $T>20\text{ MeV}$ -- before the onset of oscillations -- for the asymmetries $|L_{\alpha}|<0.1$. Shifting the production to smaller temperatures requires much higher asymmetries. The status of such asymmetries, and in particular whether they are consistent with BBN and CMB, is an open question nowadays, as we discuss in the main text and Sec.~\ref{app:QCD-uncertainties-impact}.

\subsection{Calculation of the Lyman-$\alpha$ domain}
\label{app:structure-formation}

Structure formation information can be inferred from several observables (e.g., Lyman-$\alpha$ forest, Milky Way satellites, and strong lensing), but robust constraints for non-thermal dark matter require a dedicated analysis of the full transfer function and astrophysical systematics. 
As discussed in the main text, we therefore do not derive bounds here; instead, we \emph{indicate} where Lyman-$\alpha$ forest observations may become relevant using a one-parameter ``equivalent-$m_{\rm WDM}$'' prescription~\cite{Ballesteros:2020adh}. 
We characterize each sterile-neutrino momentum distribution by an effective thermal-relic mass $m_{\rm WDM}$, defined such that the corresponding thermal relic exhibits a comparable small-scale cutoff in the \emph{linear} matter power spectrum. 
As an illustrative reference, we use the 95\% C.L.\ thermal-relic result $m_{\rm WDM}>3.1~\mathrm{keV}$ from Ref.~\cite{Villasenor:2022aiy} and express it in terms of our spectra via Eq.~(2.26) of Ref.~\cite{Ballesteros:2020adh}, derived for fermionic warm dark matter with two degrees of freedom. 
Within this approximate construction, the thermal-relic reference corresponds to the inequality
\begin{equation}
    m_{s} < 7.56~\mathrm{keV}\left( \frac{m_{\rm WDM}}{3~\mathrm{keV}}\right)^{\!\frac{4}{3}}
    \,\frac{\langle p\rangle_{0}}{T_{0}}\,
    \frac{\sqrt{\langle q^{2}\rangle}}{\langle q\rangle}\,
    \label{eq:ly-alpha-bound}
\end{equation}
Here, $q\equiv p/T_{*}$ is the comoving momentum normalized by a fixed energy scale $T_{*}$, $\langle\cdots\rangle$ denotes averages over the momentum distribution, and $\langle p\rangle_{0}/T_{0}$ is the ratio of the mean momentum today to the CMB temperature. In practice, it is convenient to use the simplification
\begin{equation}
    \frac{\langle p\rangle_{0}}{T_{0}}\,
    \frac{\sqrt{\langle q^{2}\rangle}}{\langle q\rangle}
    \;=\;
    \frac{\sqrt{\langle p^{2}\rangle_{0}}}{T_{0}},
    \label{eq:ratio-Ly-a}
\end{equation}
since $p_{0}=q\,T_{*}$ at $a=1$.

This dependence of $\sqrt{\langle p^{2}\rangle_{0}}/T_{0}$ shown in Fig.~\ref{fig:p2-to-T} yields the characteristic shape of the Lyman-$\alpha$ domain in Fig.~\ref{fig:parameterspace}. As $L_{\alpha}$ increases (equivalently, as $\sin^{2}2\theta$ decreases), the probed $m_{s}$ become slightly larger, as spectrum gets warmer.

\clearpage
\section{Comparison with the literature}
\label{app:compare}

Our study generalizes and improves a precise approach developed by Ghiglieri and Laine~\cite{Ghiglieri:2015jua}, and Venumadhav et al.~\cite{Venumadhav:2015pla} to the arbitrary lepton asymmetries. The numerical kernels of Refs.~\cite{Ghiglieri:2015jua,Venumadhav:2015pla}, which are publicly provided as \texttt{resonance-dm} and \texttt{sterile-dm}, respectively, were designed for moderately small lepton flavor asymmetries, $|L_\alpha|\lesssim 10^{-3}$. Much larger asymmetries $L_{\alpha} \gtrsim 0.01$ would require significant modifications in the description of the dynamics of active-sterile oscillations and the Universe: 
\begin{itemize}
\item For large asymmetries, active-sterile oscillations would enter the regime where they cannot be averaged in time. 
\item The chemical potentials enter the cosmological equation of state at ${\cal O}(\mu^2/T^2)$ and modify both the expansion rate $H(T)$ and the entropy density $s(T,\mu_\alpha)$ by a very sizable amount, up to $\mathcal{O}(1)$, depending on the value of $L$. 
\end{itemize}

In our work, we generalize the semi-classical Boltzmann equation for sterile neutrinos with averaged oscillations~\cite{Ghiglieri:2015jua,Venumadhav:2015pla} to one with non-averaged oscillations, which applies to arbitrary lepton asymmetries. All thermodynamic functions entering the Boltzmann system,  including the hadronic susceptibilities required by charge neutrality, and the neutrino production rates, are also computed with the full $\mu_\alpha$-dependence.

In addition, we develop a simplified approach that quickly and accurately solves the sterile neutrino evolution using the narrow-width approximation and neglecting back-reaction from sterile neutrinos on the lepton asymmetries. The full-Boltzmann approach and the simplified approach are well cross-checked with each other. 

Let us compare the results from our code with those from Refs.~\cite{Ghiglieri:2015jua,Venumadhav:2015pla}. Ref.~\cite{Ghiglieri:2015jua} (Fig.~5) and Ref.~\cite{Venumadhav:2015pla} (Fig.~10(b)) show sterile neutrino DM spectra for $m_s=7.1~\mathrm{keV}$ and a broad range of mixing angles, $\sin^{2}\theta=(0.8$--$20)\cdot10^{-11}$. For most of these couplings, sterile neutrino production occurs during the QCD transition, $T\sim150\textit{--}300~\mathrm{MeV}$ (cf.~Eq.~\eqref{eq:Tres}).

\begin{figure*}
 \centering
    \includegraphics[width=0.45\columnwidth]{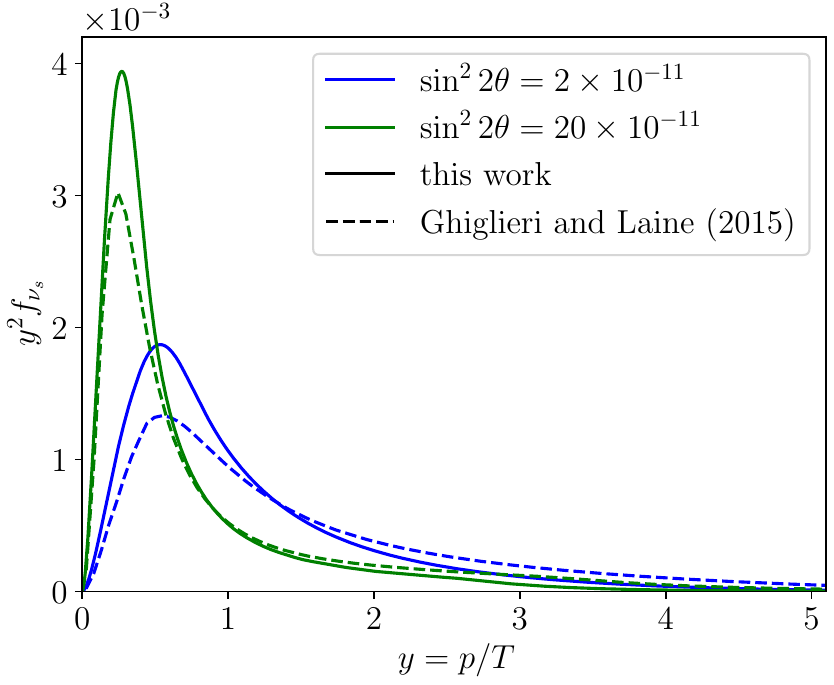}
    \includegraphics[width=0.45\columnwidth]{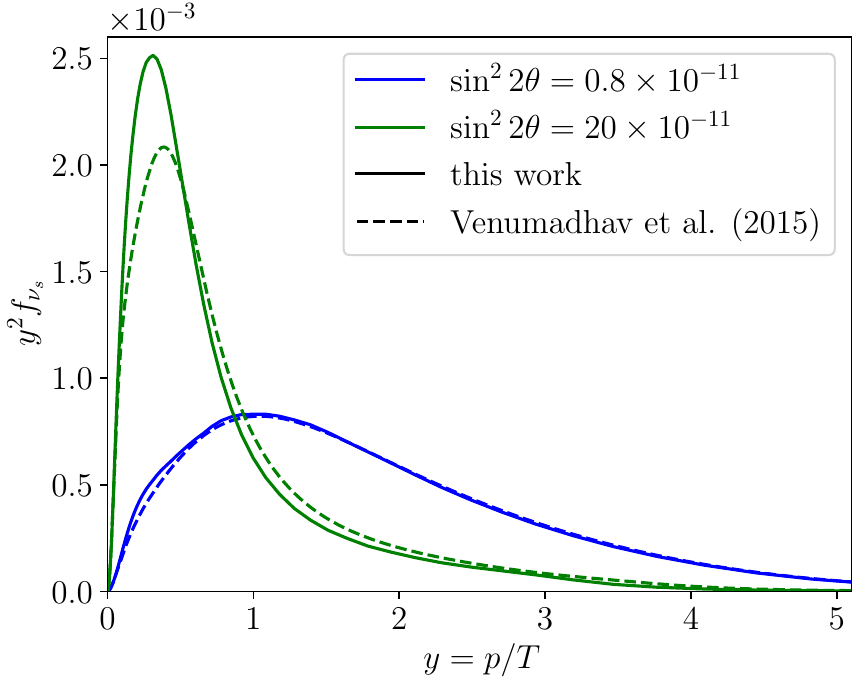}
    \vskip-6pt
    \caption{The momentum distribution of sterile neutrinos with $m_s=7.1~{\rm keV}$ in the current Universe compared to the results in Refs.~\cite{Ghiglieri:2015jua,Venumadhav:2015pla}. The case of solely $L_e\simeq (5\textit{--}8)\times 10^{-5}$ ($L_\mu\simeq(6.7\textit{--}13)\times 10^{-5}$) and the $\nu_s$ mixing with $\nu_e$ ($\nu_\mu$) are considered in the left (right) panel. The magnitude of the asymmetry is fixed for sterile neutrinos to explain all DM. Note that, in the right panel, we use the same 1000 logarithmic momentum bins of sterile neutrinos as in Ref.~\cite{Venumadhav:2015pla}, but the results have not converged numerically.}
    \label{fig:distribution_comparison}
\end{figure*}

The comparison of momentum distributions is shown in Fig.~\ref{fig:distribution_comparison}. Our results are in agreement with Refs.~\cite{Ghiglieri:2015jua,Venumadhav:2015pla} at the level of a few tens of percent. The residual discrepancies likely stem from different treatments of the QCD transition. In particular, we use the fit for the effective number of relativistic degrees of freedom from Ref.~\cite{Saikawa:2018rcs}, which incorporates lattice input. On the other hand, Refs.~\cite{Ghiglieri:2015jua,Venumadhav:2015pla} use the fit of Ref.~\cite{Laine:2006cp}, based on an extrapolation between high- and low-temperature regimes. The difference across the QCD transition is $\sim15\%$ (see the orange line in Fig.~7 of Ref.~\cite{Saikawa:2018rcs}). This interpretation is supported by the nearly excellent agreement between our results and those of Ref.~\cite{Venumadhav:2015pla} for $\sin^{2}\theta=0.8\times10^{-11}$ (the blue curves), where resonant production occurs mostly after the QCD transition.

Next, let us comment on the two studies~\cite{Gorbunov:2025nqs,Vogel:2025aut} that appeared while our work was in preparation.

Ref.~\cite{Gorbunov:2025nqs} evaluates resonant production with net zero lepton flavor asymmetries using the public \texttt{resonance-dm} package~\cite{Ghiglieri:2015jua}. It considers a single sterile neutrino mass, $m_s\simeq 7.1~\mathrm{keV}$, motivated by the $3.5~\mathrm{keV}$ line~\cite{Boyarsky:2014jta}. Providing qualitative arguments, the authors state that the net-zero lepton flavor asymmetries reopen the parameter space of sterile neutrinos. Our analysis complements and extends Ref.~\cite{Gorbunov:2025nqs} in several respects. First, the status of the $3.5~\mathrm{keV}$ line is currently disputed~\cite{Dessert:2023fen,deBlas:2025gyz}. Second, the authors restrict to $|L_{\alpha}|\lesssim \mathcal{O}(10^{-2})$, which aligns with the regime of applicability of the \texttt{resonance-dm} setup, where the impact of large asymmetries on the thermodynamics of the Universe is not included. In our study, we map the maximal sterile neutrino parameter space over a broad mass range and for large lepton flavor asymmetries, while our public framework provides the resulting spectra required for subsequent structure formation analyses. 

Ref.~\cite{Vogel:2025aut} scans a broader range of masses and mixing angles, but studies a different setup in which the initial asymmetry resides solely in the muon flavor (to weaken BBN constraints), with magnitudes $L\gtrsim 0.01$ (still smaller than the asymmetries considered in our study). Their production calculation employs \texttt{sterile-dm}~\cite{Venumadhav:2015pla}. Similarly to \texttt{resonance-dm}, the code does not incorporate large chemical potentials in the background thermodynamics and active neutrino rates. In addition, it assumes oscillations in the averaged regime, thereby missing non-averaged effects, which become crucial for the evolution of the sterile neutrinos in the presence of non-negligible asymmetries $|L|\gtrsim 0.01$. 

Another aspect of comparison with Refs.~\cite{Gorbunov:2025nqs,Vogel:2025aut} is the behavior of the sterile neutrino spectrum today as a function of the asymmetry $L = |L_{\alpha}|$ in the scenarios with constant $L$. Our results (summarized in Sec.~\ref{app:sterile-spectrum}) are that the spectrum predominantly becomes warmer with increasing $L$, which is a consequence of producing sterile neutrinos at later times and resulting reduction of post-production dilution. This agrees with~\cite{Vogel:2025aut}, which predicts a warming of the spectrum for asymmetries $L\gtrsim 10^{-4}$ (see also other studies~\cite{Kasai:2024diy,Kasai:2025xaw}). However, it differs from Ref.~\cite{Gorbunov:2025nqs}, which predicts that an increase in $L$ is associated with a colder spectrum, in the context of structure formation bounds.

To this point, we note that Ref.~\cite{Gorbunov:2025nqs} motivates the ``larger-$L$-colder-spectrum'' trend by only discussing snapshots of the distribution at fixed temperature $T$, when the production of sterile neutrinos was still happening. Then, the resonant momentum indeed drops as $1/L$ (which may be seen by inverting Eq.~\eqref{eq:Tres}), which means that at the given $T$, produced sterile neutrinos are colder for larger $L$. However, this feature does not necessarily translate to the final sterile neutrino spectrum \emph{today} (and in particular the average momentum-to-temperature ratio), which is the cumulative effect of sterile neutrino production and post-production entropy dilution (recall Fig.~\ref{fig:p2-to-T}). Both the analytic estimates and the numeric results in~\cite{Gorbunov:2025nqs}, the fixed-$T$ snapshots, miss this effect.

To summarize, both works~\cite{Gorbunov:2025nqs,Vogel:2025aut} explore particular scenarios in which lepton flavor asymmetries may open regions of sterile neutrino parameter space; however, unlike our analysis, these do not map the \emph{maximal} viable parameter space, and each carries additional limitations, related to treating large lepton flavor asymmetries. By contrast, our study treats arbitrary sterile neutrino masses and mixing angles and accommodates larger lepton flavor asymmetries up to $|L_{\alpha}|\simeq 0.1$ with a precise and reliable framework. In doing so, it furnishes a comprehensive map of the viable parameter space for resonantly produced sterile neutrino dark matter. 

\end{document}